\documentclass[useAMS,usenatbib]{mnras}
\usepackage{graphicx}
\usepackage{enumerate}
\usepackage[version=3]{mhchem} %Formules chmiques
\usepackage{subfig}
\usepackage{lscape}
\usepackage{epstopdf}
\usepackage{amssymb}
\usepackage{times}

\usepackage{ulem}
\newcommand{\St}{\mathrm{St}}

\defcitealias{2017MNRAS.469..237P}{Paper~I}

\title[Sorting fragmenting dust in discs]{Size and density sorting of dust grains in SPH simulations of protoplanetary discs II: Fragmentation
}
\author[Pignatale et al.]{F. C. Pignatale$^{1,2,3}$\thanks{E-mail:
pignatale@ipgp.fr}, J.-F. Gonzalez$^{3}$, Bernard Bourdon$^{4}$, Caroline Fitoussi$^{4}$ \\
$^{1}$
Mus\'eum national d'Histoire naturelle, UMR 7590, CP52, 57 rue Cuvier, 75005, Paris France\\
$^{2}$
{Institut de Physique du Globe de Paris (IPGP), Univ Paris Diderot, CNRS, 1 rue Jussieu, 75005, Paris, France}\\
$^{3}$ 
Univ Lyon, Univ Claude Bernard Lyon1, Ens de Lyon, CNRS, Centre de Recherche Astrophysique de Lyon UMR5574, F-69230, Saint-Genis-Laval, France\\
$^{4}$ 
Univ Lyon, Univ Claude Bernard Lyon1, Ens de Lyon, CNRS, UMR 5276 LGL-TPE, F-69342, Lyon, France \\
}

\begin{document}

\date{Accepted 4 October 2019; Received 20 September 2019; in original form 30 October 2018}

\pagerange{\pageref{firstpage}--\pageref{lastpage}} \pubyear{2019}

\maketitle

\label{firstpage}

\begin{abstract}

Grain growth and fragmentation are important processes in building up large dust aggregates in protoplanetary discs. Using a 3D two-phase (gas-dust) SPH code, we investigate the combined effects of growth and fragmentation of a multi-phase dust with different fragmentation thresholds in a time-evolving disc. We find that our fiducial disc, initially in a fragmentation regime, moves toward a pure-growth regime in a few thousands years. Timescales change as a function of the disc and dust properties.  When fragmentation is efficient, it produces, in different zones of the disc, Fe/Si and rock/ice ratios different from those predicted when only pure growth is considered. Chemical fractionation and the depletion/enrichment in iron observed in some chondrites  can be linked to the size-density sorting and fragmentation properties of precursor dusty grains . We suggest that aggregation of chondritic components could have occurred where/when fragmentation was not efficient if their aerodynamical sorting has to be preserved.  Chondritic components  would allow aerodynamical sorting in a fragmentation regime only if they have similar fragmentation properties. We find that, in the inner disc, and for the same interval of time, fragmenting dust can grow larger when compared to the size of grains predicted by pure growth. This counter-intuitive behaviour is due to  the large amount of dust which piles up in a fragmenting zone followed by the rapid growth that occurs when this zone transitions to a pure growth regime. As an important consequence, dust can overcome the radial-drift barrier within a few thousands years.

\end{abstract}


%%%%%%%%%%%%%%%%%%%%%%%%%%%%%%%%%%%%%%%%%%%%%%%%%%%%

\section{Introduction}
\label{intro}
In protoplanetary discs, collisions among dust grains can lead to sticking and, thus, to the formation of larger aggregates. However, if the relative velocity between  colliding particles becomes greater than a critical velocity, a collision would likely result in fragmentation \citep{2008ARA&A..46...21B}.

Dust growth and fragmentation are competitive and complex processes. They are regulated by the interactions between dust grains under the effect of the gas drag in discs, as well as e.g. by thermal processes like sublimation and condensation. The dust-gas interactions have been investigated in great detail in numerous studies such as \citet{1977MNRAS.180...57W,2004A&A...421.1075D,2005A&A...443..185B,2006mess.book..353C,2012MNRAS.420.2345L,2012A&A...537A..61L,2015MNRAS.452.3932B}. 

Several theoretical studies focused on the effects of grain growth and fragmentation in determining the dust behaviour and its distribution in protoplanetary discs \citep{2005A&A...434..971D,2008A&A...487..265L,2008A&A...480..859B,2008A&A...491..663D,2014MNRAS.437.3025L,2014MNRAS.437.3055L,2015P&SS..116...48G}. On the other hand, there are several experimental and numerical studies which aimed at understanding the different physical properties and thresholds of fragmentation between dust grains and between dust aggregates \citep{2008ARA&A..46...21B,2009ApJ...702.1490W,2009MNRAS.393.1584T,
2010A&A...513A..57Z,2013A&A...559A..62W,2013MNRAS.435.2371M,2014ApJ...783L..36Y}. 
There are several factors that determine the critical velocity of  a dust grain/aggregate such as its chemical composition,  porosity  and size \citep{2008ARA&A..46...21B}. However, to date, there is no unified theoretical framework which can be used to extrapolate unknown fragmentation properties of aggregates of different size and species. Nevertheless, there is a large spread of values for fragmentation thresholds for different types of  dusty grains that range from 1~$\rm{m\,s^{-1}}$ to 50~$\rm{m\,s^{-1}}$  \citep{2008ARA&A..46...21B,2009ApJ...702.1490W,2009MNRAS.393.1584T,
2010A&A...513A..57Z,2013A&A...559A..62W,2013MNRAS.435.2371M,2014ApJ...783L..36Y,
2016MNRAS.456.4328D}.

Fragmentation can constitute a crucial problem in the framework of planet formation. The so-called radial-drift barrier describes the process for which grains cannot grow large enough to decouple from the gas before the gas drag-induced drifts makes them fall onto the central star \citep{1977MNRAS.180...57W}. Fragmentation may thus damp the growth, preventing grains from becoming large enough to overcome the radial drift barrier \citep{2008A&A...480..859B,2008A&A...486..597J} making the process of planet formation inefficient. However, more recent studies have shown that, even in  high fragmentation regimes, dust particles can pile up, grow and decouple from the gas \citep[]{2015MNRAS.454L..36G,2017MNRAS.467.1984G,2017MNRAS.472.1162G}. This is due to the key role of the dust back-reaction (i.e.\ the drag of dust on gas) and its role in the formation of self-induced dust trap. Other  proposed mechanisms to solve the radial-drift barrier include photophoresis \citep{2010Icar..208..482W,2013ApJ...769...78W,2016MNRAS.458.2140C}, dead zones  \citep{2007ApJ...664L..55K,2011ARA&A..49..195A}, embedded protoplanets and particle traps \citep{2004A&A...425L...9P,2007A&A...474.1037F,2010A&A...518A..16F,
2012A&A...538A.114P,2012A&A...547A..58G,2015P&SS..116...48G}, evaporation fronts \citep{2008A&A...487L...1B}, streaming instability \citep{2007Natur.448.1022J}, high dust to gas ratios \citep{2014MNRAS.437.3037L}, radial mixing \citep{2004A&A...415.1177K}, meridional circulation \citep{2011A&A...534A.107F}, radiation pressure \citep{2014A&A...566A.117V}, or grain porosity \citep{2013A&A...557L...4K}.

In \citet{2017MNRAS.469..237P}, hereafter Paper~I, we studied the size-density sorting of
the multi-phase dust component of a protoplanetary disc in a pure-growth regime. We found that i) dust grains can be efficiently size-density sorted\footnote{Two co-located grains $i$ and $j$ with size $s_i$, $s_j$ and density $\rho_i$, $\rho_j$ are size-density sorted if they have the same aerodynamic parameter ($\zeta=s\rho$), i.e. $s_i\rho_i=s_j\rho_j$.}, and ii) changes in the chemical composition in different  disc zones can be driven by the combined effects of intrinsic density and size of the dust grains on their dynamics. We also found that the properties of the dust aggregates were in good agreement with the physical properties of chondrites  (chemically fractionated with their components size-density sorted) \citep{2003TrGeo...1..143S}. In fact, since the interactions  between dust and gas are driven by the  aerodynamic parameter, $\zeta=s\rho$, of the dust grains \citep{2006mess.book..353C}, the  dust motions  (vertical settling and radial drift) in discs act as sorting mechanisms. Sorting is considered an efficient process to fractionate the dust \citep{1998LPI....29.1457B,1999Icar..141...96K,2005M&PS...40..123L,2012Icar..220..162J} and  also the gas from the solar composition,  if total or partial separation between the two phases occurs  \citep{2016MNRAS.457.1359P}.

In \citetalias{2017MNRAS.469..237P} we focused our attention on a large T-Tauri disc ($20< R(\rm au)<400$) where the disc structure does not result in high relative velocities  between dust particles. In that case, a pure-growth regime is a very good approximation to describe the process of grain growth. This work is the direct follow-up of \citetalias{2017MNRAS.469..237P}. Here we focus on the effect of fragmentation in determining the dust content and properties in the inner region of discs. This study is driven by the fact that, in these zones, the relative velocities between particles could easily reach high values (see equation~\ref{vrelvrel}) and, thus, lead to an efficient dust fragmentation. Moreover, we  extend the  conclusions derived in \citetalias{2017MNRAS.469..237P}, and investigate if and how fragmentation can change the size, the chemical distribution  and the aerodynamic sorting of  dust grains.

The structure of the paper is as follows: in Section~\ref{method} we describe the code we use for our simulations, introduce the disc model, the chemical characterization of the dust and its growth and fragmentation properties. In Section~\ref{results} we present the results produced by our simulations comparing pure growth and fragmentation, and in Section~\ref{discussion} we detail the resulting changes in the chemical disc composition, and the properties of the dust grains and aggregates. We draw our conclusion in Section~\ref{conclusions}.

\section{Methods}
\label{method}

We compute the vertical settling, radial drift, growth and fragmentation of a multi-phase dust using our 3D two-phase (gas + dust) Smoothed Particle Hydrodynamic (SPH) code. The code with all the cited implementations has been described, tested and  discussed in detail in \citet{2005A&A...443..185B} where the two-phase (gas and dust, including the dust back-reaction on the gas) code was first presented, \citet{2008A&A...487..265L} where dust growth was included, \citet{2015P&SS..116...48G}, where fragmentation was implemented, \citetalias{2017MNRAS.469..237P} where a  chemical characterisation of grains was added, and \citet{2017MNRAS.467.1984G} where further studies of dust growth and fragmentation were performed.  \citet{2013MNRAS.433...98A} have shown that the SPH formalism naturally reproduces the expected properties of Prandtl-like turbulence. The code uses the standard SPH artificial viscosity \citep{1989JCoPh..82....1M}, and in this paper we set $\alpha_\mathrm{SPH}=0.1$ and $\beta_\mathrm{SPH}=0.0$, emulating a uniform \citet{1973A&A....24..337S} turbulence parameter $\alpha_\mathrm{SS}=0.01$ \citep[see][for a discussion]{2007A&A...474.1037F}.
No further modifications on the code are made in this paper.

\subsection{Disc Model}
\label{Disc}
We take into consideration an  inner slice ($1.87\le R\rm{(au)} \le 50$) of a typical T-Tauri disc with $M_{\rm disc}=0.02~M_{\star}$ and a radial extension of $0.5\le R\rm{(au)} \le 400$, orbiting a star of mass $M_{\star}=1~M_{\odot}$, for which we recalculate, using power-law parametrizations \citep{2005A&A...443..185B}, all the necessary quantities for the selected region. Our reference radius is $R_{0}=1$~au, where the temperature is set at $T=243.5$~K \citep{Dalessio1998,1999ApJ...527..893D}. The parametrization for the temperature and surface density follows the same power laws as \citetalias{2017MNRAS.469..237P} ($T\propto R^{-3/4}$, $\Sigma\propto R^{-3/2}$), the disc is vertically isothermal, the vertical scale height is $H = c_s/\Omega_K$ \citep[\citetalias{2017MNRAS.469..237P}]{2005A&A...443..185B,2008A&A...487..265L}. At the reference radius, $\Sigma_0=4182.25$~kg$\rm{m^{-2}}$ with a disc total mass of $M_{\rm disc}=0.00374~M_{\odot}$ within the 0.5-50~au radial range. The disc is composed of 99$\%$  gas and 1$\%$  dust by mass. The disc is flared and $H/R =0.031$ at $R_{0}$. The choice of  $R=1.87$~au as the inner limit is made because this is the location of the water snowline where $T\sim150$~K and $P\sim10^{-2}$~pascal \citep{Lewis440}. This will allow us to  ignore the effects of the evaporation of ice on the disc and on the dust chemistry.
 
\subsection{Grain species, growth and fragmentation}
\label{chemistry}

We report in Table~\ref{frac-abundance2} our fiducial  dust compositions with the relative intrinsic densities, $\rho_{\rm d}$. Abundances are adapted from \citet{2003ApJ...591.1220L}.

\begin{table*}							
\begin{tabular} {l| c| c|c|c| c| c|}	
\hline						
Symbol	&	Dust species	&	$\rho_{\rm d}$  &	 $v_{\rm frag}$ & Nominal Abundance	&  Real abundances 	&  \\
	&		& ($\rm{g\,cm^{-3}}$)	 &	 ($\rm{m\,s^{-1}}$) & 	$(\%)$	 & (particles number) & \\
\hline	
\hline	
Fe	&	wustite	&	5.74	& 35 &	6.25 &	7365 & \\
	&	sulfides		&	4.55	& 42 &	6.25 & 7593& \\					
\hline	
Si	&	silicates 	&	3.2	&  36 &	20.4 & 24971 & \\
\hline	
ice	&	ice 	&	1	&  56 	&	67.1 & 85071& \\
\hline	
np$_{dust}$ & & & & & 125000 & \\
\hline	
Fe/Si	&	&		&		& 	0.61 & 0.599  & \\
rock/\ce{H2O}&	 	&		&		& 0.49&  0.469 &   \\
\hline	
															
\end{tabular}							
\caption{ Initial dust composition for our fiducial disc. In the first column we report the symbols used for our considered species, that are listed in the second column. In the third, fourth and fifth columns we  report their  density,  fragmentation velocity (see Section~\ref{chemistry}) and nominal abundances. In the sixth column we report the  number of dust particles at the time of the dust injection. We also report the resulting nominal ``solar'' ratios for Fe to Si and rock (Fe+Si) to \ce{H2O} and the  ratios determined by the dust injection.  \label{frac-abundance2}} 
\end{table*}

Grain growth has been implemented in our SPH code  in \citet{2008A&A...487..265L} where they followed the prescription derived by \citet{1997A&A...319.1007S}. The resulting growth profiles has been extensively discussed in previous work \citep[\citetalias{2017MNRAS.469..237P}]{1997A&A...319.1007S,2008A&A...487..265L,2015P&SS..116...48G,2017MNRAS.467.1984G}. To summarize, grain dynamics is regulated  by the dust stopping time, $t_{\rm s}$,
\begin{equation}
t_{\rm s} = \frac{s_{\rm d} \rho_{\rm d}}{c_{\rm s} \rho_{\rm g}}= \frac{\zeta}{c_{\rm s} \rho_{\rm g}} ,\
\label{stoppingtime}
\end{equation}
where $c_{\rm s}$ is the sound speed, $\rho_{\rm g}$ is the gas density,  $\rho_{\rm d}$ and $s_{\rm d}$  are respectively the intrinsic density and size of the dust particle, and $\zeta=s_{\rm d}\rho_{\rm d}$  is the aerodynamic parameter. Moreover, growth is also a function of the relative velocities between particles, $v_{\rm rel}$, and of the dust total density \citep[]{1997A&A...319.1007S,2008A&A...487..265L,2017MNRAS.467.1984G}.
Fragmentation has been implemented in our code and discussed in  \citet{2015P&SS..116...48G,2017MNRAS.467.1984G}.  They defined a threshold velocity, $v_{\rm frag}$, which is compared to  $v_{\rm rel}$. If  $v_{\rm rel}<v_{\rm frag}$ grains grow, while if $v_{\rm rel}>v_{\rm frag}$, grains fragment  with fragmentation modelled as a negative growth. $v_\mathrm {rel}$ depends on the Stokes number
\begin{equation}
\St = \Omega_\mathrm{K} t_\mathrm{s},
\label{stokesnumber}
\end{equation}
where $\Omega_\mathrm{K}$ is the Keplerian frequency, as

%At given disc location and condition (such as $\Omega,\alpha,c_s$), $v_{\rm rel}$ is proportional to the Schmidt number, Sc, \citep{1997A&A...319.1007S}. Sc can be approximated to  $Sc=1+St$  \citep{2008A&A...487..265L} where St is the Stokes number:
%\begin{equation}
%St = \frac{\rho_{\rm d} s_{\rm d}}{\Sigma_{g}}\frac{\pi}{2}  \,.
%\label{stokesnumber}
%\end{equation}
%
%From the definition of  $v_{\rm rel}$ in \citet{1997A&A...319.1007S} and  Sc, it is found that
\begin{equation}
v_{\rm rel} \propto \frac{\sqrt{\St}}{1+\St}
\label{vrelvrel}
\end{equation}
\citep[see][]{2017MNRAS.467.1984G}. It is thus possible to study the evolution of $v_{\rm rel}$ as a function of the particle size, as the Stokes number is proportional to $\zeta$ and thus (when $\rho_\mathrm{d}$ is fixed) to the particle size.

As  pointed out in Section~\ref{intro}, to date, there is no  theoretical derivation which can summarize all the aspects involved in dust fragmentation. As such, in order to find characteristic  values for  $v_{\rm frag}$ for the chosen species, we follow the approach of \citet{2014ApJ...783L..36Y}. For the sake of  simplicity we assume compact spherical grains for which the fragmentation velocity  is a function only of the chemical composition and it does not depend on the grain size and shape. This approach has already been discussed and justified in  \citet{2015P&SS..116...48G} and \citetalias{2017MNRAS.469..237P}.  \citet{2014ApJ...783L..36Y} define  $v_{\rm frag}$ using the following expression:
\begin{equation}
v_{\rm frag}= k \sqrt{E_{\rm br}/m},
\label{frag1}
\end{equation}
where $k$ is a numerical factor set at 15 for collision between particles with the same mass and size. This is our case, since in our SPH code, collisions occur between particles with the same physical properties. $E_{\rm br}$ is the energy which is needed to separate two grains which are in contact at equilibrium, and $m$ is the mass of the grain.  $E_{\rm br}$ is defined as
\begin{equation}
E_{\rm br} = 23 [\gamma^{5}s_{\rm d}^{4}(1-\nu^{2})^2/\xi^{2}]^{1/3},
\label{frag2}
\end{equation}
where $s_{\rm d}$ is the size of the grain, $\xi$ the Young modulus, $\nu$ is the Poisson's ratio, and $\gamma$ the surface energy which is defined as
\begin{equation}
\gamma \sim 0.1244T_{m}^{1.015},
\label{frag3}
\end{equation}
where $T_{m}$ is the melting temperature. In case the value of the Young modulus for a given species is unknown, the value of $\xi$ can be determined by
\begin{equation}
1/\xi = (1/3)[(1/3K)+(1/\mu)],
\label{frag4}
\end{equation}
where $K$ is the bulk modulus and $\mu$ the shear modulus \citep{2014ApJ...783L..36Y}.

%84.23-pag 191,0.36

For silicates and ice we take the velocity of fragmentation derived by \citet{2014ApJ...783L..36Y}, respectively 36~$\rm{m\,s^{-1}}$ and 56~$\rm{m\,s^{-1}}$. For \ce{FeO} (wustite) we take the values of the  Young modulus ($\xi=130$~GPa) and Poisson ratio ($\nu=0.36$) from \citet{krzyzanowski2010oxide}, while the melting temperature ($T=1650$~K) is taken from \citet{patnaik2003handbook}. These values return a $v_{\rm frag}$ of 35 $\rm{m s^{-1}}$. To represent sulfides we choose pyrrhotite (\ce{Fe$_{(1-x)}$S}), with $0<x<0.2$) for which values of the bulk modulus ($K=53.8$~GPa), shear modulus ($\mu=34.7$~GPa) and Poisson ratio ($\nu=0.23$) are available in \citet{mavko2009rock}. The melting temperature ($T=1461$~K) is taken from \citet{rolls1972melt}. These values allow to calculate the Young modulus and then $v_{\rm frag}$ which equals 42~$\rm{m\,s^{-1}}$.
A summary of the $v_{\rm frag}$ values is reported in Table~\ref{frac-abundance2}. In Section~\ref{caveat} we will discuss the limitations of these determinations.

\subsection{Simulations}
\label{simulations}

We run two simulations each with  250,000 SPH particles, for a total time of $t=3000$~yr. The  first simulation is characterized by pure-growth (G), while in the second simulation we allow grains to grow and fragment (GF). The gas disc (125,000 SPH particles) is relaxed for $t\sim165$~yr, that is $\sim1000$ times the keplerian timescales at the reference radius, 1~au.  Similarly to \citetalias{2017MNRAS.469..237P} the dust particles are then injected on top of the gas particles \citep{2005A&A...443..185B}. Particle intrinsic density is assigned following the abundances reported in Table~\ref{frac-abundance2}\, which also shows the effective particle distribution between species.

In Appendix~\ref{resolutiontest}, we  test our results against higher resolution (400,000 SPH particles) simulations and also verify the solution criterion $h<c_{s}t_{s}$,  where $h$ is the smoothing length, proposed by \citet{2012MNRAS.420.2345L}. We demonstrate that resolution does not affect the results and that the resolution criterion is satisfied for 250,000 particles.

\section{Results}
\label{results}

\subsection{Global evolution}
\label{globalevol}

\begin{figure*}
{\includegraphics[width=1.5\columnwidth]{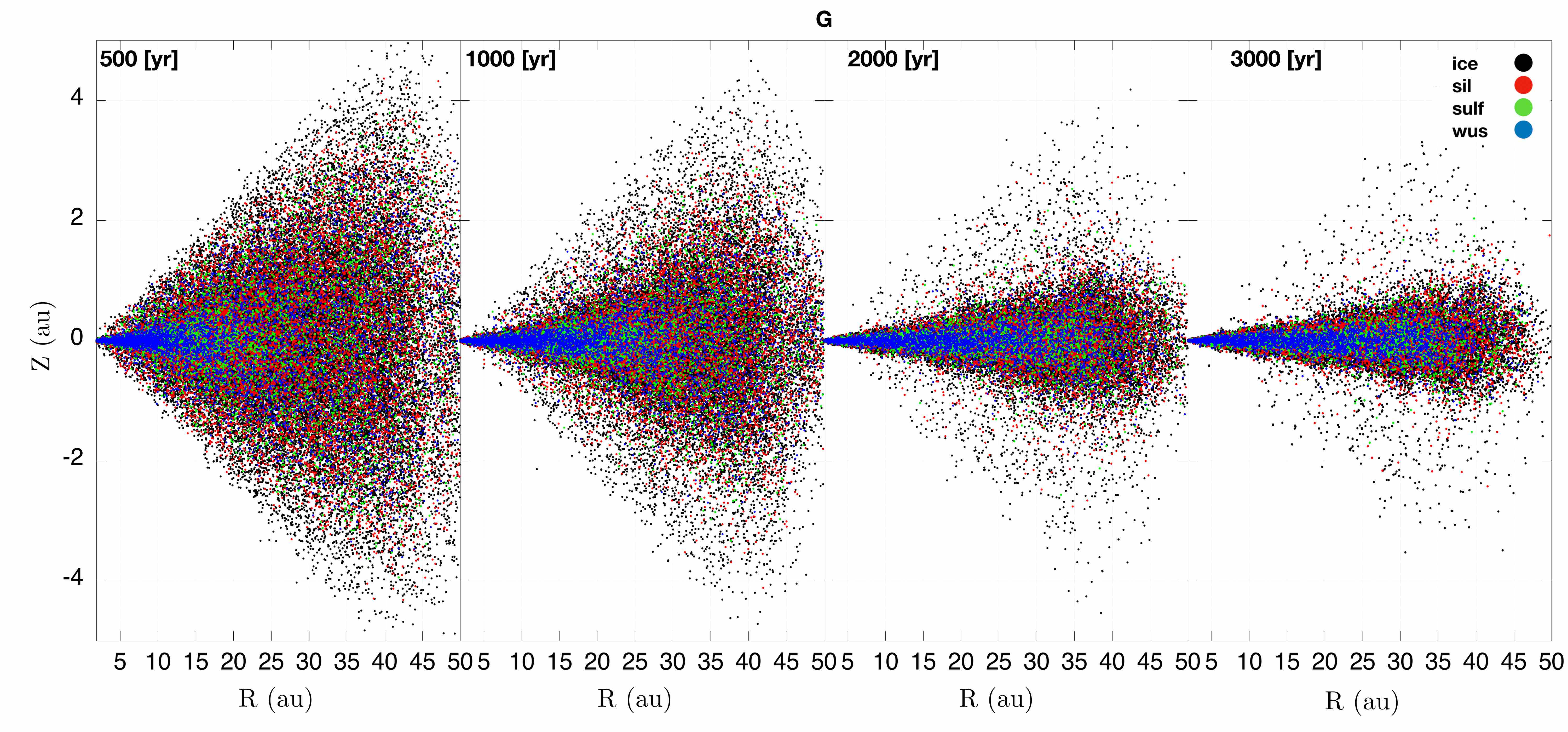}} \\
{\includegraphics[width=1.5\columnwidth]{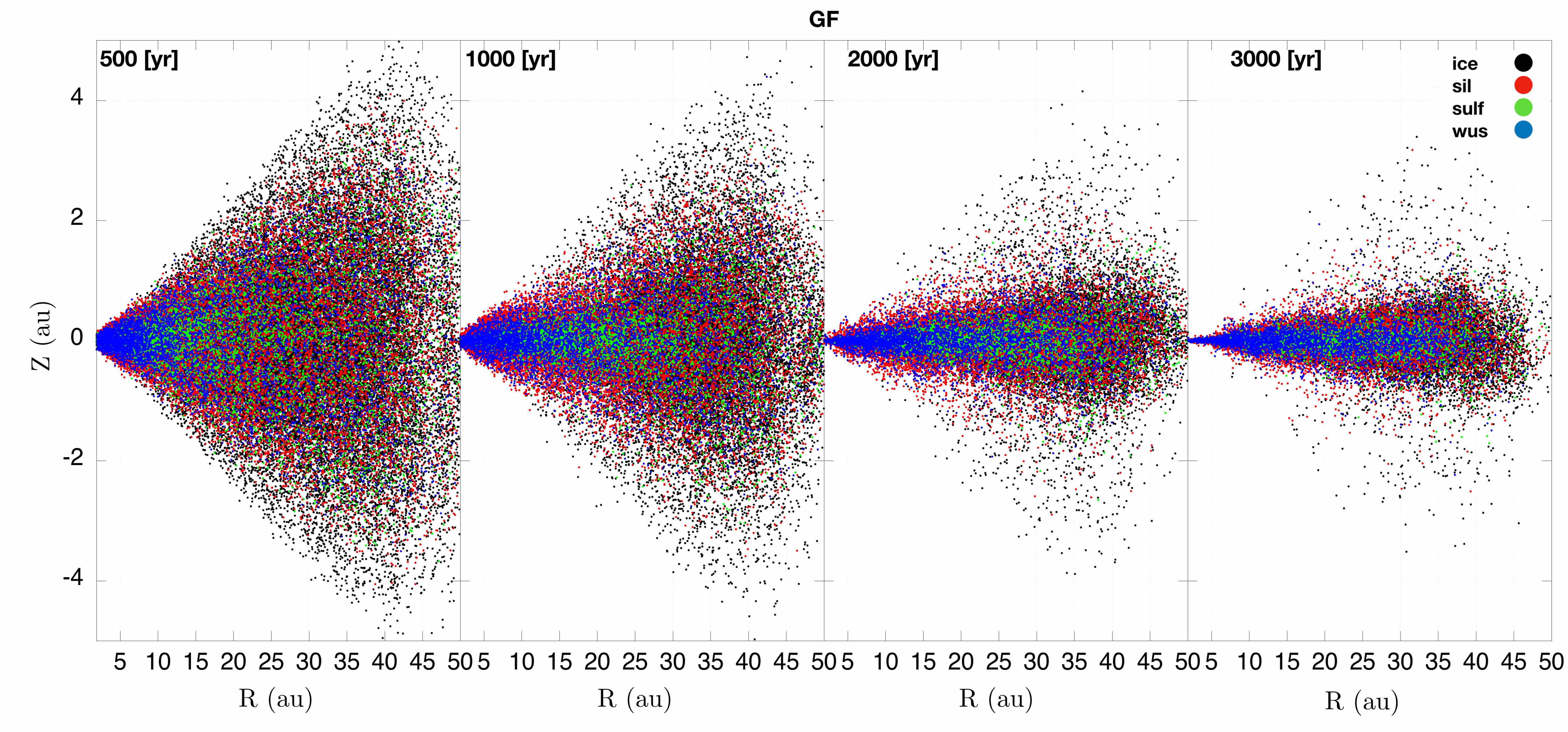}} 
%{\includegraphics[width=1.0\columnwidth]{R-Z-growth.png}}
%{\includegraphics[width=1.0\columnwidth]{R-Z-frag.png}} \\
\caption{Dust distribution in the disc, in the $(R,Z)$ plane,
for different chemical species for growth, G, (top panel) and growth+fragmentation, GF, (bottom panel). Species are superimposed for ease of reading. Pure-growth and fragmentation lead to a different dust distribution, expecially in the inner zone of the disc. \label{fig1}}
\end{figure*}

Figure~\ref{fig1} shows the dust distribution in the disc, in the $(R,Z)$ plane, at four evolutionary times, 500, 1000, 2000, 3000~yr, for G (left box) and the GF (right box) simulations. Different colors represent different chemical species. Chemical species are superimposed (wustite on top of sulfides on top of silicates on top of ice) for ease of visualization. In the G case we find a behaviour similar to \citetalias{2017MNRAS.469..237P}: (i) when grains are small, the rate at which particles vertically settle is regulated by their intrinsic densities with heavier grains experiencing a faster settling compared to the lighter grains; (ii) at later stages, the evolving size takes over in dictating the dust dynamics, but (iii)  particles with different densities still have different radial drift timescales.

\begin{figure}
{\includegraphics[width=1.0\columnwidth]{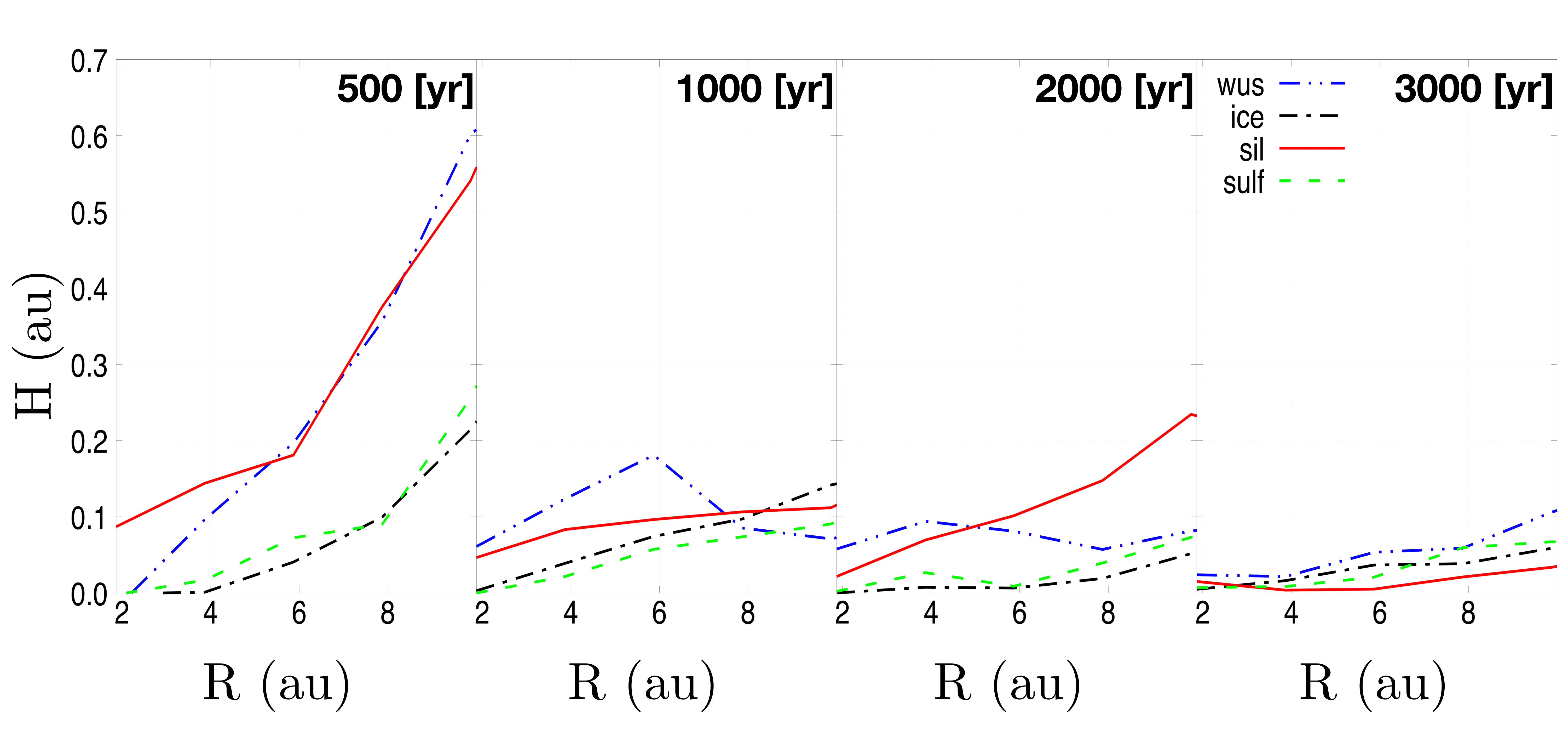}}
%{\includegraphics[width=1.0\columnwidth]{R-Z-ice-10.png}}
\caption{Scale height $H$ for different species at the selected time. Ice and sulfides settle more efficently when compared with wustite and silicates. This is due to the different fragmentation thresholds that make the growth and, thus, the settling of wustite and silicate less efficient. This behaviour is opposite to that found when only pure growth is taken into account. \label{fig2}}
\end{figure}

In the GF case the behaviour is different. In the inner part of the disc where $1.87\le R{\rm (au)} \le 10$, ice particles settle quicker than other heavier particles.  For ease of viewing,  we show in Fig.~\ref{fig2} the scale height $H$ of the different species in the inner 10~au. The disc surface becomes ice-poor. This is the opposite of what we find in the  G case, where the disc surface appears ice-enriched  until later stages.  This is due to the inefficient settling of wustite and silicates in the GF case because fragmentation keeps these grains from growing larger/settling faster. The outer part of the disc, $R>10$~au,  shows a trend which is similar in both cases: a layered structure, already found in \citetalias{2017MNRAS.469..237P}, with iron-enriched midplane and ice-enriched  surface. 

\begin{figure*}
{\includegraphics[width=1.5\columnwidth]{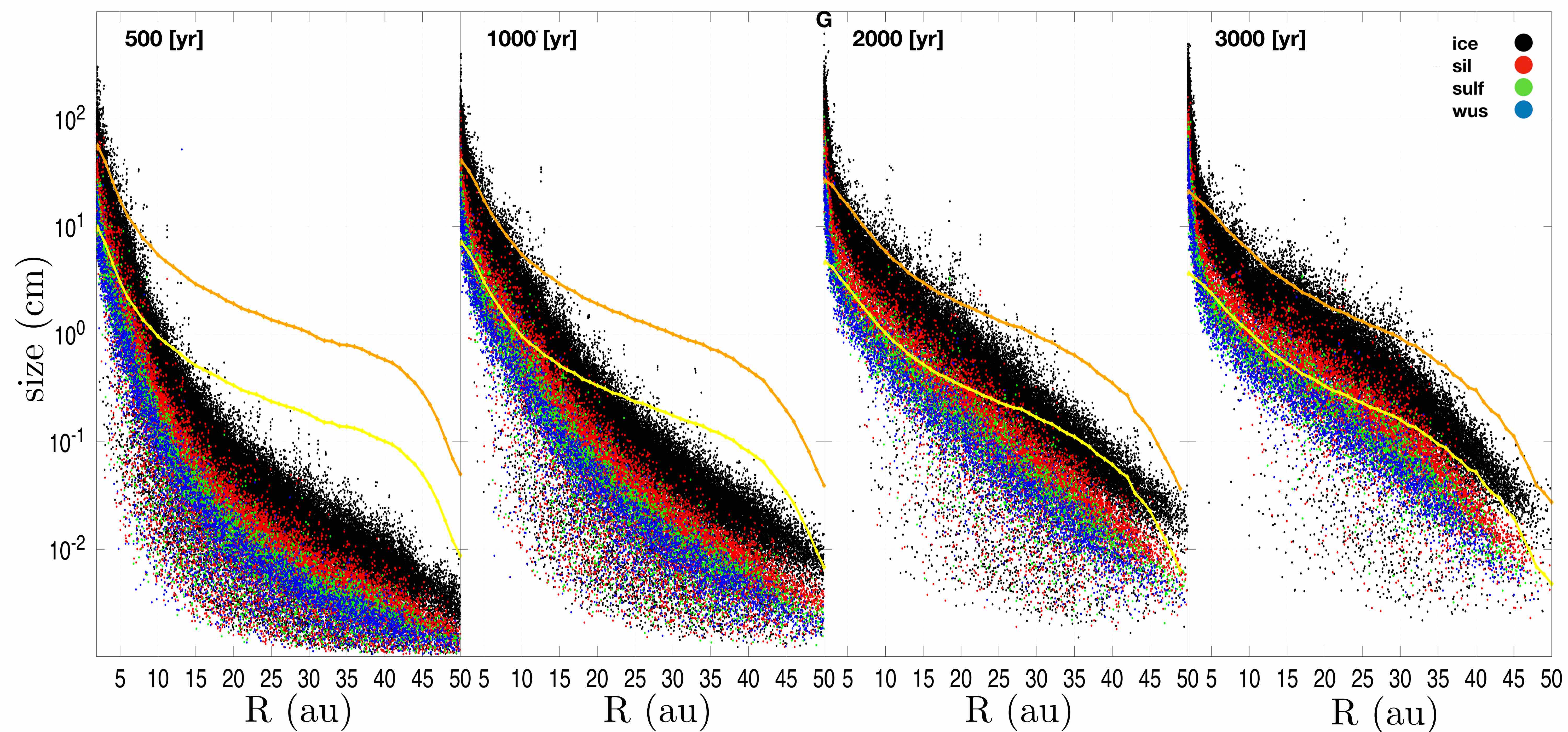}}\\
{\includegraphics[width=1.5\columnwidth]{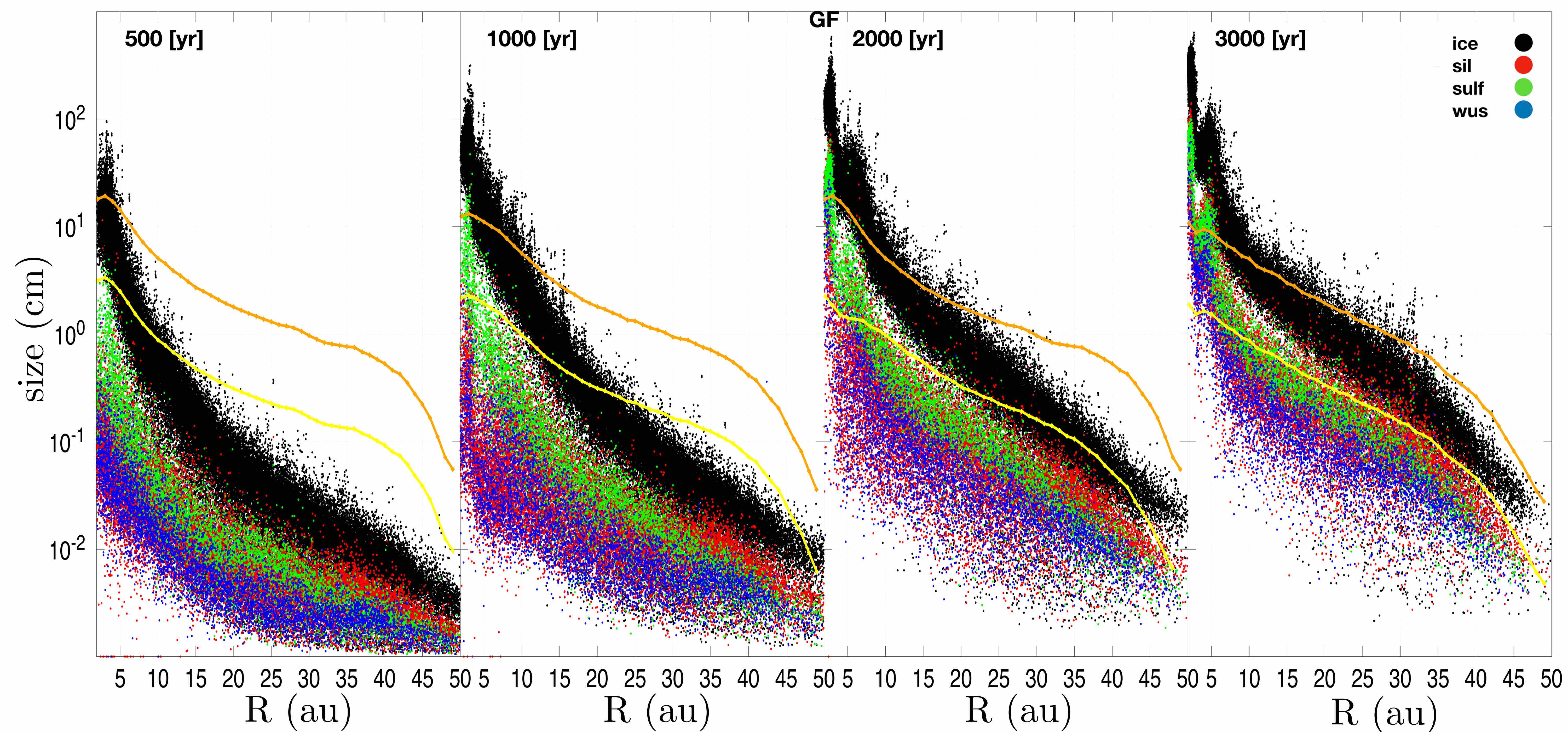}}
%{\includegraphics[width=1.0\columnwidth]{R-size-growth.png}}
%{\includegraphics[width=1.0\columnwidth]{R-size-frag.png}} \\
\caption{Size distribution in the disc as a function of radius, for different chemical species in cases G (top panel) and GF (bottom panel). Species are superimposed for easy comparison. Dust species grow at different rates according to their density and fragmentation regime. The orange and yellow lines represent the optimal drift size in the midplane for ice, $s_{\rm opt}^{\rm ice}$, and wustite, $s_{\rm opt}^{\rm wus}$, respectively, calculated with equation~\ref{optimalone2}. \label{fig3}}
\end{figure*}

In Fig.~\ref{fig3} we plot the time evolution of the grain size for the same evolutionary stages reported in Fig.~\ref{fig1}. The orange and yellow lines represent  the optimal drift size (the particle size at which the drift is most efficient, i.e.\ for which $\St=1$) in the midplane for ice, $s_{\rm opt}^{\rm ice}$, and wustite, $s_{\rm opt}^{\rm wus}$, respectively, calculated using the following equation:
\begin{equation}
s_{\rm opt}^{\rm i} =  \frac{\Sigma_{\rm g}}{\sqrt{2\pi} \rho_{\rm i}} \,,
\label{optimalone2}
\end{equation}
\citep{2007A&A...474.1037F,2008A&A...487..265L}.   $s_{\rm opt}^{\rm sulf}$ for sulfides  and  $s_{\rm opt}^{\rm sil}$ for silicates lay in between as they are functions of the intrinsic density,  $\rho_{\rm i}$.

The profile of the size distribution in the G case shows a trend  which is similar to our previous results, as the resulting size distribution is a function of the different intrinsic densities of the dust species, with an average size decreasing from ice to silicates, sulfides and wustite. In the GF case we see that sulfides grow larger than silicates in the inner regions. Moreover, in the disc zone where $5\le R{\rm (au)} \le 10$, the size of the ice and sulfide particles is also larger when compared to the respective values which result from the pure-growth simulation. This behaviour, which at first glance can seem counter-intuitive, will be explained in detail in Section~\ref{discussion}. In the outer part of the disc the resulting growth profile is not so different when both cases are compared.

\begin{figure*}
{\includegraphics[width=1.5\columnwidth]{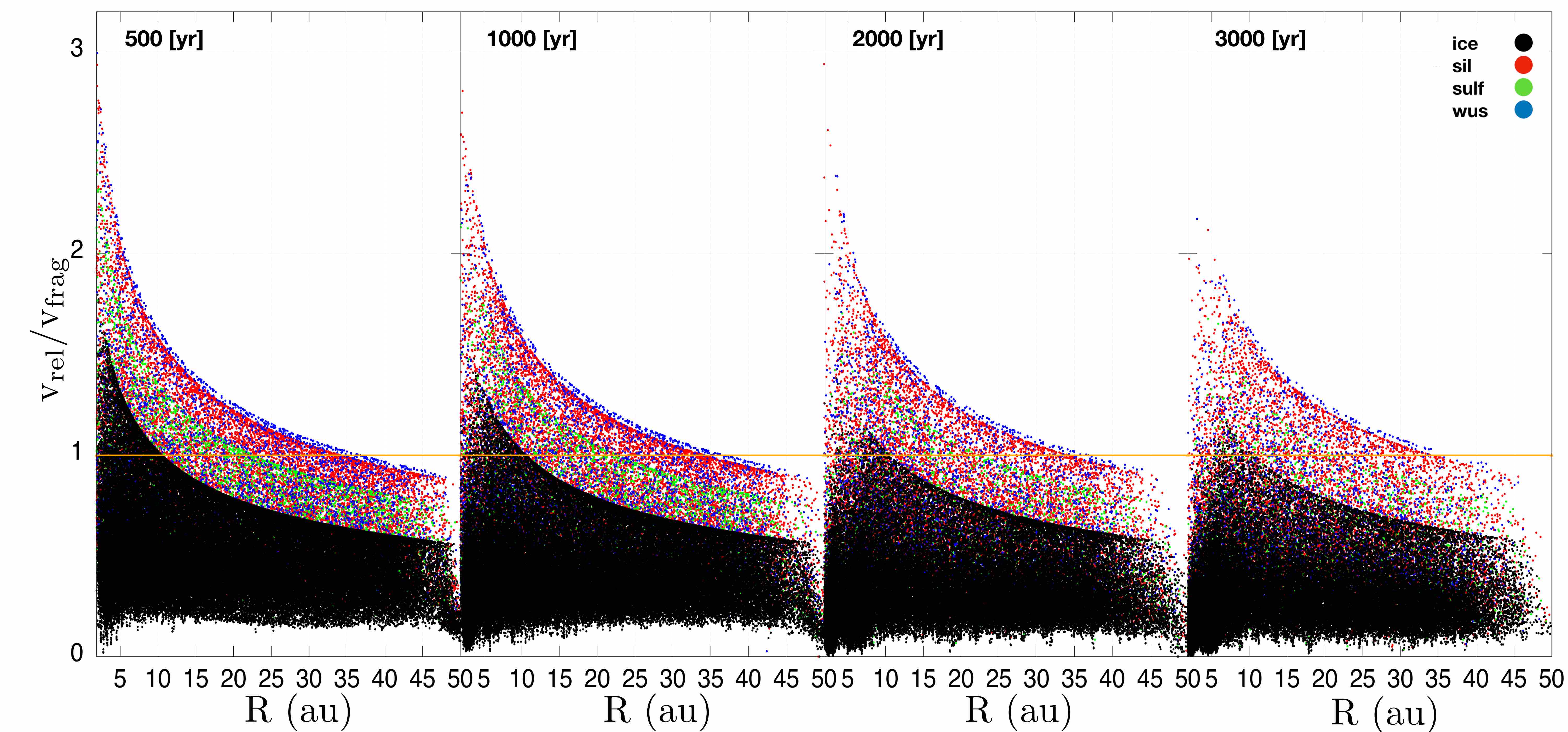}} \\
%{\includegraphics[width=1.5\columnwidth]{R-vfrag.png}} \\
\caption{Ratio of the relative velocity $v_{\rm rel}$ to the fragmentation threshold $v_{\rm frag}$ for different species at different evolutionary times in the GF simulation. While other species have particles over the fragmentation threshold, ice tends to be closer to the pure growth regime after a few thousand years. \label{fig4}}
\end{figure*}

Moreover, growth and fragmentation affect the radial drift in different ways. It can be seen that, in the  G case, silicate and wustite particles reach their optimal drift size in $t\sim1000$~yr in the inner disc and within 2000-3000~yr in the outer disc. In the  GF case, these timescales increase.

In Fig.~\ref{fig4} we plot the ratio between the relative velocity of particles, $v_{\rm rel}$, and their respective fragmentation velocities, $v_{\rm frag}$, (see Table~\ref{frac-abundance2}). At early stages of the simulation, the fragmentation threshold is reached for all species up to large distances (up to $R\sim35$~au). In the inner disc we have $v_{\rm rel}/v_{\rm frag}$ values that can reach over twice the threshold. However, in the case of ice grains, fragmentation occurs within $R\sim15$~au and only for the first $t\sim2000$~yr. After this time, ice can be considered in the regime of pure growth while other species keep experiencing fragmentation. Beyond $R\sim35$~au, the disc is basically in a pure-growth regime as all the $v_{\rm rel}$ are under their respective fragmentation velocities. This can explain the similarity of the results between the G and GF simulations in the outer disc. 

\begin{figure}
{\includegraphics[width=1.\columnwidth]{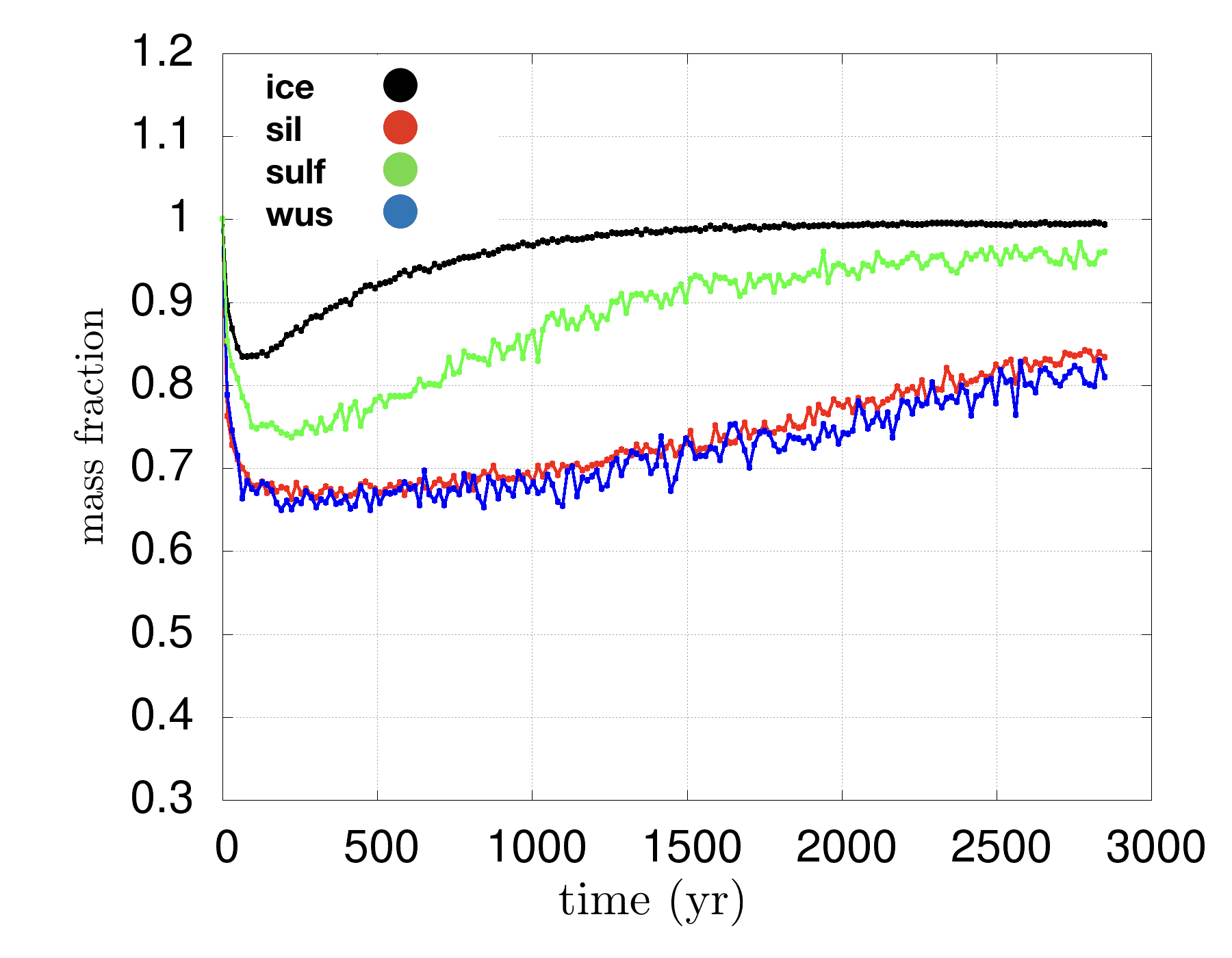}}\\
{\includegraphics[width=1.\columnwidth]{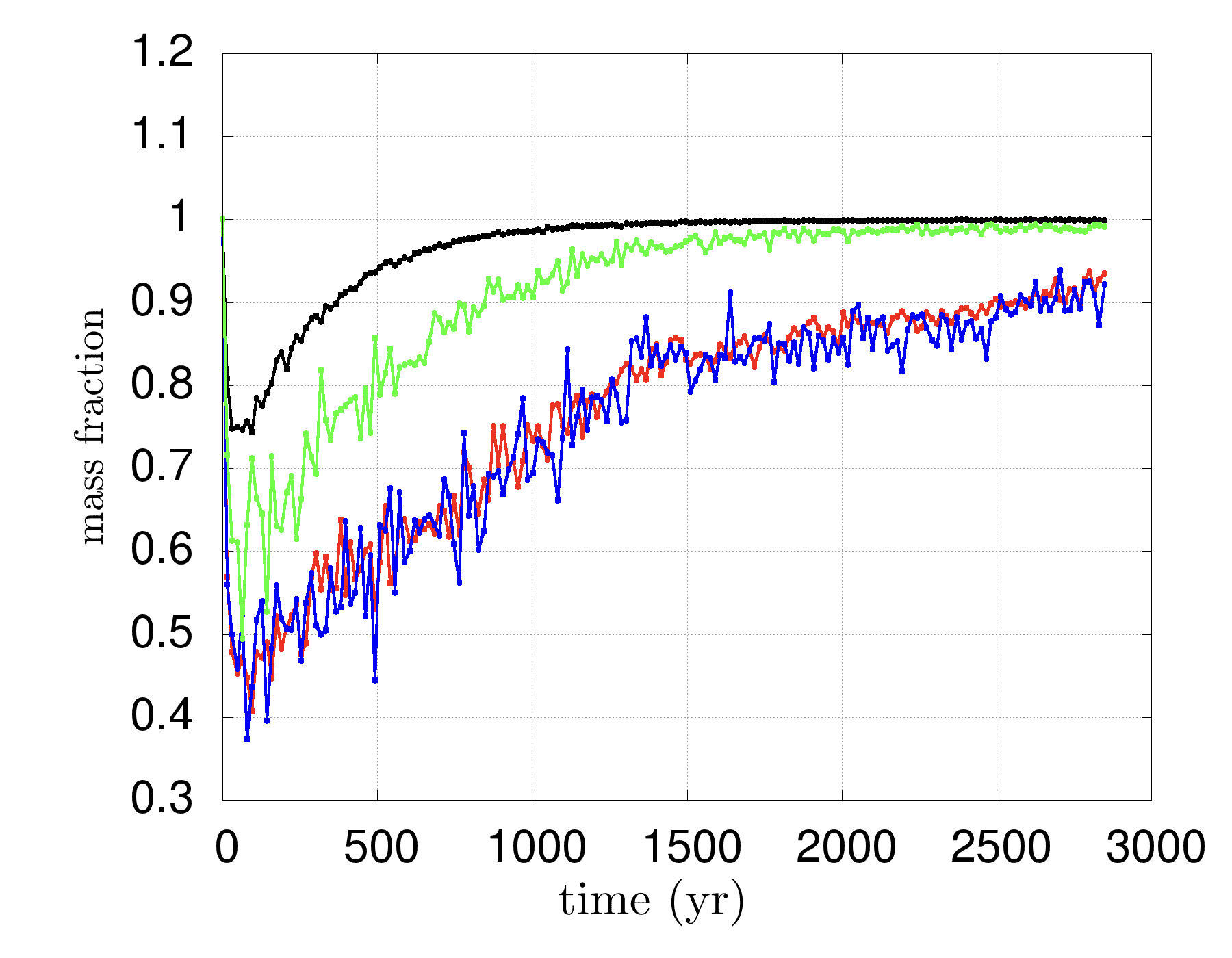}}\\
{\includegraphics[width=1.\columnwidth]{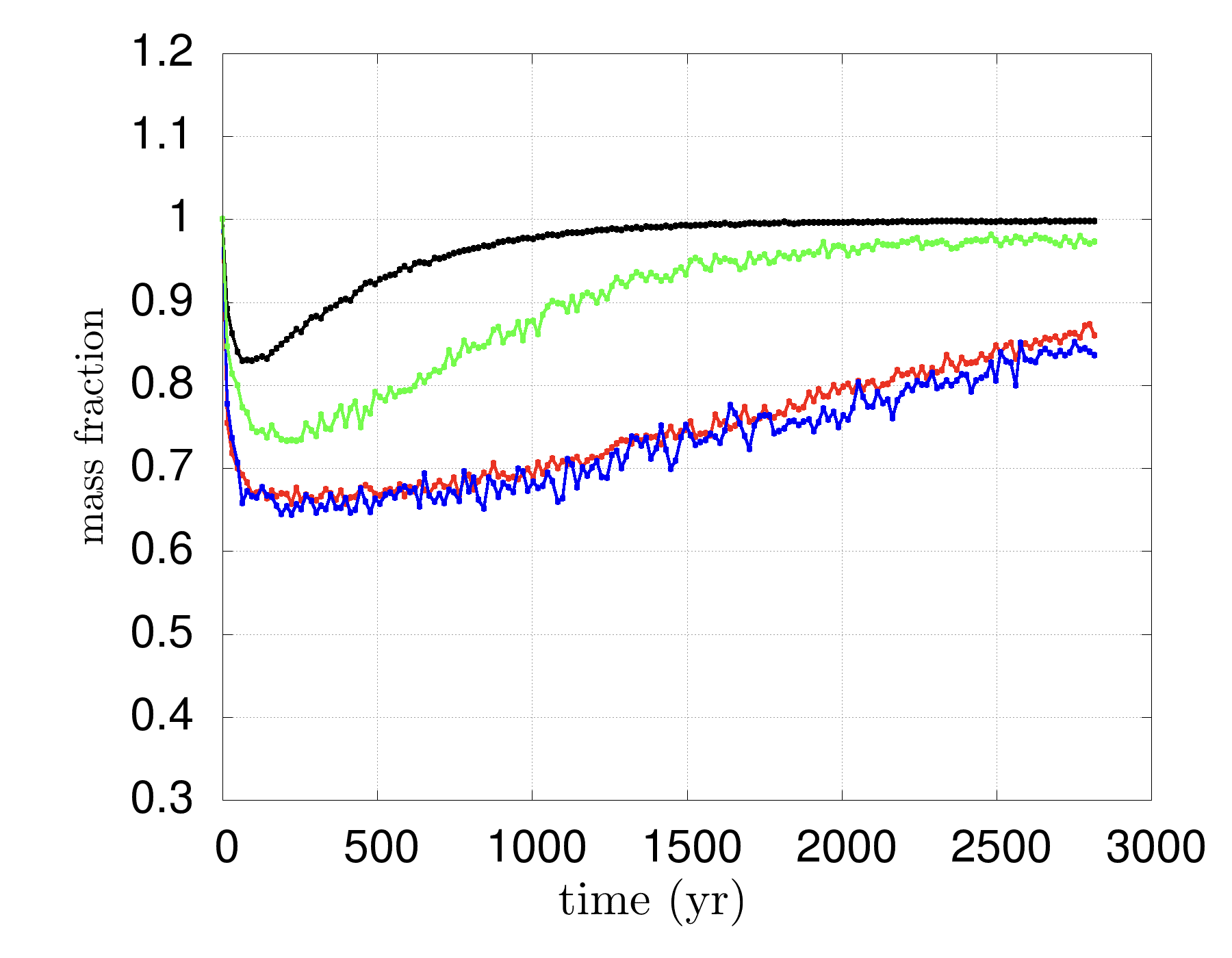}}\\
%{\includegraphics[width=1.\columnwidth]{R-vrel-norm-mid.png}}\\
%{\includegraphics[width=1.\columnwidth]{R-vrel-norm-all.png}}\\
\caption{Mass fraction of dust under the fragmentation threshold for different species and as a function of time for the inner 10~au of the disc. From top to bottom: disc surface, midplane, surface plus midplane. \label{fig5}}
\end{figure}

In the following, we will mainly focus on the inner 10~au of the disc where the differences between the G and GF simulations are more evident. Hereafter, similarly to \citetalias{2017MNRAS.469..237P}, we define the surface of the disc as the location for which $\lvert Z\rm{(au)}\rvert>0.1$ and the midplane the location where ${\lvert Z\rm{(au)}\rvert}<0.1$. The choice of a flat boundary between the midplane and the surface and that of the (although arbitrary) midplane's thickness is suggested by theoretical and observational evidence. \citet{2008A&A...487..265L,2014MNRAS.437.3055L}, \citetalias{2017MNRAS.469..237P} (and references therein) showed that the distribution of large grains ($\geq$ mm) follows an approximately flat radial distribution. Flat mm-dust distributions are also observed in protoplanetary discs \citep[e.g.][]{2016ApJ...816...25P}. In Paper I we considered a midplane thickness of  $\pm1$~au with a disc size in the order of $R\sim100$~au, thus with a ratio of 0.01. This value is  consistent with the millimeter dust scale height  of 1~au at 100~au for HL-Tau \citep{2016ApJ...816...25P} evidencing a flat mm-dust profile. This ratio is preserved, for consistency, in this work. Note that the chosen  thickness can encompass the total dust scale-height  in the innermost region of the disc where $R\sim1$~au.  This does not affect the results as the settling is very efficient (\citetalias{2017MNRAS.469..237P} and \citet{2008A&A...487..265L}) and as we are interested in exploring the differences between the G and GF cases for which we fix a common  set of parameters.

 In Fig.~\ref{fig5} we report, for  different species and for the first 10~au, the evolution with time of the mass fraction of each species which is under the fragmentation threshold. From top to bottom, we report values for the disc surface, midplane and the whole disc (surface and midplane). Figure~\ref{fig5} shows that the mass fraction of ice material which is under the fragmentation threshold reaches 0.9 after just $t\sim$500~yr and almost 1 after $t\sim$1500~yr. Sulfides reaches 0.9 at $t\sim$1000~yr and approach 1 at $t\sim$2500~yr. When the whole disc is considered, silicates and wustite particles do not reach 0.9 within the considered evolutionary time. We will demonstrate that this distinct behaviour has important implications in determining the dust content and the aerodynamic sorting of the dust within the inner disc region. 

\subsection{Disc surface and settling}
\label{discsurface}

\begin{figure*}
{\includegraphics[width=1.\columnwidth]{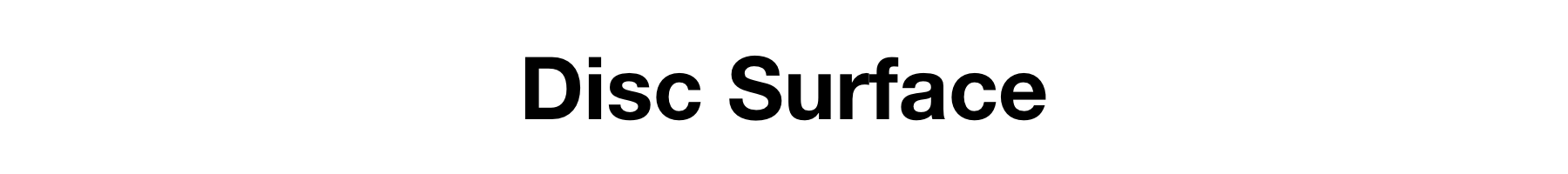}}
{\includegraphics[width=1.\columnwidth]{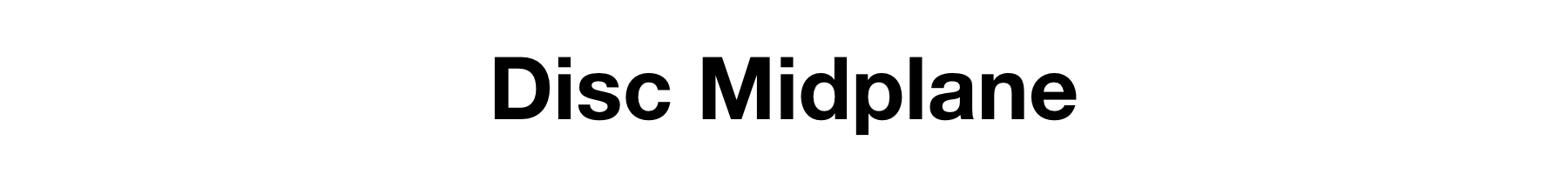}}\\
{\includegraphics[width=1.\columnwidth]{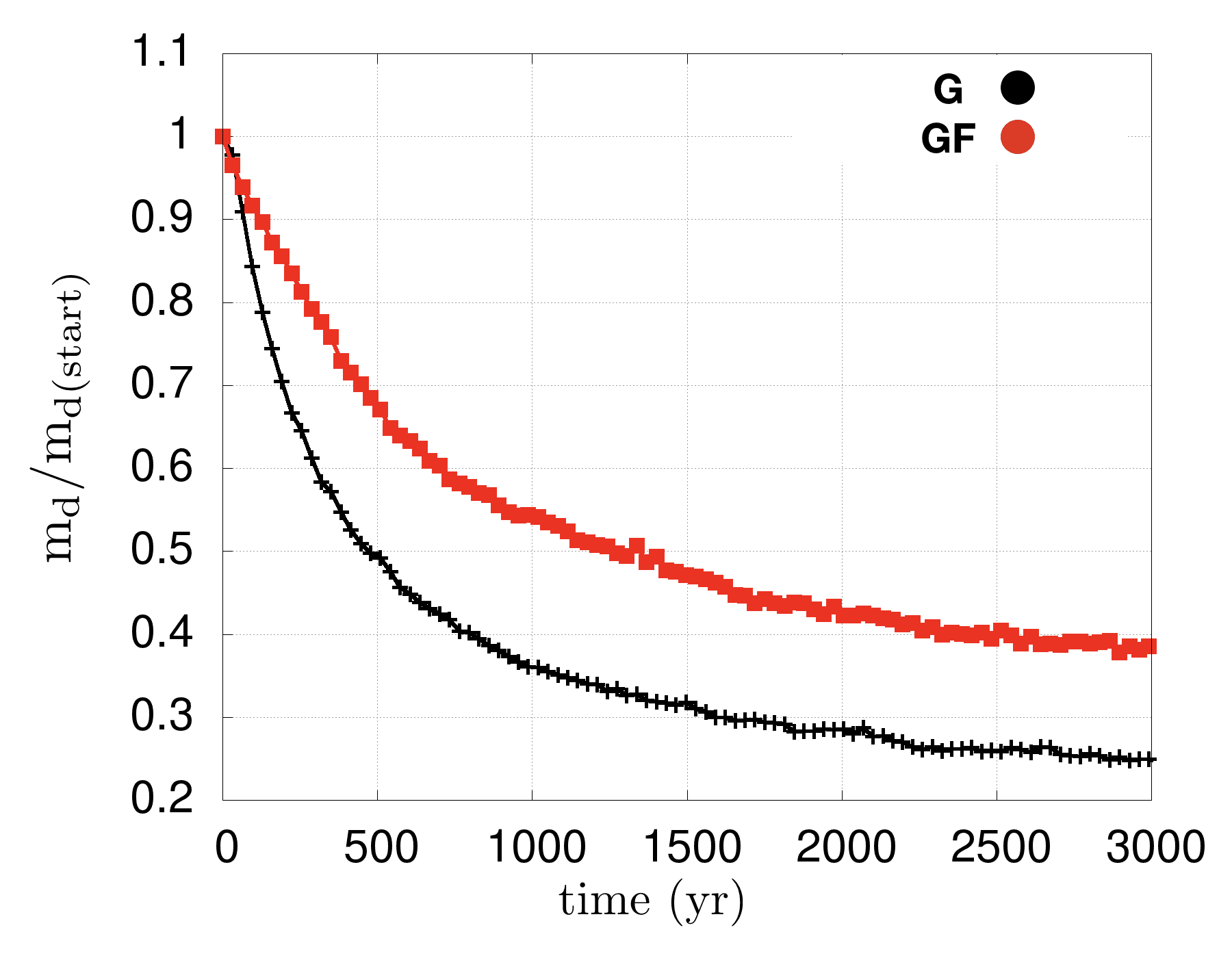}}
{\includegraphics[width=1.\columnwidth]{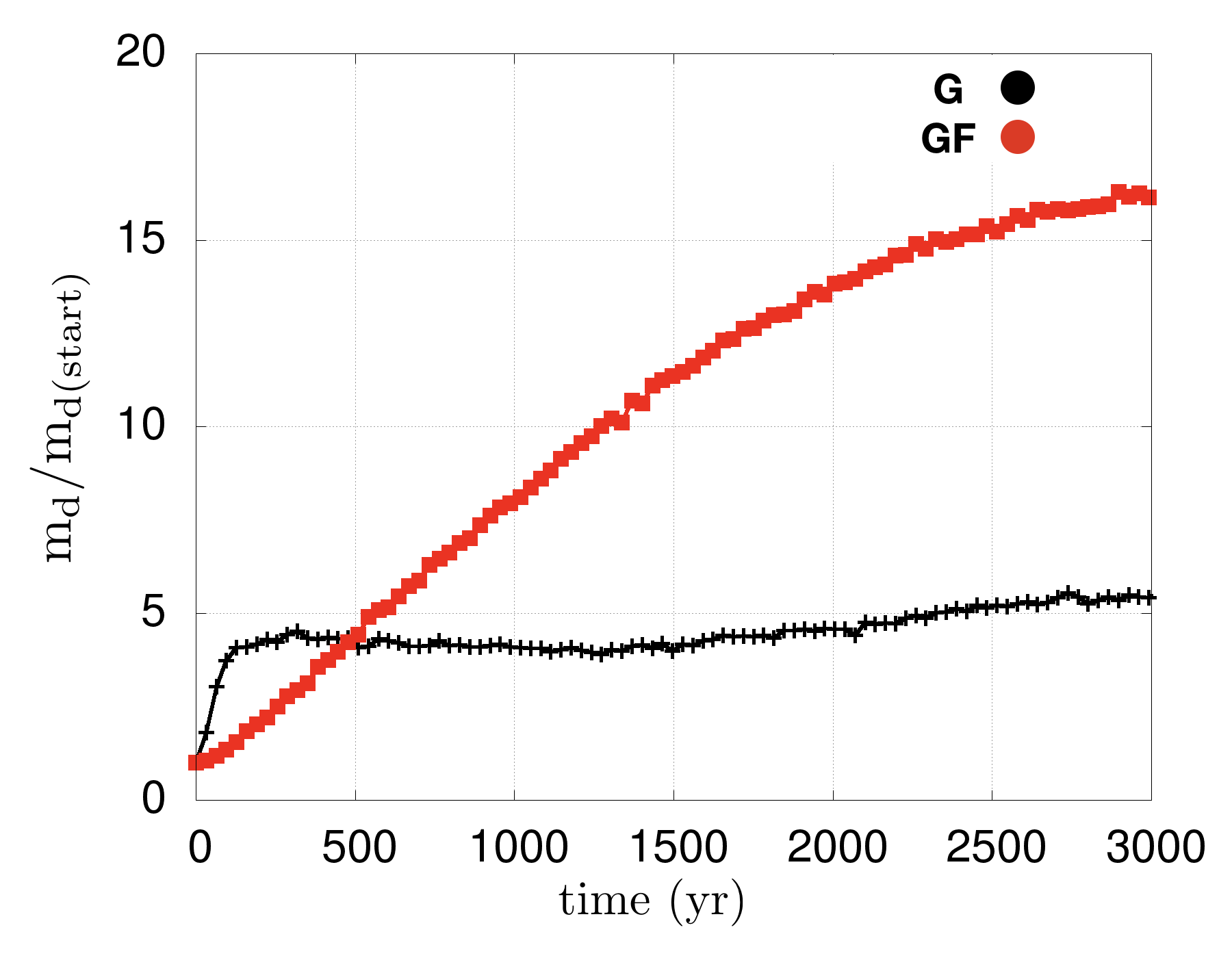}}\\
{\includegraphics[width=1.\columnwidth]{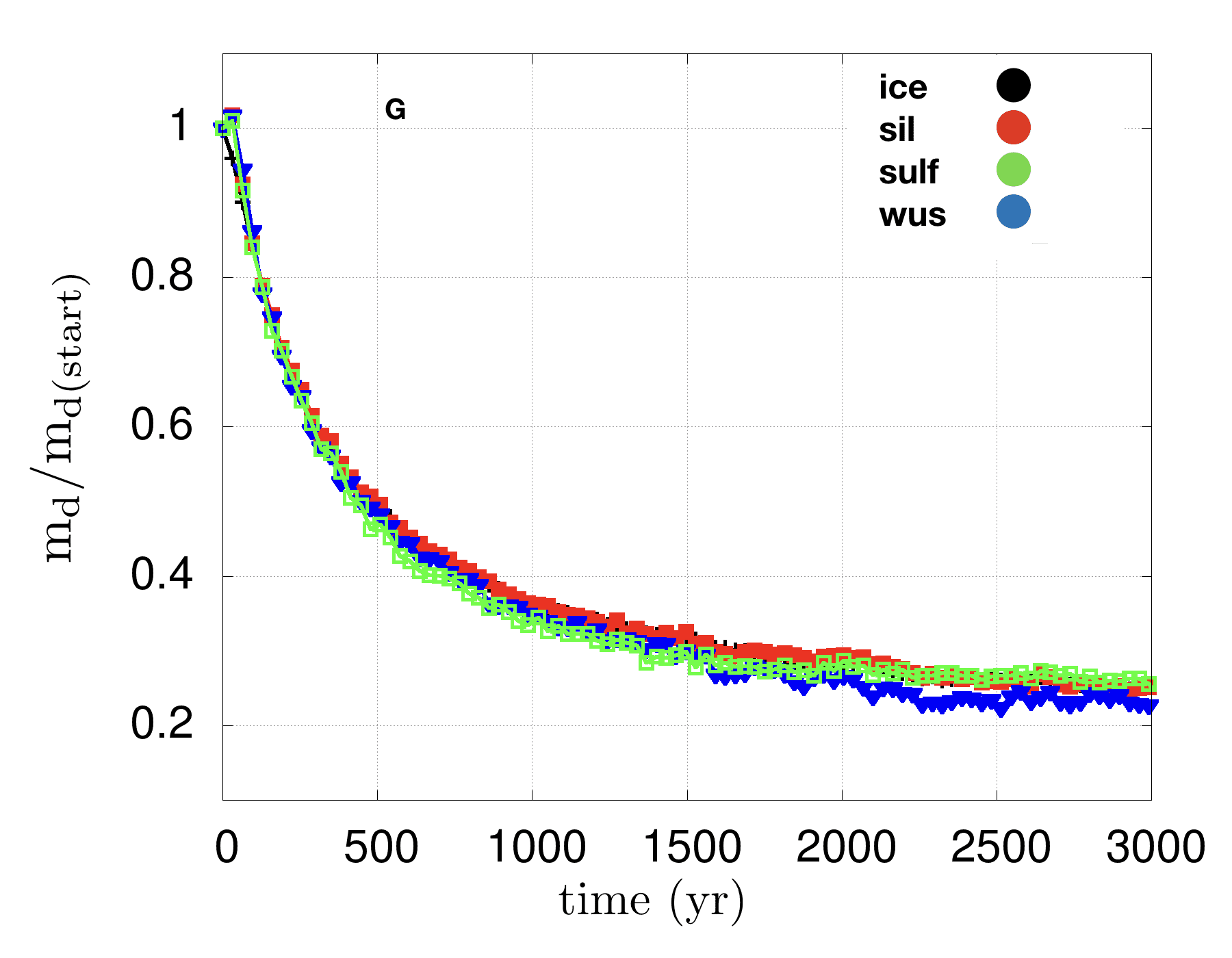}}
{\includegraphics[width=1.\columnwidth]{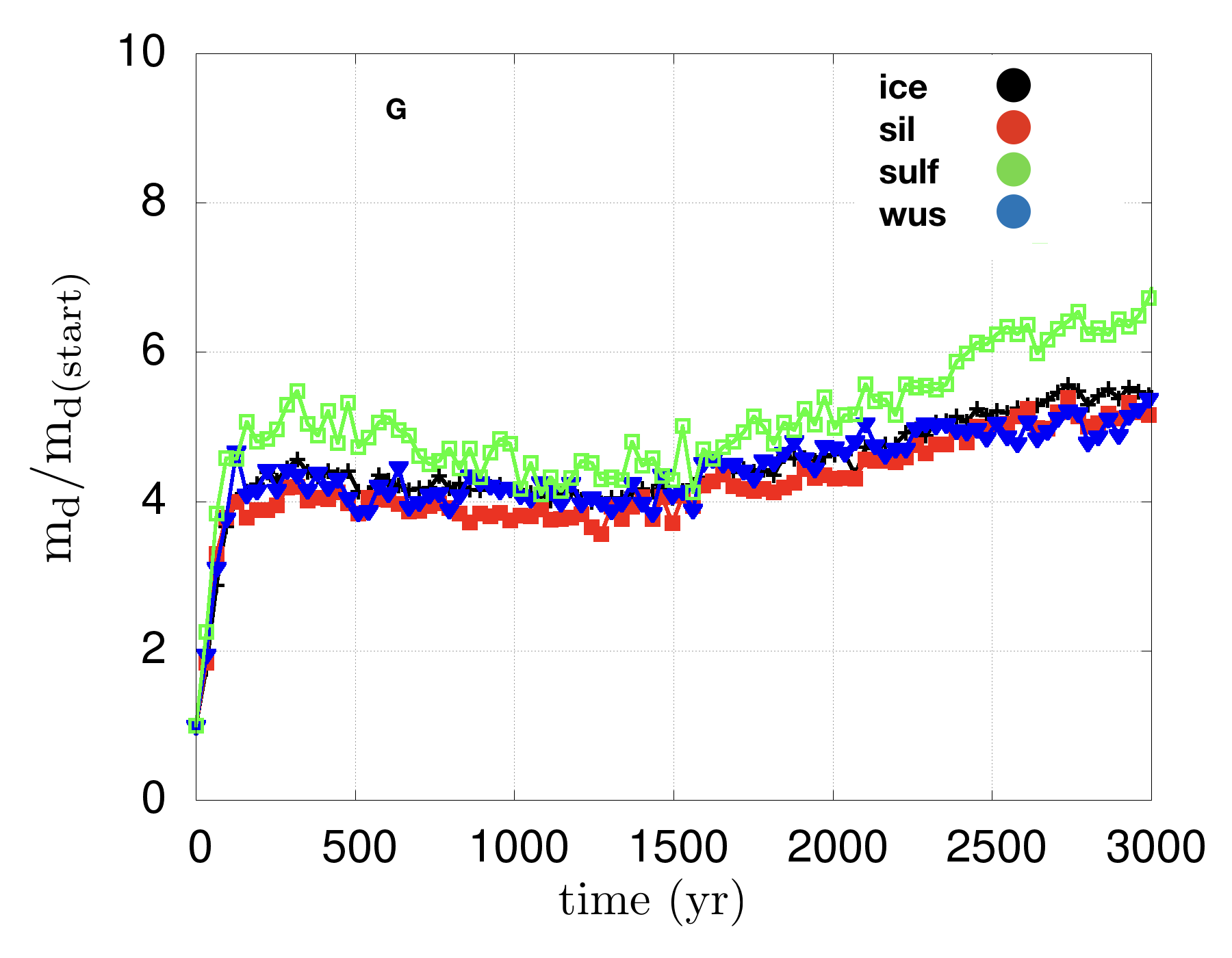}}\\
{\includegraphics[width=1.\columnwidth]{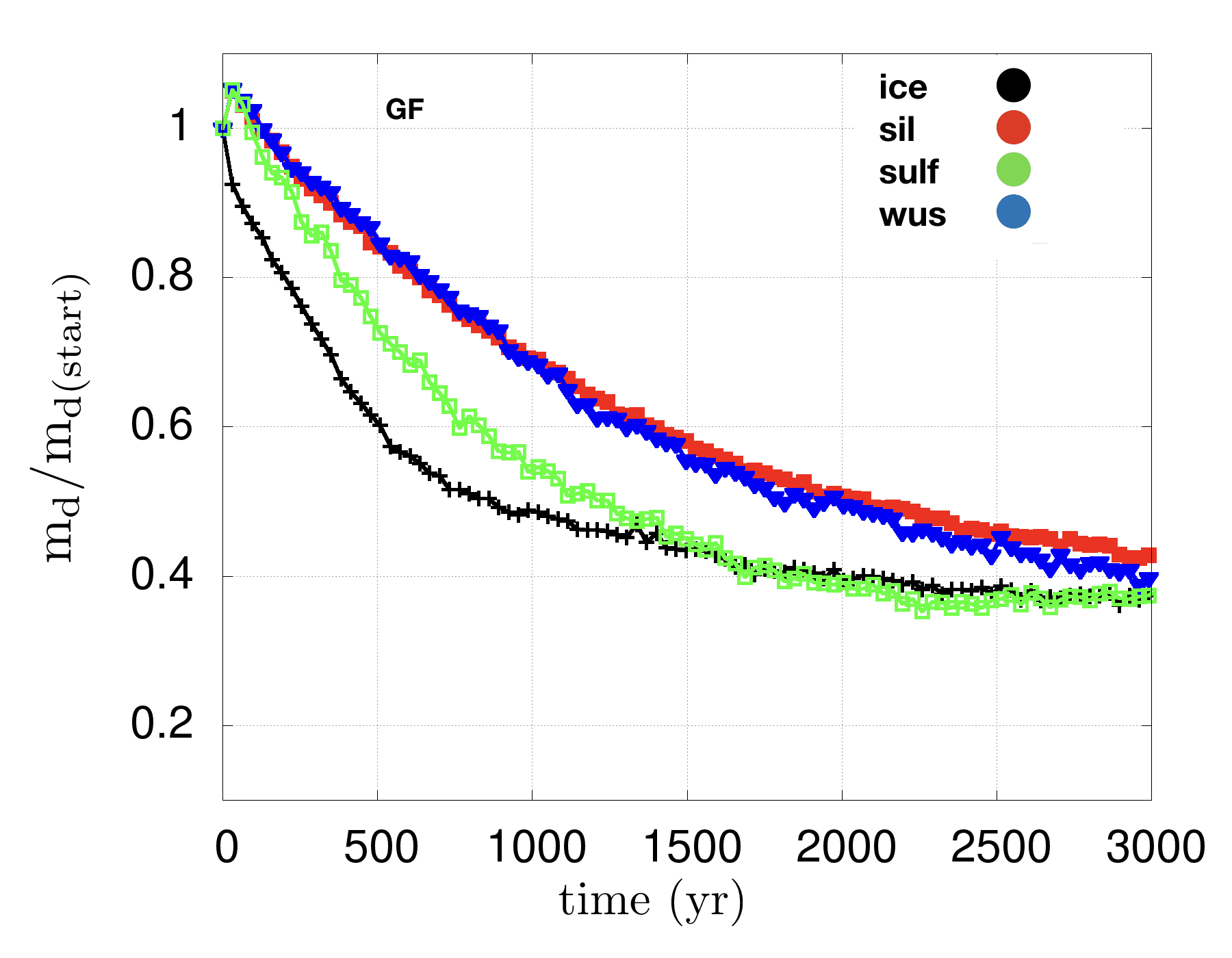}}
{\includegraphics[width=1.\columnwidth]{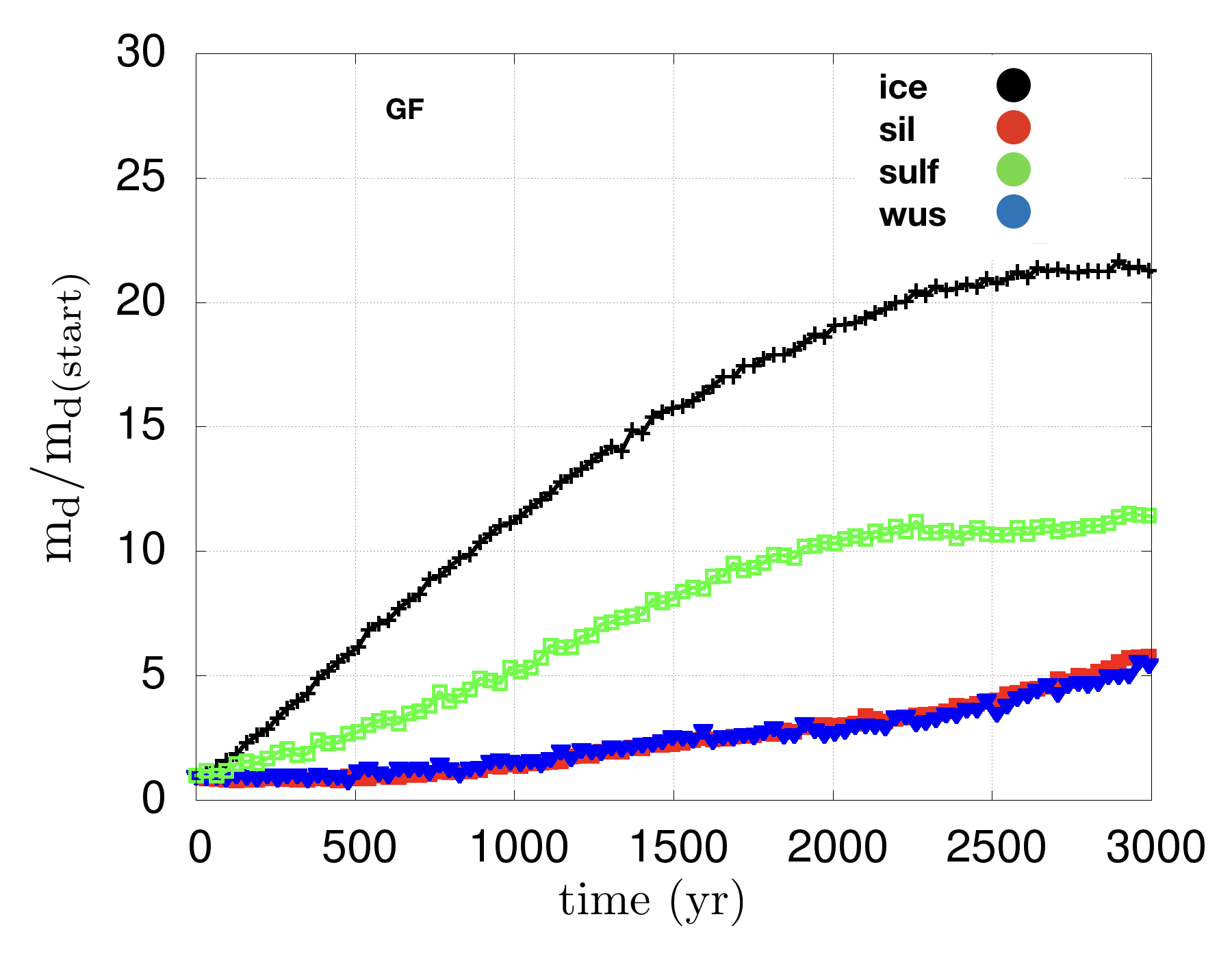}}\\
\caption{Time evolution of the dust mass in $1.87\le R{\rm (au) \le 10}$ relative to the initial dust mass in the disc surface $\lvert Z\rm{(au)}\rvert>0.1$ (left) and disc midplane ${\lvert Z\rm{(au)}\rvert}<0.1$ (right). The top panel compares the total dust mass while the middle and bottom panels show the curves for the individual species in the G and GF cases, respectively.
\label{fig9}}
%\caption{Left column: disc surface where $1.87\le R{\rm (au) \le 10}$ and $\lvert Z\rm{(au)}\rvert>0.1$. Right column: disc midplane where $1.87\le R{\rm (au) \le 10}$ and $-0.1<Z\rm{(au)}<0.1$. Top: time evolution of the total dust mass content compared to the initial mass in the plotted region. The middle and the bottom panels show the same curves for individual species in the G and GF cases, respectively.
%\label{fig9}}
\end{figure*}

In Fig.~\ref{fig9}, left column, we report the time evolution of the total dust mass content compared to the initial mass present in the disc surface where $1.87\le R{\rm (au) \le 10}$ and $\lvert Z\rm{(au)}\rvert>0.1$: globally in the G and GF cases (top), and for single species in the G (middle) and  GF (bottom) cases. In G dust settles toward the midplane at a higher rate. The rate of settling for each single species in G is similar, with differences driven by their intrinsic densities as explained in \citetalias{2017MNRAS.469..237P}. In GF, ice and sulfides settle with higher rates than wustite and silicates.

\begin{figure*}
{\includegraphics[width=1.\columnwidth]{FIG17a.png}}
{\includegraphics[width=1.\columnwidth]{FIG17b.png}}\\
{\includegraphics[width=1.\columnwidth]{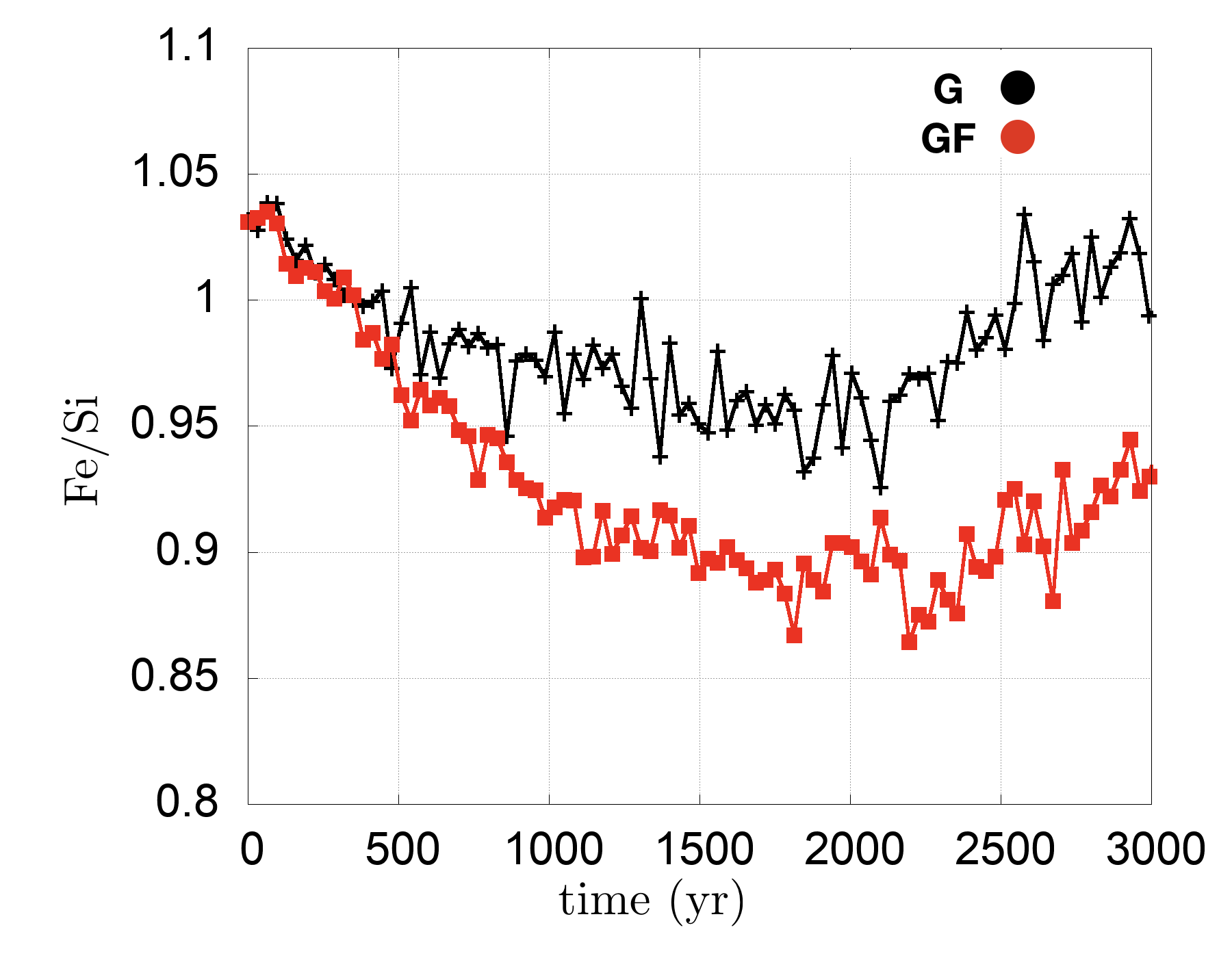}}
{\includegraphics[width=1.\columnwidth]{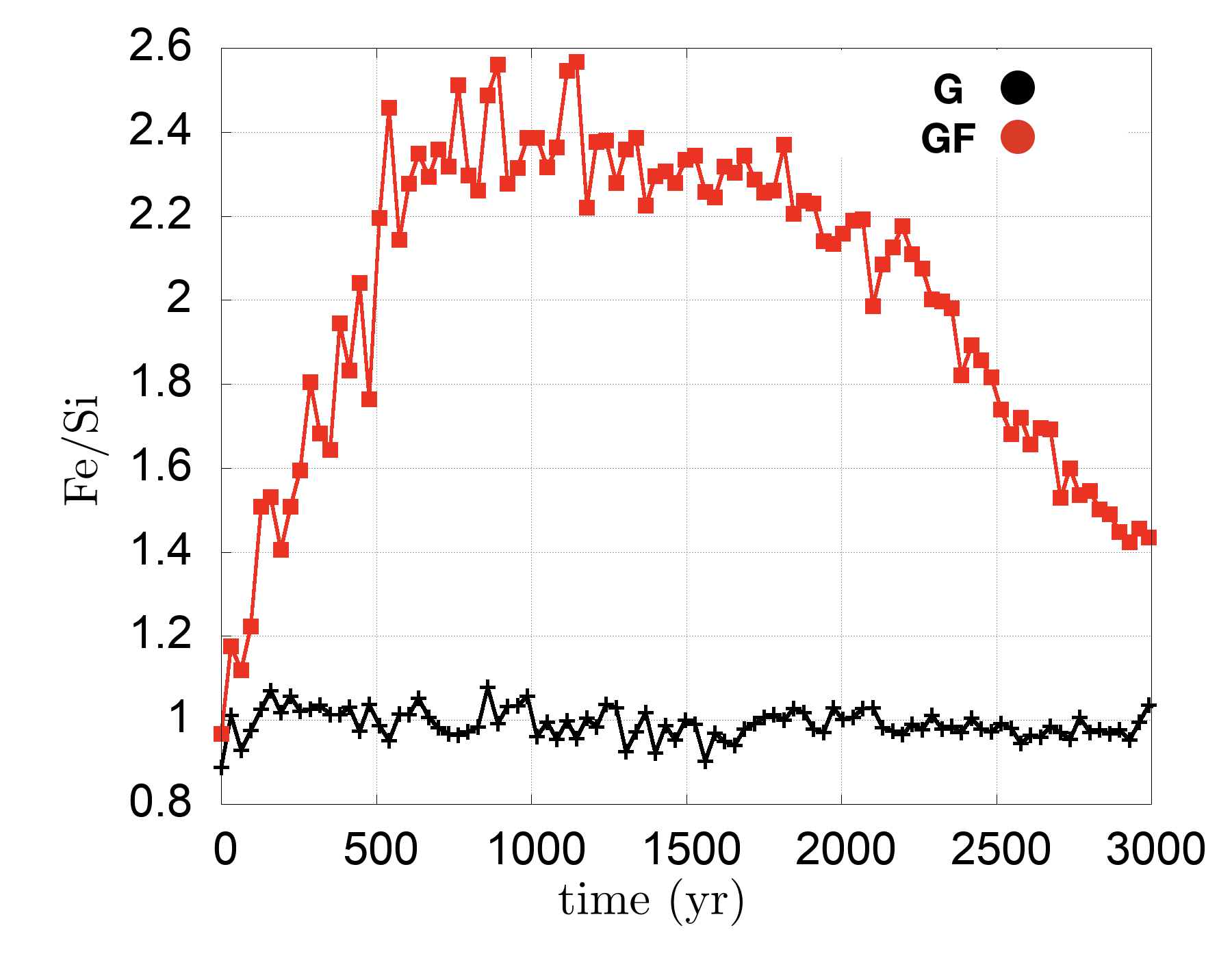}}\\
{\includegraphics[width=1.\columnwidth]{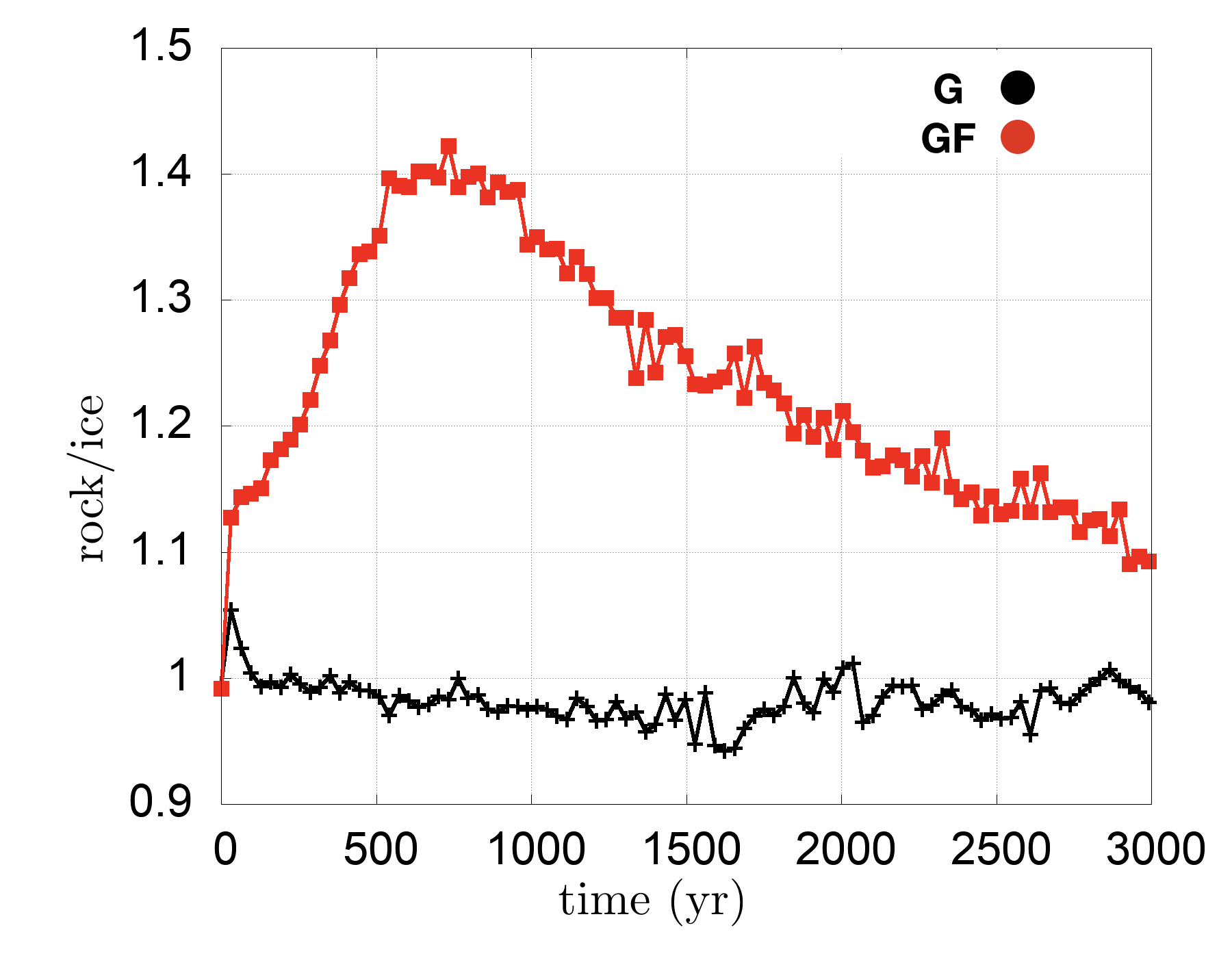}}
{\includegraphics[width=1.\columnwidth]{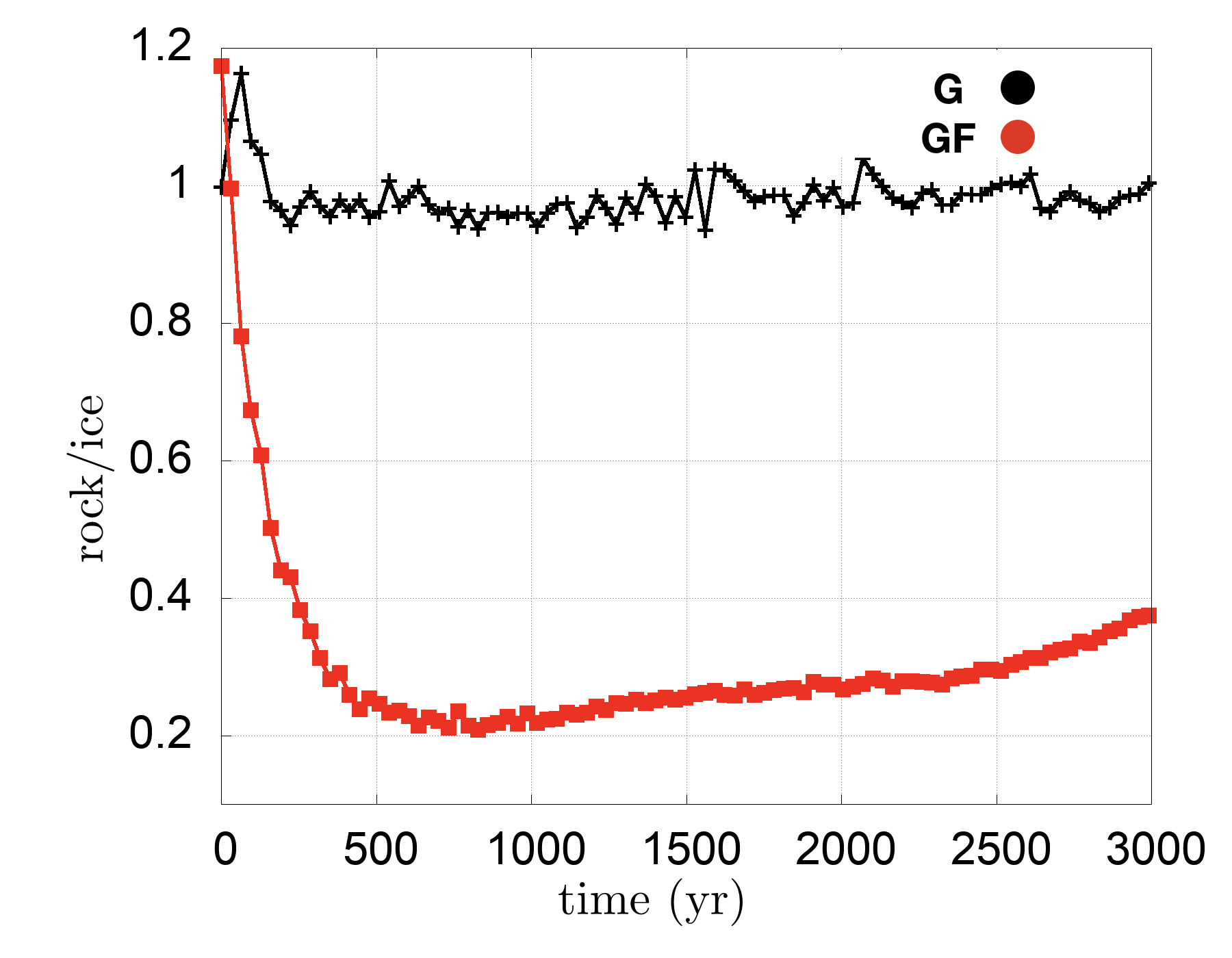}}\\
\caption{Fe/Si (top) and rock/ice (bottom) ratios for G (black) and GF (red). Left: disc surface. Right: disc midplane. \label{fig6}}
\end{figure*}

In Fig.\ref{fig6}, left column, we report the \ce{Fe}/\ce{Si} (top) and rock/ice (bottom) ratios for the inner disc surface ($1.87\le R{\rm (au) \le 10}$) and $\lvert Z\rm{(au)}\rvert>0.1$). Similarly to \citetalias{2017MNRAS.469..237P}, these ratios are the ratios between the number of particles of a given species (\ce{Fe}, \ce{Si}, rock=(Fe+Si), \ce{H2O}) which are populating a given region at a given time.  For example, the  \ce{Fe}/\ce{Si} ratio is
$ \ce{Fe}/\ce{Si}=n_{\rm (wus+sul)}/n_{\rm sil}$. All values are then normalized to the initial values at the time of injection, called from here on ``solar'', and reported in Table~\ref{frac-abundance2}. Similarly to the case of pure growth, GF fractionates the disc surface in its iron content, with respect to the initial solar value. It is interesting to note that in the GF case, the slope of the \ce{Fe}/\ce{Si} ratio is steeper. On the other hand, the rock/ice ratio shows an opposite behaviour when G and GF are compared. While for G it goes from ``solar" to ``sub-solar" values, showing faster depletion of rocky particles, in GF we see an abrupt decrease of ice particles with the rock/ice ratio reaching $\sim$1.4 times the ``solar" value before decreasing smoothly after $t\sim1000$~yr.

\begin{figure*}
{\includegraphics[width=1.\columnwidth]{FIG17a.png}}
{\includegraphics[width=1.\columnwidth]{FIG17b.png}}\\
{\includegraphics[width=1.\columnwidth]{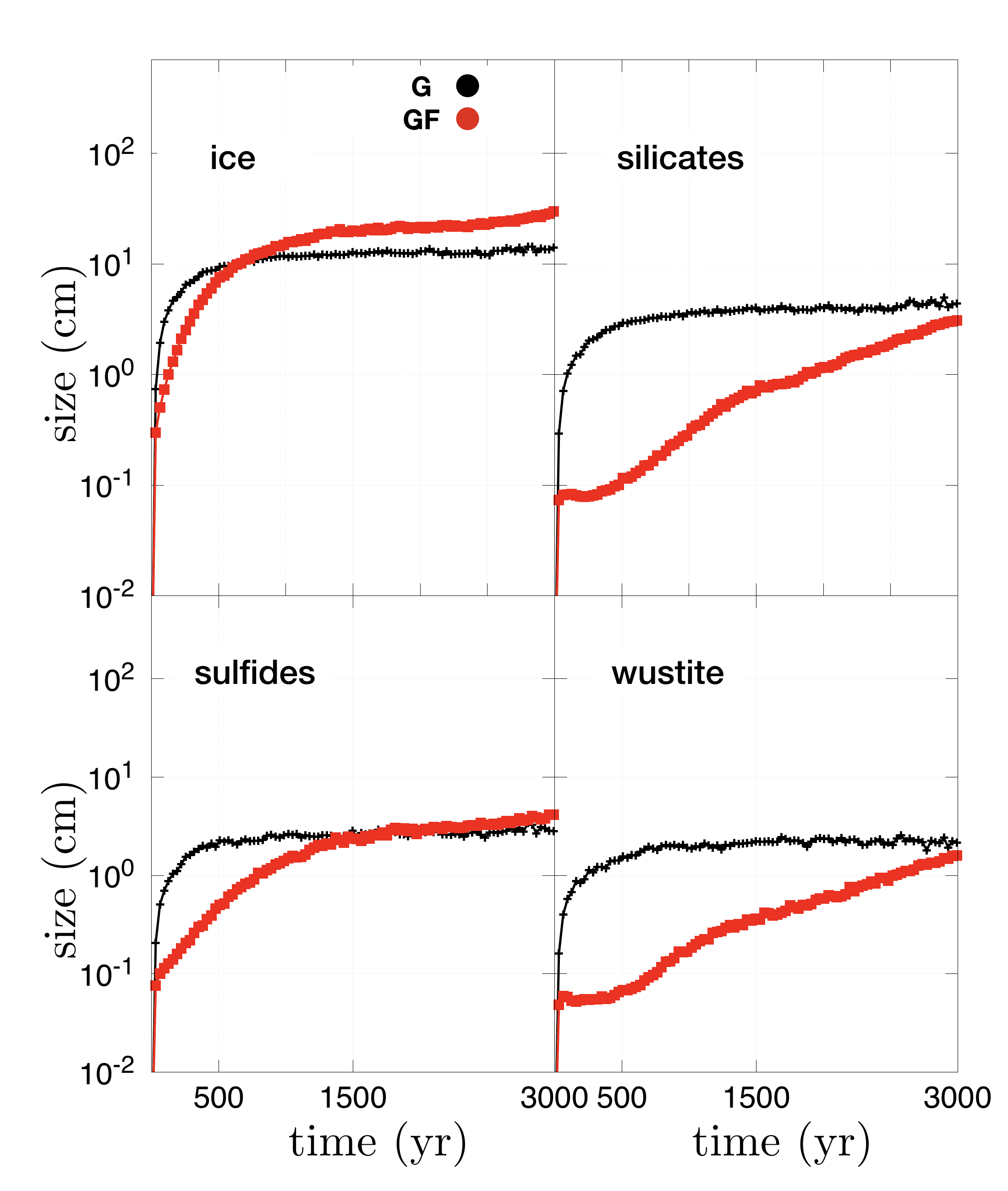}}
{\includegraphics[width=1.\columnwidth]{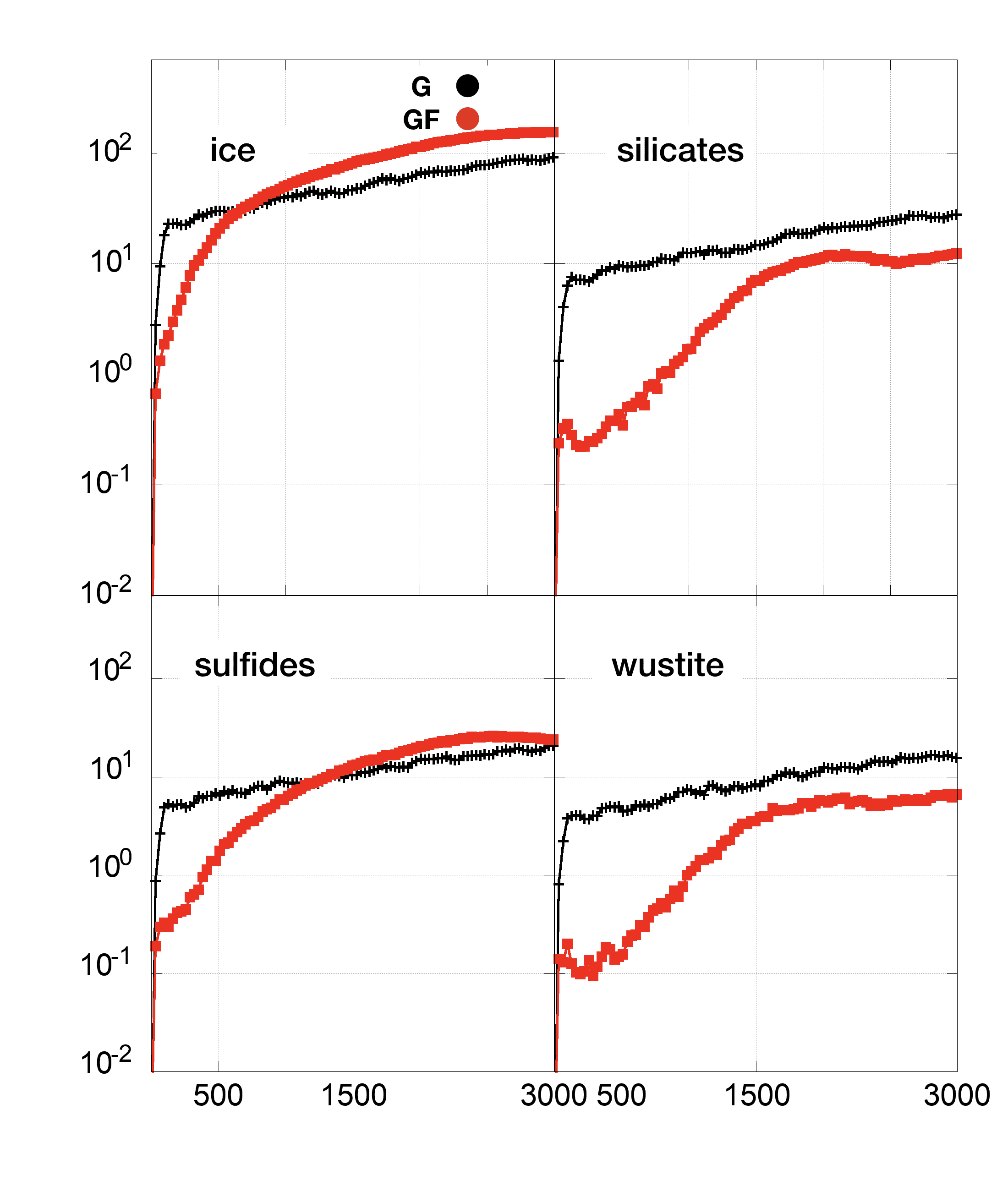}}\\
%{\includegraphics[width=1.0\columnwidth]{R-size-species.png}}\\
\caption{Time evolution of the average size of different species, for G (black) and GF (red). Left column: disc surface. Right column: disc midplane.} \label{fig7}
\end{figure*}

In Fig.~\ref{fig7}, left box, we plot the time evolution of the average size of each species in the disc surface in the G and GF cases. In the G case grains grow rapidly to reach cm-dm size within a few hundreds years, while, as expected, GF results in a slower growth rate. However, it is interesting to note that, after $t\sim500$~yr for ice, and $t\sim1500$~yr for sulfides, grains start to grow larger than in the G case. For silicate and wustite particles, the growth rate is slower for all the considered evolutionary time, with the average size  closer to the values predicted by pure growth only after  $t\sim3000$~yr.  

\begin{figure*}
{\includegraphics[width=1.\columnwidth]{FIG17a.png}}
{\includegraphics[width=1.\columnwidth]{FIG17b.png}}\\
{\includegraphics[width=1.\columnwidth]{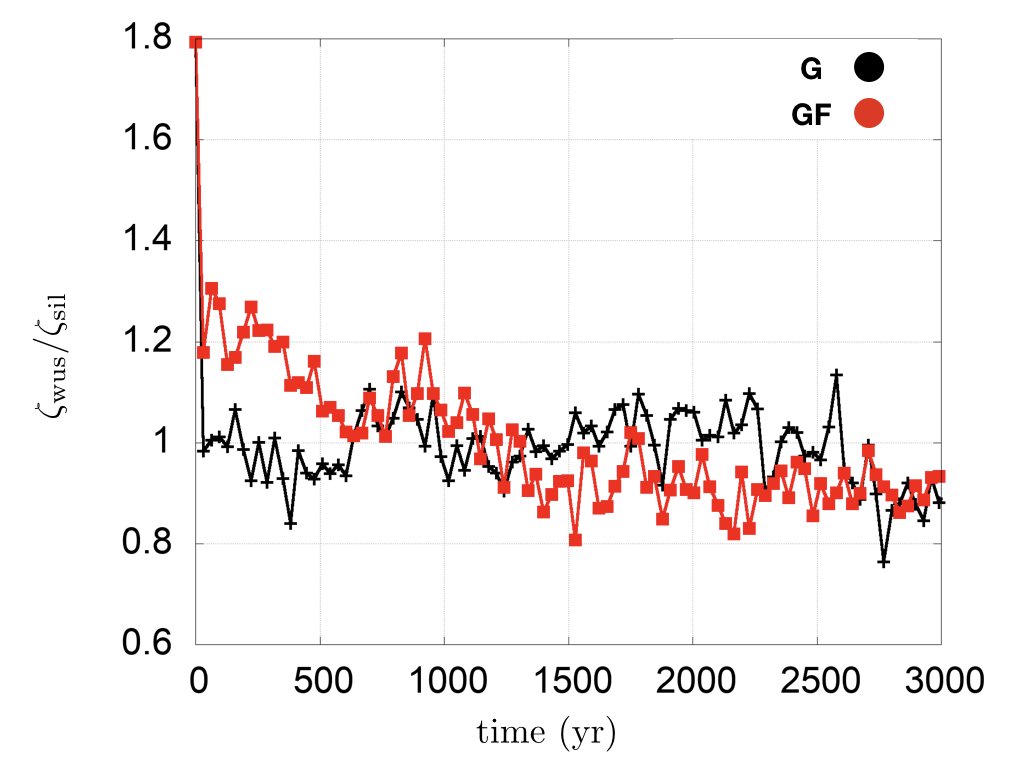}}
{\includegraphics[width=1.\columnwidth]{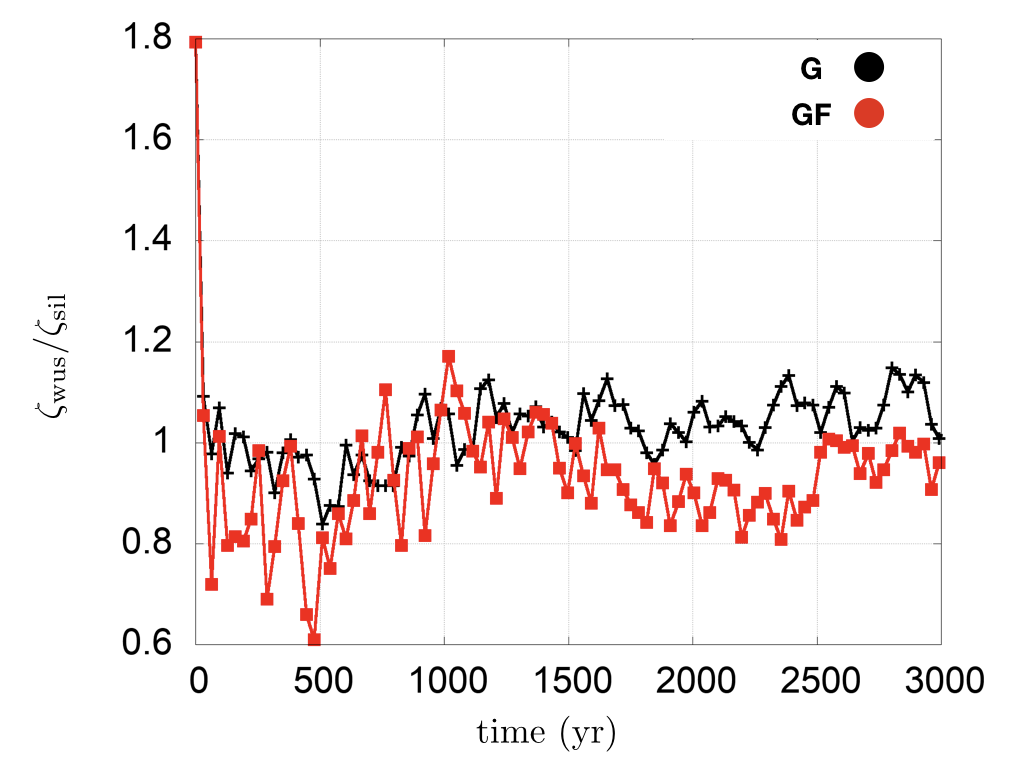}}\\
{\includegraphics[width=1.\columnwidth]{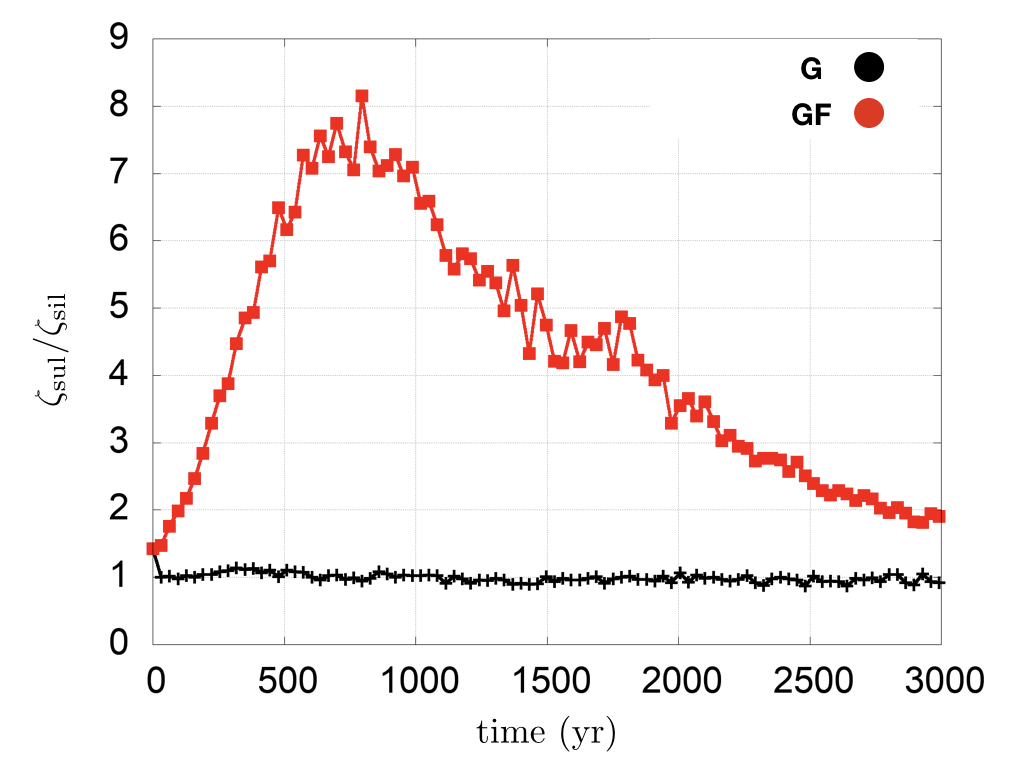}}
{\includegraphics[width=1.\columnwidth]{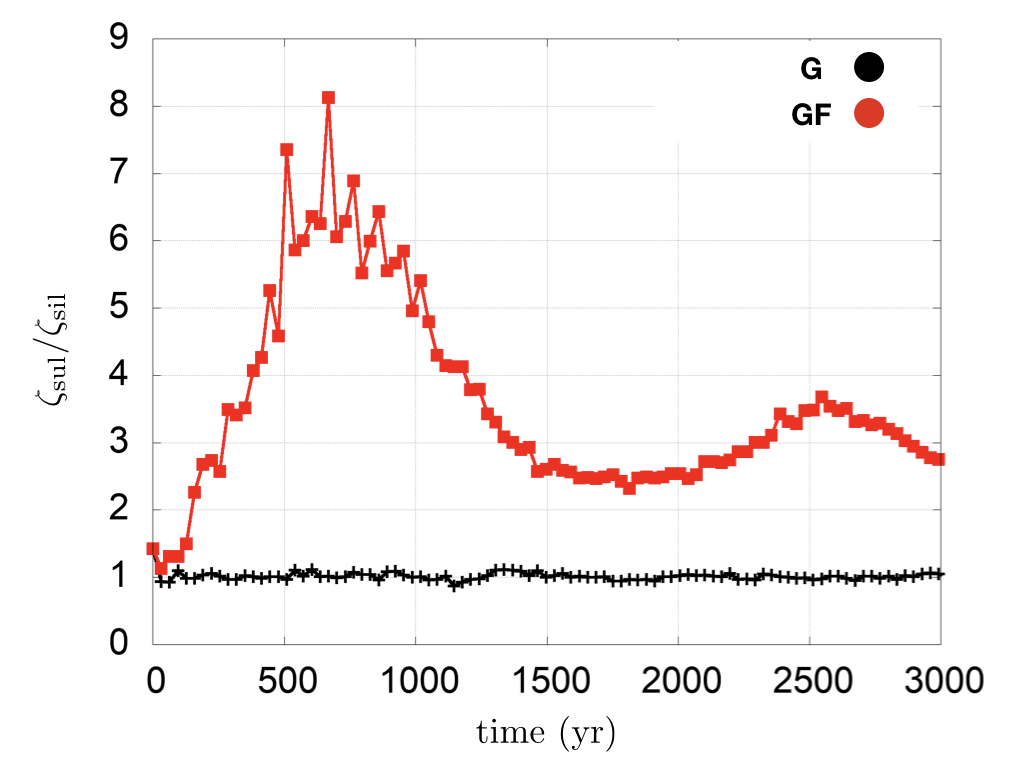}}\\
%{\includegraphics[width=1.0\columnwidth]{sortingFeO10.png}}\\
%{\includegraphics[width=1.0\columnwidth]{sortingFeS10.png}}\\
\caption{Time evolution of the ratio of the aerodynamic parameter of wustite, $\zeta_{\rm wus}$, to that of silicates, $\zeta_{\rm sil}$, (top row), and of that of sulfides, $\zeta_{\rm sul}$, to that of silicates (bottom row), in the disc surface (left column) and disc midplane (right column) for the G (black) and GF (red) cases. The ratio between the aerodynamic parameters of wustite and silicate grains quickly evolve towards unity ($\sim$100~yr), indicating that the two species are size-density sorted. For the sulfides,  instead, we observe that particles evolve away from aerodynamic sorting with silicates before turning back towards sorting at later stages. \label{fig8}}
\end{figure*}

In Fig.~\ref{fig8}, left column, we now focus on the single dust species and report the ratios between the averaged aerodynamic parameters, $\zeta$, of wustite and silicates, $\zeta_{\rm wus}/\zeta_{\rm sil}$ (top) and  sulfides and silicates, $\zeta_{\rm sul}/\zeta_{\rm sil}$ (bottom). We see that the $\zeta_{\rm wus}/\zeta_{\rm sil}$ ratio returns a similar trend when G and GF are considered. The wustite and silicates particles tend to become size-density sorted, ($\zeta_{\rm wus}/\zeta_{\rm sil}\sim1$), almost immediately in the case of pure growth and after $t\sim500$~yr in the case of fragmentation. The case of $\zeta_{\rm sul}/\zeta_{\rm sil}$ is different: in the G simulation the size-density sorting is evident, while GF initially ``unsorts'' particles for $t\sim1000$~yr before moving them toward a size-density sorting. 

\subsection{Disc midplane and radial drift}
\label{discmidplane}

In this section we discuss the the inner disc midplane where $1.87\le R{\rm (au) \le 10}$ and $-0.1<Z\rm{(au)}<0.1$.

In Fig.~\ref{fig9}, right column, we report the time evolution of the total dust mass content compared to the initial mass present in the disc midplane in the G and GF cases (top), and for single species in G (middle), and GF (bottom). We see that in G, after an initial increase, the $m_{\rm d}/m_{\rm d(start)}$ ratio remains constant around a value of $\sim4$ before increasing slowly after $t\sim1500$~yr. In GF, within the first  $t\sim500$~yr, the $m_{\rm d}/m_{\rm d(start)}$ ratio increases slower when compared with G. However, it keeps increasing with values reaching 15 times the initial mass. When the behaviour of the single species is taken into account we see that, in G, species behave similarly: a first increment due to the vertical settling, and a second increment ($t\sim1500$~yr) due to the radial drift of the particles from the outer disc ($R>10$~au) which is becoming more efficient (see also \citetalias{2017MNRAS.469..237P}). In GF, there is a dramatic increase of the mass of ice and sulfides, while the increase of the mass of other species is characterized by a lower rate, more similar to G.

Similarly to the previous Section we plot in Fig.~\ref{fig6}, right column, the \ce{Fe}/\ce{Si} ratio, (top), and the rock/ice ratio, (bottom), as a function of time. The  \ce{Fe}/\ce{Si} ratio in G varies within 10\% of the initial ``solar''  value, with a similar trend found in \citetalias{2017MNRAS.469..237P}. In GF the \ce{Fe}/\ce{Si} ratio reaches 2.4 times the ``solar''  value after $t\sim1000$~yr, before dropping toward lower values. The rock/ice ratio shows an important depletion in the case of GF compared to G.  This is because, as we saw from the disc surface, the ice is efficiently populating the midplane.

Figure~\ref{fig7}, right box, shows the time evolution of the average size of each species in the disc midplane in the case of G (black) and GF (red). We find a behaviour which is similar to the disc surface, with an efficient growth at early stages in the G case. Ice and sulfide grains then reach larger size in the case of fragmentation, in the same timescales found for the disc surface.

Finally, in Fig.~\ref{fig8}, right column, we report the evolution of the ratios between the aerodynamic parameters of wustites and silicates, and between sulfides and silicates. We see that, similary to the disc surface, there is a size-density sorting between wustite and silicates and an ``un-sorting'' between sulfides and silicates for the first 1000~yr. In this latter case, the  curve has two  peaks within the considered  time period.

\section{Discussion}
\label{discussion}

In \citetalias{2017MNRAS.469..237P} we investigated the dust properties and disc chemical composition which resulted in the different aerodynamical sorting of the dust species in a pure-growth regime. In this Section we repeat that analysis comparing our two simulations, G and GF.

\subsection{Time evolution of  relative velocities}
\label{relvel}

At the end of  Section~\ref{globalevol} we showed that the mass fraction of different species that is over/under the respective fragmentation thresholds changes with time. This is because $v_{\rm rel}$ varies with time due to the change in St as grains grow/fragment (see equation~(\ref{vrelvrel}) and Section~\ref{method}).

As the particle size increases,  $v_\mathrm{rel}$ reaches a maximum and then decreases ($v_\mathrm{rel}\propto s_\mathrm{d}$ for small sizes and $v_\mathrm{rel}\propto s_\mathrm{d}^{-1}$ for large sizes, see equations~\ref{stokesnumber} and~\ref{vrelvrel}). This explains the fact that the dust in the disc can naturally transition from a fragmentation regime to a pure-growth regime (see also Sections 2.3, 2.4 and Appendix A in \citet{2017MNRAS.467.1984G}). The values of $v_{\rm rel}$  will be determined by the gas drag and by the disc local conditions, but its overall behaviour will remain unchanged. The timescales of the fragmentation regimes, will be then determined by the disc and dust properties.

\subsection{Effect of growth and fragmentation in sorting dust}
\label{discdifferences}
The differences found between G and GF can be explained by the different response of  single dust species to the growth and fragmentation. The evolution of dust can be followed from the disc surface.  In G, where  growth proceeds similarly for all the species, it is the intrinsic density which initially drives the vertical settling as shown in \citetalias{2017MNRAS.469..237P}. In the GF case, the \ce{Fe}/\ce{Si} ratio has a steeper decrease because sulfides have a higher fragmentation thresholds ($v_{\rm frag}=42~\rm{m\,s^{-1}}$) compared to silicates and wustite, ($v_{\rm frag}=36~\rm{m\,s^{-1}}$ and $v_{\rm frag}=35~\rm{m\,s^{-1}}$ respectively). As such, they experience less fragmentation (Fig.\ref{fig4} and~\ref{fig5}), grow larger than silicates (see Fig.~\ref{fig7}) and thus decouple from the gas more efficiently. This contributes to a more efficient depletion of iron-rich particles in the disc surface. The values of the rock/ice ratios can be explained with a similar argument. Ice particles in our calculations have the highest fragmentation threshold ($v_{\rm frag}=56~\rm{m\,s^{-1}}$). As a consequence, they are less sensitive to fragmentation, thus they grow and decouple from the gas, settling toward the disc midplane and depleting the disc surface of ice. This can be clearly seen in Figs.~\ref{fig2} and~\ref{fig9} (bottom left)  where ice and sulfides, in GF, settle toward the midplane at higher rates compared to silicates and wustite.

Fragmentation properties can also explain the sorting and ``unsorting'' of the single species found in Fig.~\ref{fig8}. \citetalias{2017MNRAS.469..237P} and references therein showed that, in the G case,  grain growth is regulated, if  other parameters such as temperature, gas densities, turbulence are kept fixed, by their aerodynamic parameter, $\zeta$. This process distributes all the grains toward a significant size-density sorting. The GF case alters this behaviour as there is another important factor to consider: $v_{\rm{frag}}$. Silicates and wustite have similar $v_{\rm{frag}}$ and thus respond to the growth-fragmentation in an overall similar way. The difference in  $v_{\rm{frag}}$  between sulfides and silicates is more significant and thus they do not respond to the growth-fragmentation in the same way. This is the reason  sulfides and silicates are not aerodynamically sorted at the early evolutionary stages. 

The evolution of the  \ce{Fe}/\ce{Si} and the rock/ice ratios in the midplane is directly connected to the behaviour of the dust in the disc surface: the fast rate of settling of ice (Fig.~\ref{fig9}, left column) decreases the rock/ice ratio, (see Fig.~\ref{fig6}, right column). The rapid depletion of sulfides in the disc surface increases instead the \ce{Fe}/\ce{Si} ratio in the midplane. The evolution of the size-density sorting between wustite and silicates, and sulfides and silicates follow the same explanation as for the disc surface (see Fig.~\ref{fig8}): wustites and silicates aerodynamically sort since the early stages given their similar $v_{\rm{frag}}$, while sulfides and silicates do not as they have different $v_{\rm{frag}}$.

Our simulations return  other very interesting results: (i) in GF, ice and sulfide grains grow to an average larger size compared with the average size reached by the two species when pure growth is considered (see Fig.~\ref{fig7}), and (ii) after a first stage in which sulfides and silicates are aerodynamically ``unsorted'', they start to move toward a size-density sorting (see Fig.~\ref{fig8}). Moreover, in the midplane, at $t\sim2500$~yr, we saw that the curve has a second peak.

Let us first focus on the two evolutionary stages at which the average size of  ice ($t\sim500$~yr) and sulfides ($t\sim1000$~yr) particles overtake the corresponding average size resulting from G  (see Fig.~\ref{fig7}). At these stages the amount of dust in the disc surface is larger than in the case of pure growth (see Fig.~\ref{fig9}, top left). A similar situation in found in the midplane (see Fig.~\ref{fig9}, top right). This is mainly due to the fact that fragmentation does not allow grains to grow large enough to decouple from the gas and start drifting and then accreting into the central star. Moreover, at $t\sim500$~yr most of the ice particles evolved under the fragmentation thresholds (90\%) and thus, into a pure-growth regime (for sulfides the 90\% threshold is reached at $t\sim1000$~yr) (see Figs.~\ref{fig4} and \ref{fig5}). When ice and sulfides evolve under the fragmentation threshold, there is more dust compared to the G case. Indeed, in Fig.~\ref{fig9} (top) we see that after $t\sim500$~yr the $m_{\rm d}/m_{\rm d(start)}$ ratio is $\sim0.7$ of the initial amount in GF and  $\sim0.5$ in  G (given the overall efficiency of dust settling). At $t\sim1000$~yr the $m_{\rm d}/m_{\rm d(start)}$ ratio is $\sim0.55$ of the initial amount in GF and  $\sim0.35$ in  G. Since the efficiency of growth is proportional to the dust mass present around the dust particle \citep[]{2008A&A...487..265L}, when ice and sulfides turn into the pure-growth regime, there is more dust available and, as a consequence more collisions and thus, more growth.
\begin{figure*}
{\includegraphics[width=1.8\columnwidth]{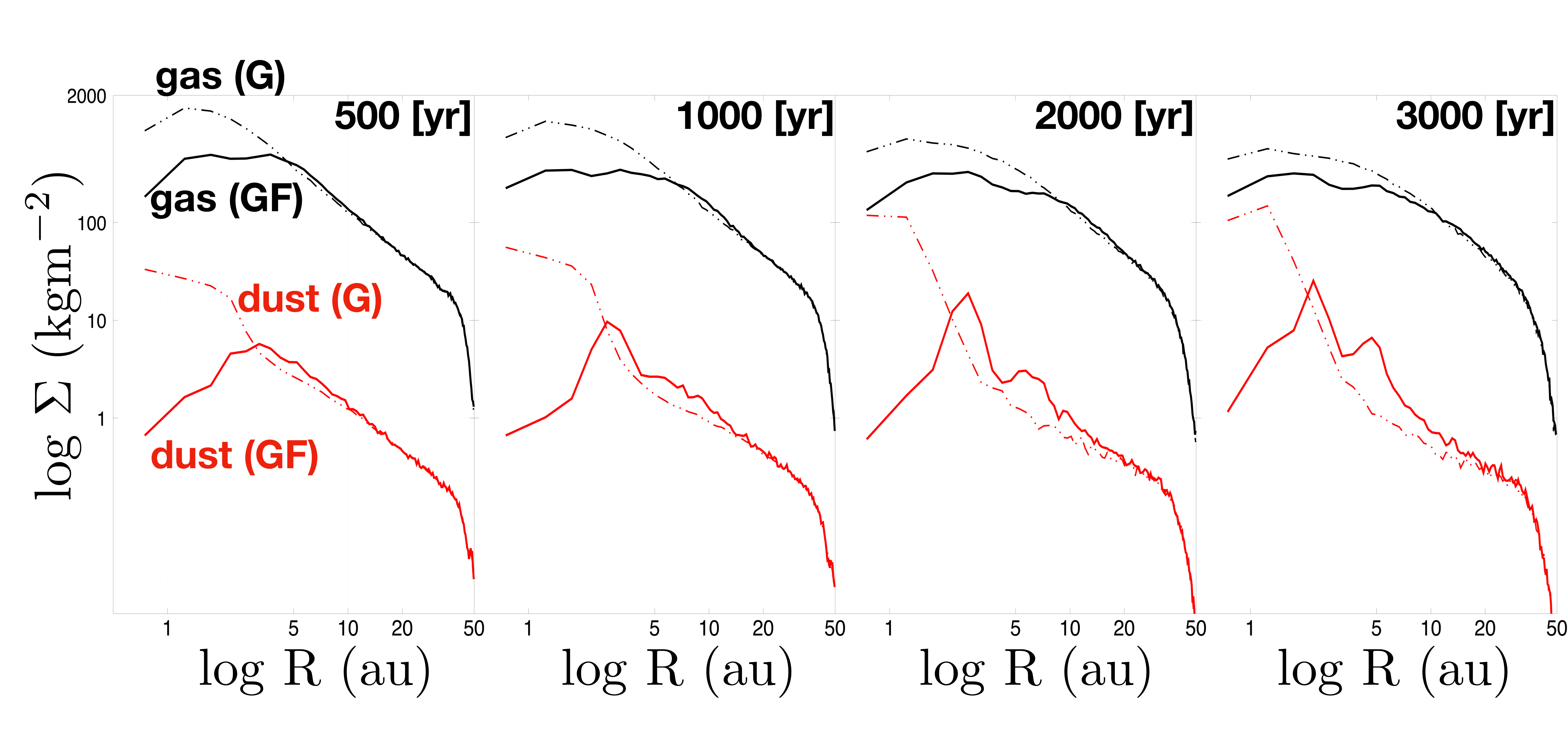}}
%{\includegraphics[width=1.\columnwidth]{FIG14b.png}}\\
%{\includegraphics[width=1.0\columnwidth]{growth-sd.png}}
%{\includegraphics[width=1.0\columnwidth]{frag-sd.png}}\\
\caption{ Gas (black) and dust (red) surface density for G  (dash-dotted lines) and GF (solid line). As dust accumulates in the inner disc due to fragmentation, the effect of the back-reaction of the dust on gas becomes more evident as it causes two enhancements of the gas pressure, i.e. two self-induced dust traps, at the locations of the two dust accumulations.\label{fig14}}
\end{figure*}

Nevertheless, there is another important aspect which has to be taken into account. In Fig.~\ref{fig14} we report the time evolution of the gas and dust  surface density for G and GF. In G the dust and gas profiles do not change dramatically with time. Dust drifts toward the inner disc past our inner disc limit and eventually onto the star, or decouples from the gas, piling up at the inner disc edge. Indeed, the inner boundary of our simulations at 0.5~au mimics a central cavity with a sharp drop of gas density at the very inner rim. The gas density has a maximum just outside this drop, where grains having decoupled accumulate. Their back-reaction further enhances this gas maximum. These results are in very good agreement with what has been found in earlier work \citep[\citetalias{2017MNRAS.469..237P}, and references therein]{2008A&A...487..265L}. In GF we see that dust piles up at two locations in the inner region ($R\sim2$ and $\sim5$~au) but not at the inner edge where the gas density, similarly to the G case, has a maximum at early times. These two locations are self-induced dust traps and form via a mechanism explained and investigated in detail by \citet{2017MNRAS.467.1984G,2017MNRAS.472.1162G} and summarized here. Drifting dust grains reach a location where their relative velocity is larger than their fragmentation velocity. As a consequence, dust starts to fragment and slow down its drift. As the dust is accumulating, given the back-reaction of dust on gas, the gas profile is affected in correspondence of the dust peaks, generating a gas pressure maximum, i.e.\ a self-induced dust trap at the same two locations, $\sim2$ and $\sim5$~au, in the GF curves of Fig.~\ref{fig14}. In their Appendix~B, \citet{2017MNRAS.467.1984G} showed that the location of a self-induced dust trap is a function of the fragmentation velocity: $r_\mathrm{trap}\propto v_\mathrm{frag}^{-2/q}$, where $q$ is the exponent of the power law for the temperature profile. For our disc model, $r_\mathrm{trap}\propto v_\mathrm{frag}^{-8/3}$. Here, the trap at 2~au is caused by the accumulation of ice, with $v_\mathrm{frag}=56$~m\,s$^{-1}$, while that at 5~au is attributed to the other species, which all have more similar fragmentation velocities of the order of 40~m\,s$^{-1}$. Indeed, the expected location of the trap for this value is $r_\mathrm{trap}(\mathrm{other})=r_\mathrm{trap}(\mathrm{ice})\times (v_\mathrm{frag}(\mathrm{other})/v_\mathrm{frag}(\mathrm{ice}))^{-8/3} \sim4.9$~au, in agreement with the observed value.  Dust grains of different species drifting from the outer disc are then trapped at either of these locations and cannot drift further inwards, preventing the dust pile-up at the inner disc edge that was seen in the G case. Furthermore, even though the inner boundary at 0.5~au can alter the density profiles, its effect on our results is small since we only consider the disc regions outside the snowline at 1.87~au. As the growth is a function of the dust density \citep[\citetalias{2017MNRAS.469..237P}]{2008A&A...487..265L}, we would expect  an overall increase of the average size of dust particles at $R\sim2$ and $\sim5$~au, after the first 2000~yr, when most of the relative velocities between dust particles have decreased below the fragmentation threshold (see Fig.\ref{fig5}). We do indeed observe this in the bottom-right panel of Fig.~\ref{fig3}. Note that a similar behaviour can be expected with other species-dependent trapping mechanisms, such as snow lines.

\begin{figure*}
{\includegraphics[width=1.0\columnwidth]{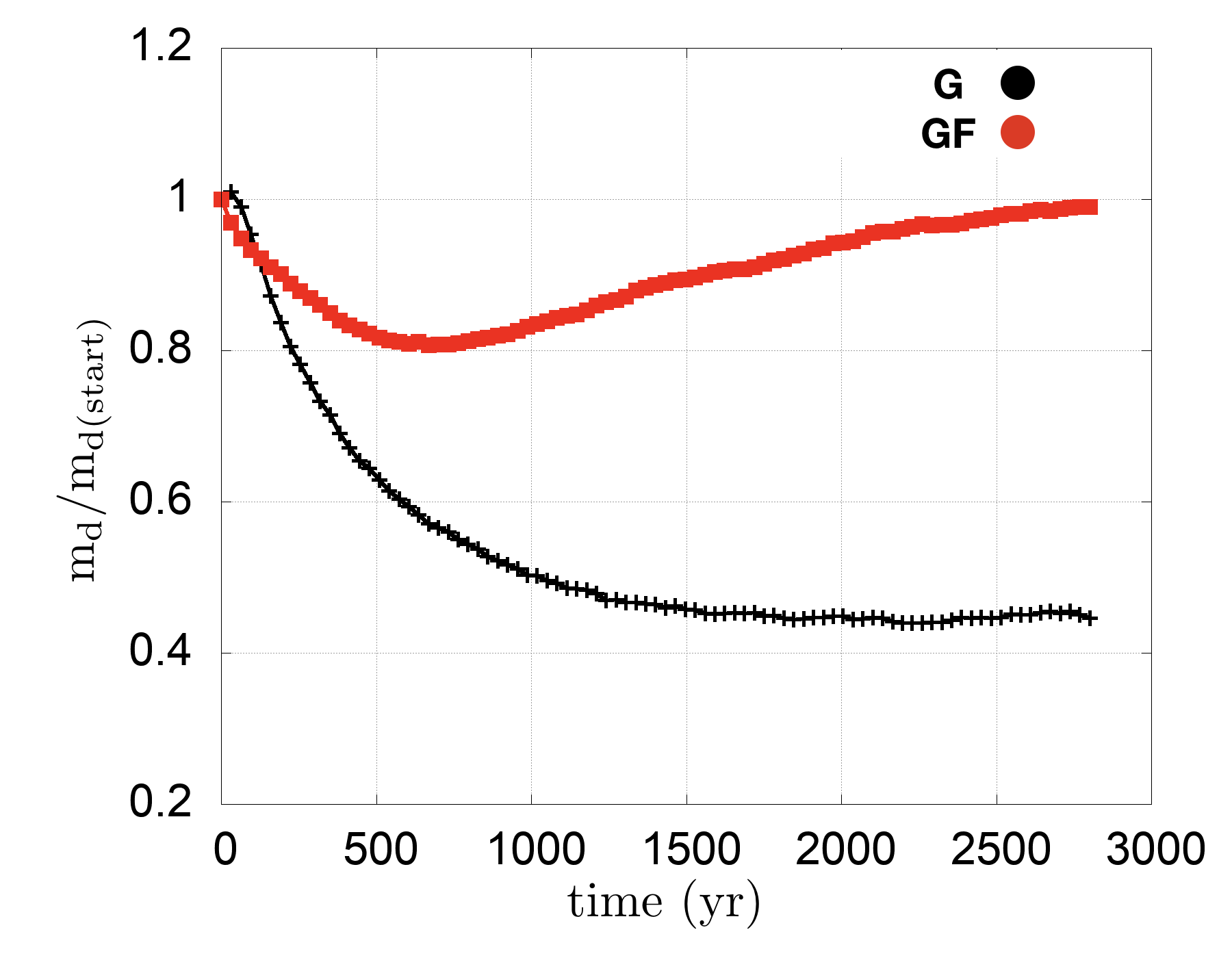}}
{\includegraphics[width=1.0\columnwidth]{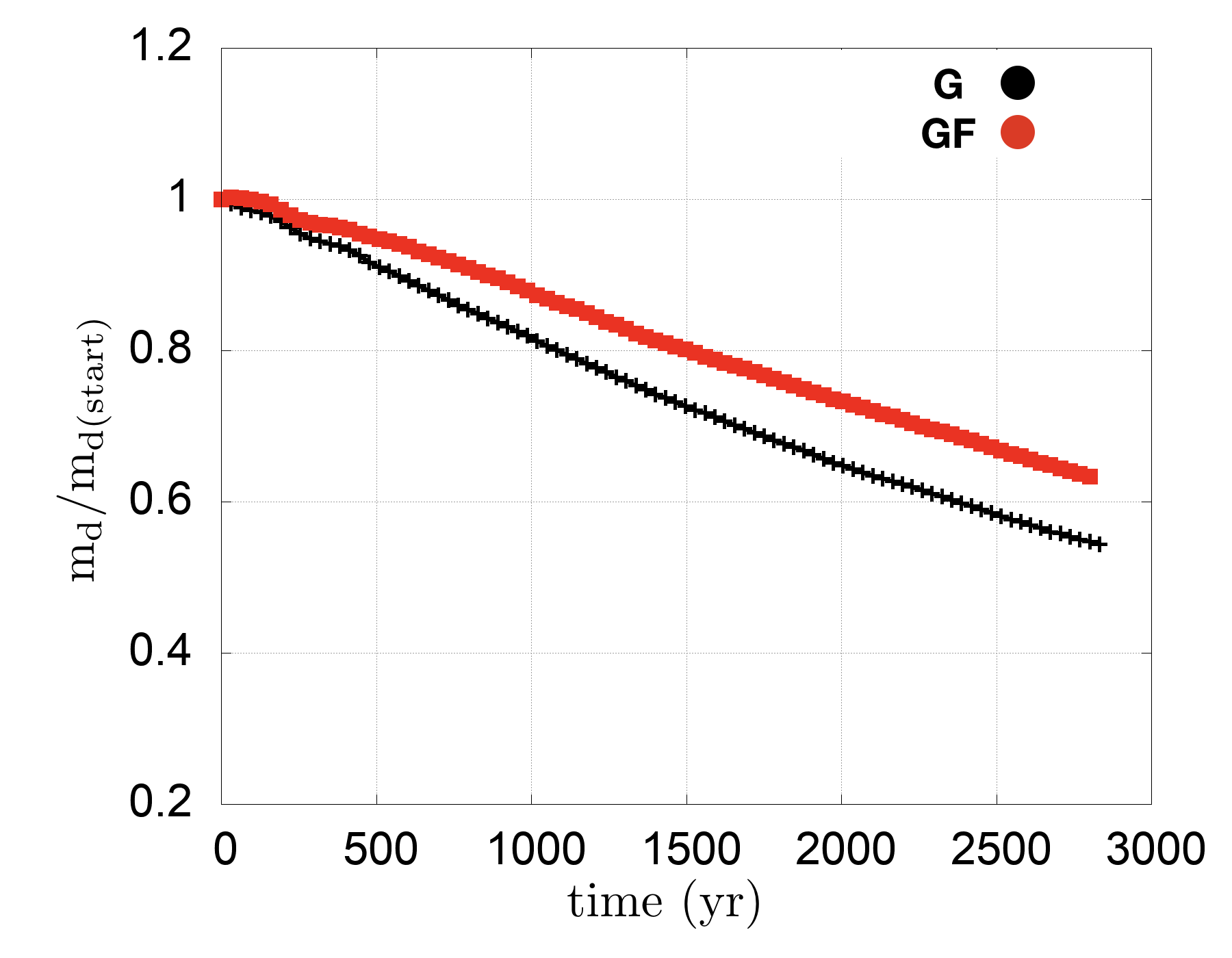}}\\
%{\includegraphics[width=1.0\columnwidth]{dust-inner.png}}
%{\includegraphics[width=1.0\columnwidth]{dust-outer.png}}\\
\caption{Time evolution of the total dust mass normalized to the initial mass in the whole (surface plus midplane) inner disc where $1.87\le R{\rm (au) \le 10}$ (left), and in the whole (surface plus midplane) outer disc where $10\le R{\rm (au) \le 50}$ (right). In the case of fragmentation, dust accumulates in the inner disc zone. \label{fig15}}
\end{figure*}

Figure~\ref{fig15} shows the time evolution of the total dust mass in the whole (surface plus midplane) inner ($1.87\le R{\rm (au) \le 10}$) disc (left) and in the outer ($10\le R{\rm (au) \le 50}$) disc (right). Figure~\ref{fig15} (left) illustrates the balance between accretion of the dust into the very inner region $R\le1.87$~au and the radial drift of particles which come from the outer disc ($R>10$~au).  Figure~\ref{fig15} (right) shows that the rate of depletion (or drift) from the outer disc is not so different when pure-growth and fragmentation are taken into account, as the fragmentation is not so efficient in that zone of the disc. However, as expected, fragmentation slows the drift of particles as the optimal drift size is reached within longer times (see Fig.\ref{fig3}). Figure~\ref{fig15} (left) clearly shows that in the G case dust ``accretes'' efficiently onto the star until the accretion rate reaches an equilibrium with the material that is moving inward from the outer disc and a balance between accretion and the initial drift from the outer disc is reached within $t\sim3000$~yr: grains grow, reach the optimal drift size and drift. In GF we see that, after a first stage in which accretion is emptying the dust content in the inner disc, at $t\sim1000$~yr, the trend is inverted and the amount of dust increases. After $t\sim3000$~yr the dust mass in the disc, in GF, is again comparable to the initial mass, while in G the dust mass in the disc is  $\sim0.4$ times the initial value. 

In GF, after  $t\sim500$~yr  most (90\%) of the ice and, after $t\sim1000$~yr, most of  sulfide particles are not  in the fragmentation regime  anymore (see Figs.~\ref{fig4} and~\ref{fig5}) and they grow larger compared to G. These particles are well above their optimal drift size (see Figs.~\ref{fig3}), and thus they decouple from the gas and pile up efficiently in the inner disc region. Moreover,  dust particles which come from the outer disc move in an environment in which the relative velocities are very close to the fragmentation threshold. This is the case of wustite and silicates particles which, after $t\sim1500$~yr are still in the fragmentation regime. Silicates and wustite particles fragment and thus, stop their drift and pile up. 

The plateau around $t\sim1500$~yr in in Fig.~\ref{fig7}  shows that  the average size  profile of wustite and silicates become flatter (the average size now takes  in account the smaller and fragmenting grains which are drifting from the outer disc) and then start to grow again around $t\sim2000$~yr, when it is clear (see Fig.~\ref{fig5} (middle)) that these species are into the pure-growth regime. At $t\sim2000$~yr, $\sim90\%$  of silicate and wustite grains are under the fragmentation threshold and start a pure growth which results in larger average sizes in the inner disc $5<R\rm{(au)<10}$ (see Fig.~\ref{fig3}). The drift also explains the second peak of the $\zeta_{\rm sul}/\zeta_{\rm sil}$ curve seen in Fig.~\ref{fig8} for the midplane: it occurs when the silicate particles, which are drifting from the outer disc, enter the inner disc zone experiencing fragmentation, and, thus, ``unsort'' with the sulfides.

\subsection{Later evolutionary stages}
\label{latestages}
\begin{figure*}
{\includegraphics[width=1.0\columnwidth]{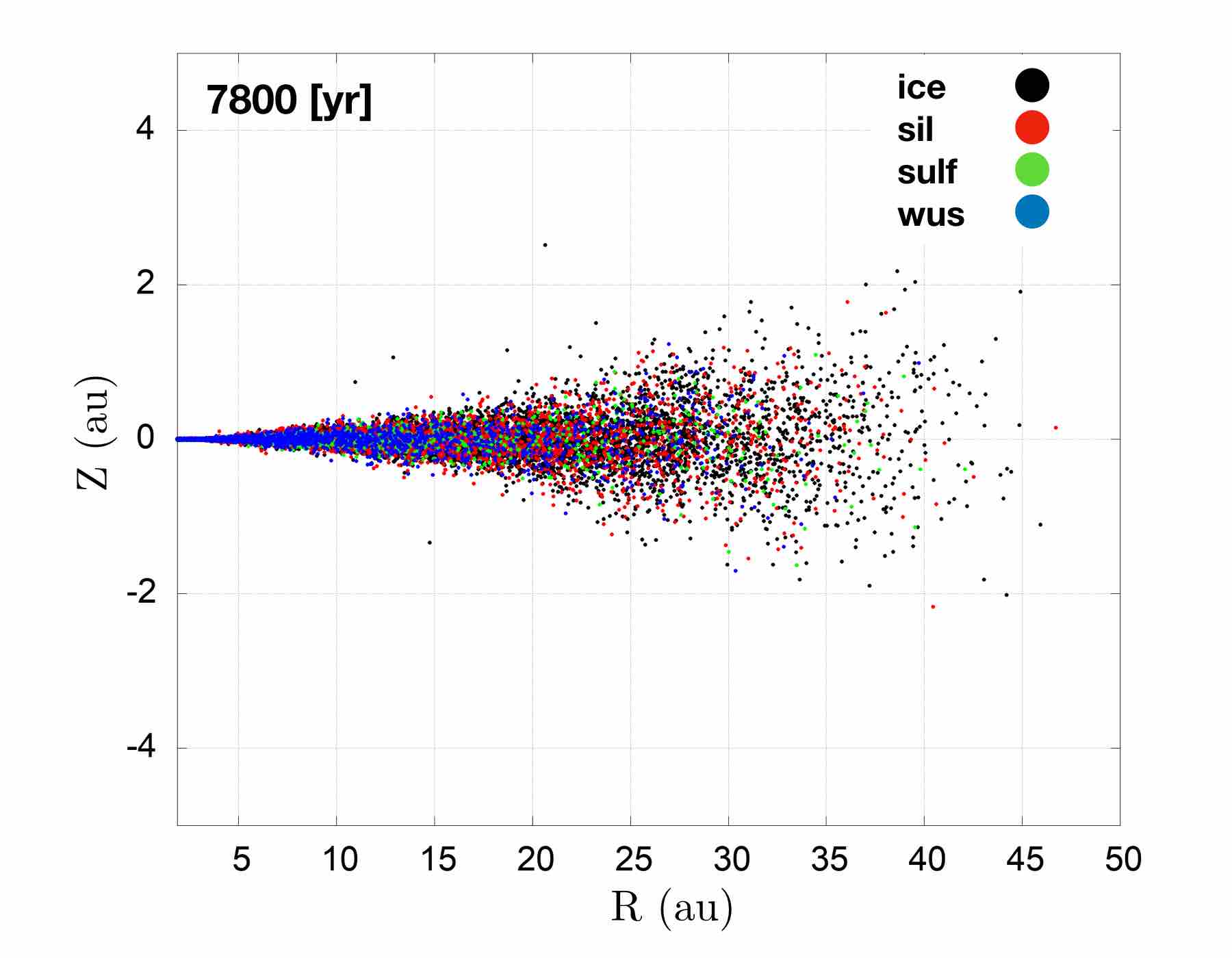}}
{\includegraphics[width=1.0\columnwidth]{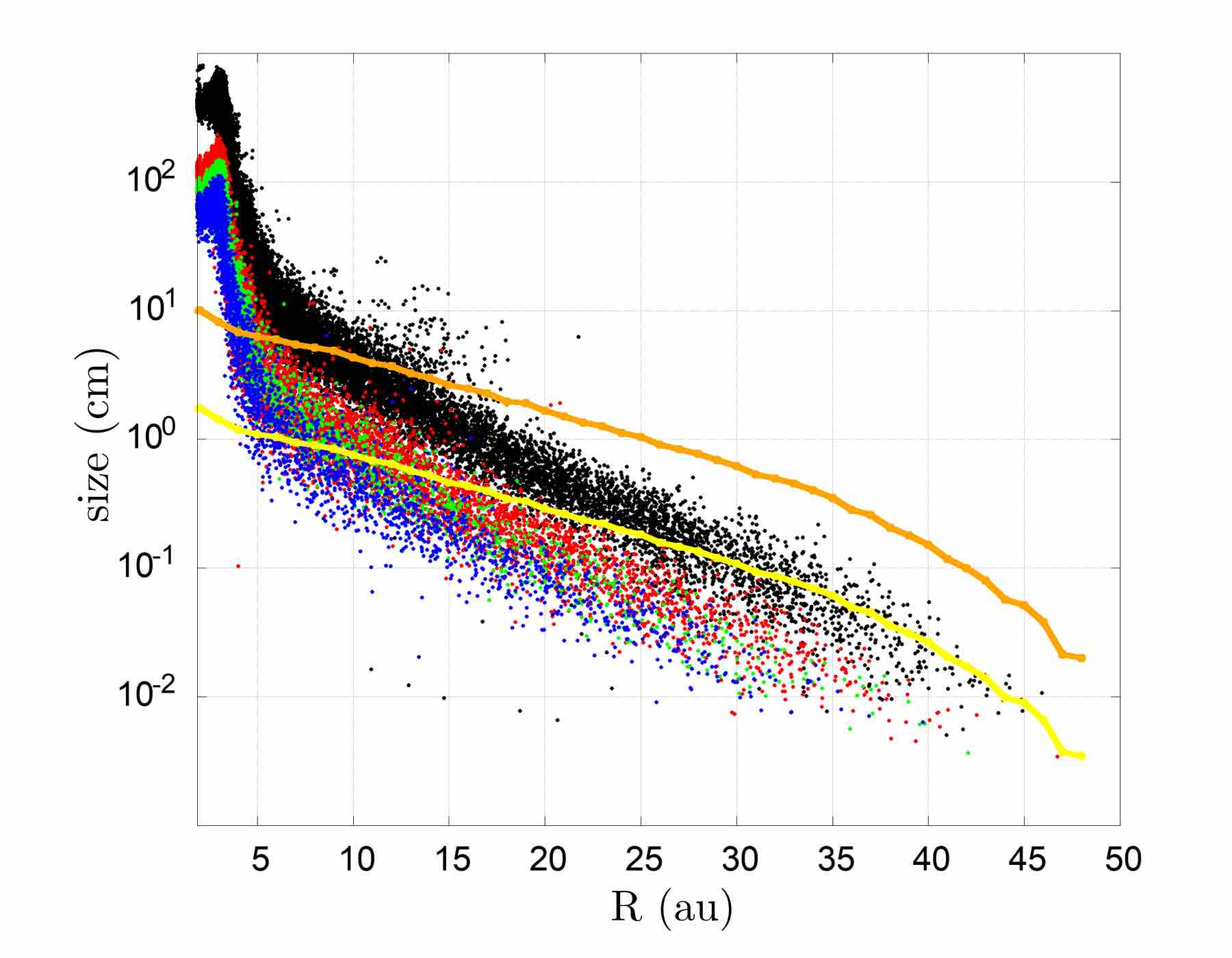}}\\
{\includegraphics[width=1.0\columnwidth]{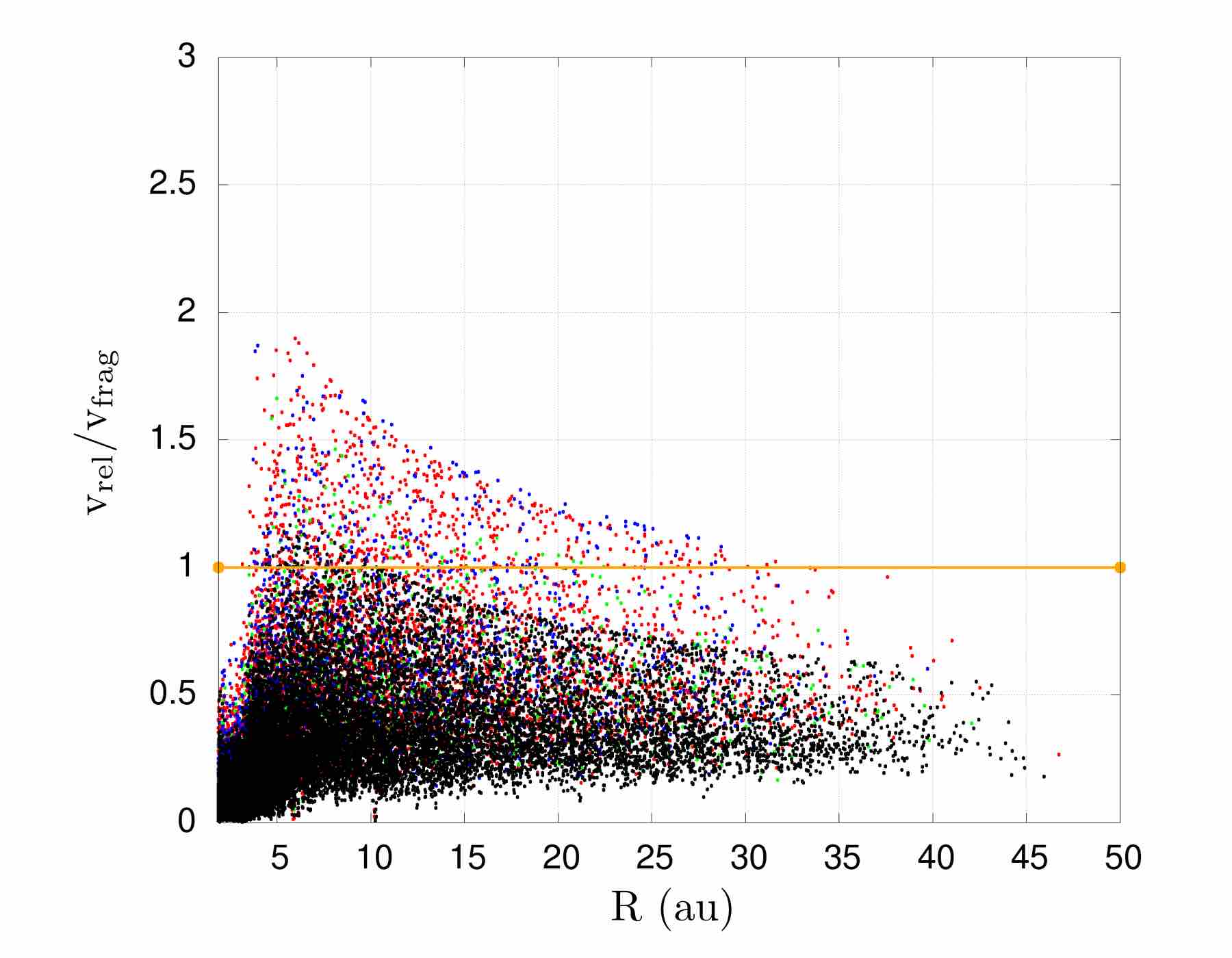}}
%{\includegraphics[width=1.0\columnwidth]{final1.png}}\\
%{\includegraphics[width=1.0\columnwidth]{final5.png}}
{\includegraphics[width=1.0\columnwidth]{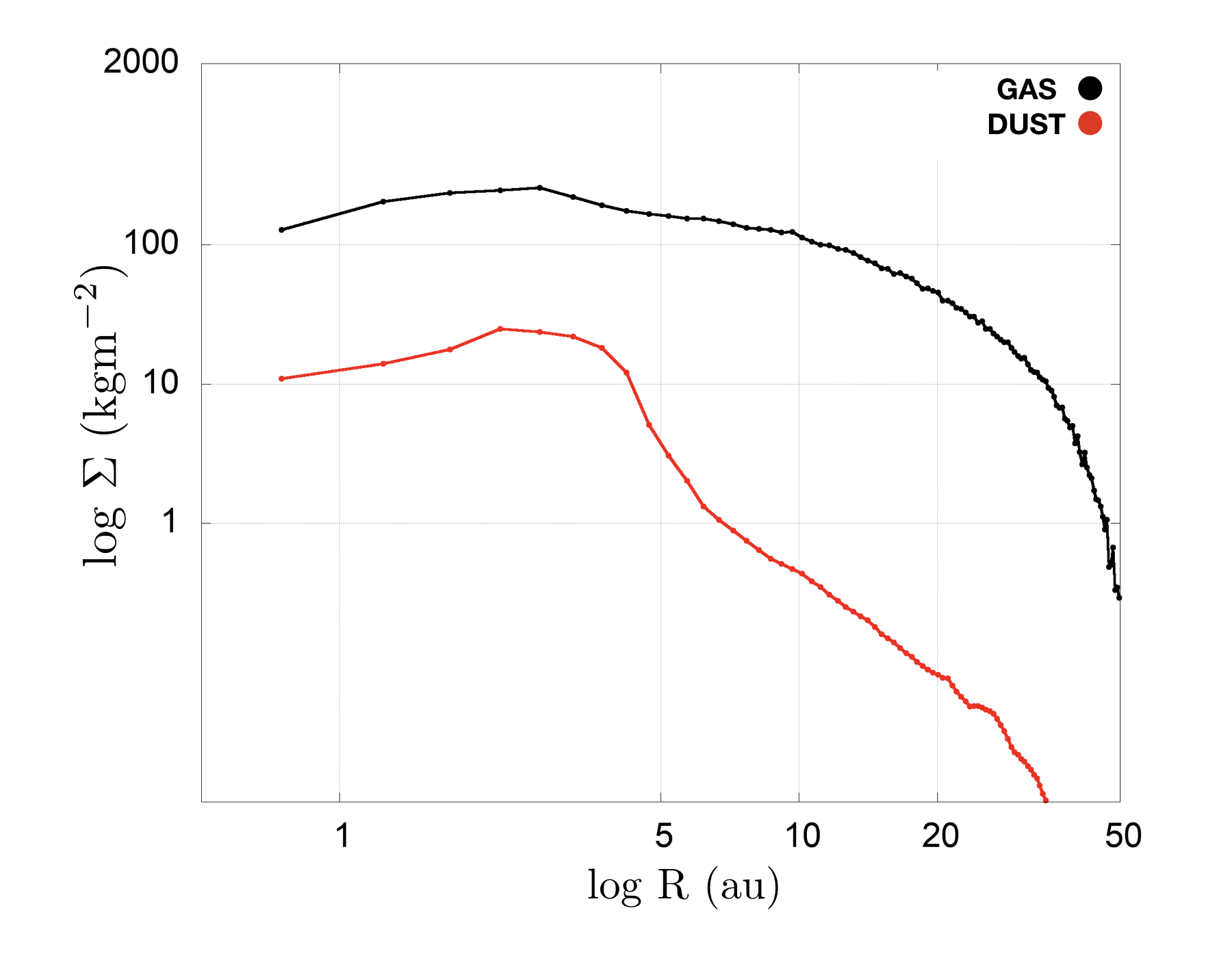}}\\
%{\includegraphics[width=1.0\columnwidth]{final3.png}}
{\includegraphics[width=1.0\columnwidth]{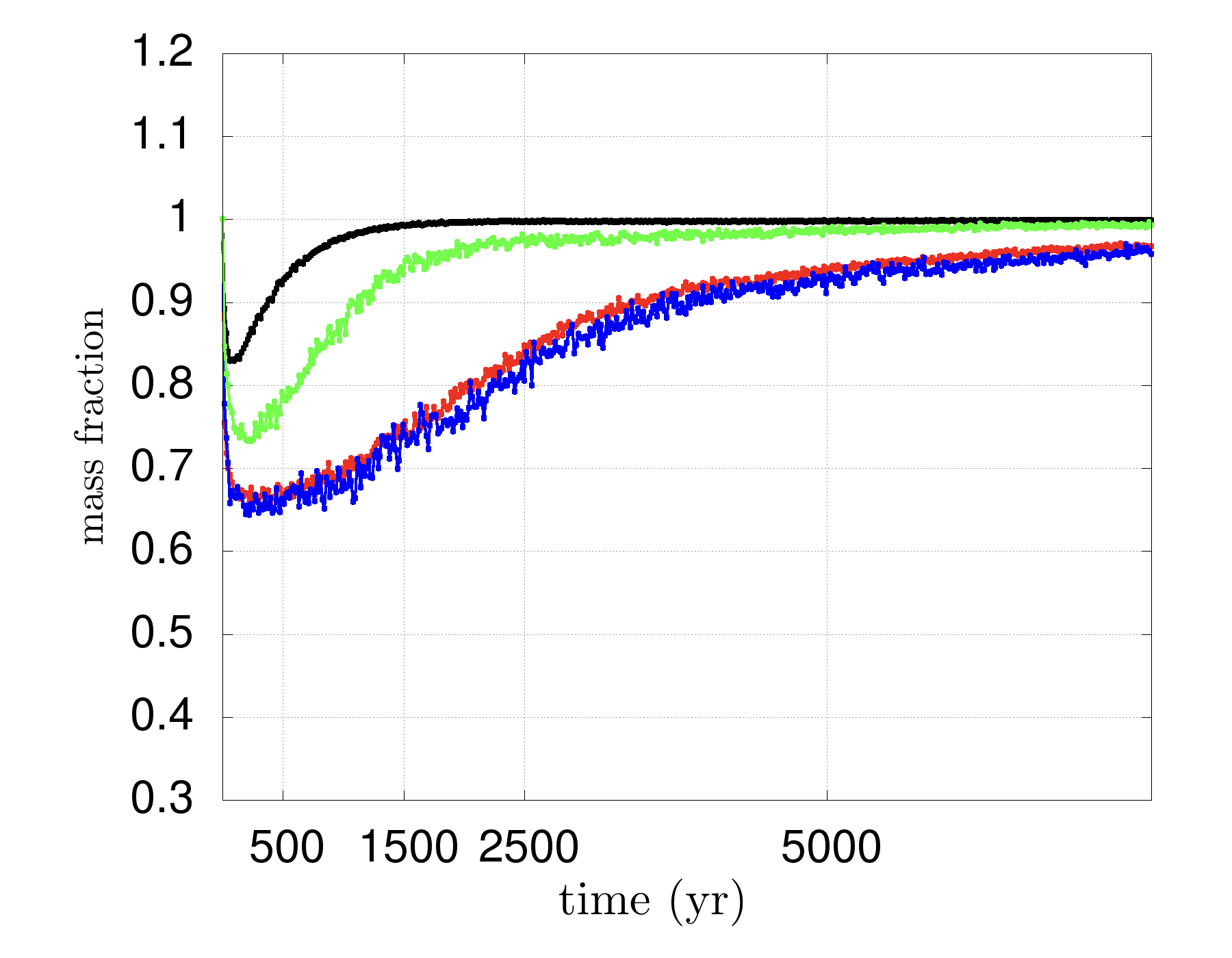}}\\
%{\includegraphics[width=1.0\columnwidth]{final4.png}}\\
\caption{Final stage of the GF simulation. Top: dust distribution ($R$ vs $Z$, left), size distribution ($R$ vs size, right), Middle: $v_{rel}/v_{frag}$ (left), gas and dust surface densities (right). Bottom: mass fraction of the dust  (by species) which is under the fragmentation threshold within 10~au. At this evolutionary stage ($t\sim7800$~yr) most of the dust is in the growth regime. A self-induced dust trap can be clearly seen at $\sim3.5$~au. \label{fig16}}
\end{figure*}

In order to investigate the evolution of the GF simulation within longer timescales, we further evolved our GF simulation for over $t\sim7000$~yr. In Fig.~\ref{fig16} we report all the quantities illustrated in the previous Sections for this final stage. We can see that our disc evolved toward a state for which most of the dust is under the fragmentation threshold.  The continued dust drift from the outer disc has increased the dust pile-up in the outer self-induced dust trap and both traps have merged into a single, broader one at $\sim3.5$~au. Most of the disc is now in a pure growth regime, and this stage compares well overall with results found in \citetalias{2017MNRAS.469..237P}, except for the location of the dust pile-up.

\subsection{Small grains and planetesimal formation}
\label{steady}
\citet{2005A&A...434..971D} investigated the effect of  growth and fragmentation of dust in protoplanetary discs. They pointed out that if growth is efficient and dominating on the fragmentation, it should deplete the small grains quickly with derived timescales that are not compatible with the IR spectra of discs. In fact, observations do show the presence of sub micron and micron size grain on the disc surface. \citet{2005A&A...434..971D} suggested that  growth and fragmentation result in a steady size distribution with a constant replenishment of smaller grains. \citet{2010A&A...513A..79B,2012A&A...539A.148B} showed that   fragmentation  is a required and important mechanism to  prevent strong radial drift and thus depletion of dust that can make the process of planetesimal formation difficult.   \citet{2012A&A...539A.148B}  also found that inner discs are generally fragmentation-dominated while outer discs are growth-dominated and then conclude that discs have to be in a strong turbulence regime, as drift alone cannot support efficient fragmentation over long evolutionary timescales. \citet{2011ApJ...728...20S} pointed out that the sublimation of the ice content of composite large grains that drift from the outer region  can release large quantities of rocky material whose smaller size would allow a re-coupling to the gas phase. This would replenish the inner disc with smaller grains. \citet{2017A&A...608A..92D} then  suggest that this dust would not be free to drift in the inner disc but it will pile-up close to the snow line and enhance  planet formation.

Our calculations show some analogies with the effects found in previous work: as the drift becomes efficient, the dust moves in the inner disc, fragments, and piles up at specific locations that depends on the density of the grains and their fragmentation properties. In our case, the dust pile-up is caused by the back-reaction of the dust onto the gas, that is not included in \citet{2010A&A...513A..79B,2012A&A...539A.148B}.
 
 However, we do not find an overall  steady state distribution of grains size within the considered time. This is because, according to our disc model, the relative velocities of the dust species evolve as a function of time, transitioning under the considered fragmentation threshold. Only in the very inner disc that we consider, a small percentage of dust is still in the fragmentation regime. However, we do expect the effects of fragmentation to have a variable duration  as the considered threshold velocities and disc parameters change.

In order to preserve a detectable quantity of smaller grains in the inner surface of old  discs,  fragmentation regimes should last longer in these regions  (or growth should be inefficient). On the other hand,  in  discs like GG Tau and  TW Hya where size sorting and stratification is detected \citep{2007A&A...469..963P,2000ApJ...534L.101W,2003ApJ...596..597W,2007ApJ...664..536H,2008ApJ...678.1119H,2012ApJ...744..162A,2014A&A...564A..93M} planetesimal and planet formation are also thought to be underway \citep{2017ApJ...837..132V}. This is also thought to occur in relatively younger objects such as HL Tau \citep{2015MNRAS.453L..73D,2015ApJ...812L..38T}.

Nevertheless, dating of achondrite meteorites (samples of already differentiated planetesimals) shows that accretion of their parent bodies occurred as early as 1 Myr after the formation of the CAIs \citep{2009GeCoA..73.5150K,2014M&PS...49..772S}.  To match  disc observations (dust size-sorting and  gaps by forming planets) with Solar System's achondrite ages,  fragmentation and growth regimes have to be  efficient at the same time, but, in different locations.

In Fig.~\ref{fig5} it can be seen that the dust in the disc surface remains, after $t\sim1000$~yr, slightly more fragmenting  than the dust located in the disc midplane. This is probably a consequence of the average smaller size of the particles in the disc surface (Fig.~\ref{fig7}) that keeps them in the fragmentation regime as the relative velocities are a function of the particle size.

Although limited by resolution and model constraints, our 3D simulations  suggest that  two different growth regimes  characterize the disc surface and the midplane. This could explain the observable discrepancies between the sub-micron and micron sized dust  in disc  surfaces and planetesimal formation and differentiation in the midplane. Moreover, if aggregates of different composition and size have different fragmentation properties we can speculate the presence in the disc of different fragmentation lines (or zones) that would act similarly to the disc particle traps or the snow/sublimation lines where dust can pile up efficiently (see section~\ref{discdifferences}). 

%CAIs, are the first solids that formed in our parent protoplanetary disks and recent work  suggests that their precursor were produced during the collapse of the parent cloud  of the Solar System  \citep[and reference therein]{2018ApJ...867L..23P} given the timescales of collapsing \citep{2011ARA&A..49...67W}. 

%Nevertheless, it is now widely accepted that dust accretion occur  from the disc surface rather than from the midplane that is largely inactive \citep{1996ApJ...457..355G,2010ApJ...722.1437B,2013A&A...552A..71F}.

\subsection{Aerodynamical sorting in chondrites}
\label{chondrites}

In this section, we explore whether the combined effects of transport and fragmentation described in previous sections can be used for explaining the chemical variations in the composition of chondrites and planetary objects. 

{In \citetalias{2017MNRAS.469..237P} we suggested that  size-density sorting that occurred via gas-dust interaction in an evolving disc could have played an important role in determining the physical and compositional properties seen in chondrites \citep{1998LPI....29.1457B,1999Icar..141...96K}.

One salient compositional feature of chondrites is the variation in the abundance of metallic iron/iron-sulfide (see Fig.2 and the Urey-Craig diagram in Fig.3 of \citetalias{2017MNRAS.469..237P}, also reported in \citet{2006mess.book..803R}). The relative abundance of reduced iron phases is controlled by the redox conditions whereby oxidizing conditions lead to more abundant FeO with limited metallic iron. For example CI chondrites and H chondrites have similar $\rm{Fe_{tot}/Si}$ ratios (where $\rm{Fe_{tot}}$ refers to the total iron content) but the amount of reduced iron is much greater in H chondrites. A second important parameter that may be influenced by aerodynamic sorting is the abundance of Fe (present as Fe or FeS) relative to silicates where iron is incorporated as FeO. An example is given by enstatite chondrites: EH and EL chondrites are similarly reduced but the EL chondrites have a low metallic iron content relative to EH chondrites. In what follows, the variations of metal content relative to silicates will be discussed as the behaviour of metallic iron can be easily extrapolated from our simulations. It is assumed here that the fragmentation threshold  of metallic particles are on the order of $\sim100~\rm{m\,s^{-1}}$ as suggested by \citet{2014ApJ...783L..36Y} or higher than the silicates values as experimental evidences suggest. This would allow a more evident separation between the Fe-rich particles and the Si-rich particles, with a behaviour of Fe-metal closer to that of the ice.

The existence of sorting according to size-density has already been described in chondrites \citep{1998LPI....29.1457B,1999Icar..141...96K,2001ApJ...546..496C}. First, it was shown that there is a relatively narrow size distribution for chondrules in ordinary chondrites. Furthermore, it was shown that the size of chondrules decreases from LL to L and H chondrites \citep{1989Metic..24..179R} (see also Fig.1 in \citetalias{2017MNRAS.469..237P}). This work was extended to metal grains by \citet{1999Icar..141...96K} who determined the size distribution of both chondrules and metals in the same ordinary chondrites H, L and LL. Overall, the LL chondrules are larger than L and H chondrules, while the metallic grains are slightly smaller. Similarly, \citet{2003GeoRL..30.1420S} showed that the metal grains and chondrules are larger in EL chondrites relative to EH chondrites, while EL chondrites are overall depleted in metallic iron, for a similar redox state. An additional observation was that the total iron content of H and EH chondrites is overall similar to that of CI chondrites, indicating that the EL, L and LL chondrites reflect a depletion of iron relative to the solar composition.  It was argued by \citet{2014Icar..232..176J} that these features could indicate sorting of metal grains relative to silicates represented by chondrules. 

It has long been known that the variations of $\rm{Fe_{tot}/Si}$ ratios in ordinary chondrites reflect a fractionation between metallic iron and silicates in protoplanetary discs \citep{1970GeCoA..34..367L,1973GeCoA..37.1603L}. However, the mechanism responsible for this fractionation has remained elusive. While the possibility of aerodynamic sorting was suggested, this mechanism has not been explored quantitatively. \citet{1972RvGSP..10..711W} argued that the separation between metal and silicates in ordinary chondrites was clearly an early process, as the refractory siderophile elements  are not equally enriched in the metal of H, L and LL chondrites, indicating that the process of metal separation took place at high, albeit different temperatures.  Furthermore, \citet{2017LPI....48.2046H} have also shown that the timing of Hf-W fractionation, itself indicative of metal-silicate separation between the three ordinary chondrite groups was estimated to be around $\sim$2 Myr after CAIs (Calcium Aluminium rich Inclusions), also suggesting an early process.

\citet{1999Icar..141...96K} have argued that the observed patterns in ordinary chondrites could result from aerodynamic sorting. Interestingly the ratio of $\zeta_\mathrm{metal}/\zeta_\mathrm{chondrules}$ ranges between 1.49 and 0.84 from LL to H chondrites, with the H chondrites having a $\zeta_\mathrm{metal}/\zeta_\mathrm{chondrules}$ closer to 1. This could mean that metal and silicates were transported jointly yielding little $\rm{Fe_{tot}/Si}$ fractionation, which is consistent with the $\rm{Fe_{tot}/Si}$ of H chondrites similar to that of CI chondrites ($\sim$ solar composition). In contrast, the LL chondrites have a higher $\zeta_\mathrm{metal}/\zeta_\mathrm{chondrules}$ and are characterized by metallic iron content (typically 3\%) lower than other ordinary chondrites (8-15\%). Thus, the LL chondrites represent a reservoir with a low $\rm{Fe_{tot}/Si}$. These observations may result from the process described in section~\ref{discdifferences} and depicted in Figures~\ref{fig6} and~\ref{fig8}, showing variations in the $\rm{Fe_{tot}/S}i$ in the disc surface compared with the midplane. A limitation to this reasoning could arise if the observed $\rm{Fe_{tot}/Si}$ fractionation was not related to the observed metal and chondrule grains but to an earlier generation of precursor grains, in which case the aerodynamic properties could have been different. Recent $^{207}$Pb--$^{206}$Pb and $^{26}$Mg--$^{26}$Al chronology of chondrules \citep{2017SciA....3E0407B} suggests that most primary chondrules formed $<1$~Myr after the beginning of the solar system. However, it could be argued that these ages are not consistent with numerous $^{26}$Mg--$^{26}$Al observations (e.g. \citet{2009Sci...325..985V}). While the timing of these processes may have to be better defined, the observations in chondrites are indeed suggestive of the process described in our model. 

In contrast with ordinary chondrites, the CH and CB chondrites are globally enriched in metallic iron \citep{2007AREPS..35..577S}, while being characterized by a higher $\rm{Fe_{tot}/S}i$ ratio than CI chondrites. Thus, a mechanism similar to that outlined here could have enriched metallic iron or iron sulfide relative to silicates. At a larger scale, Mercury is similarly reduced to EH chondrites \citep{2011Sci...333.1847N} but it has a higher abundance of metallic iron and a higher $\rm{Fe_{tot}/S}i$ ratio than CI chondrites.  The unusual compositional features of Mercury may also stem from the process described in this study and \citetalias{2017MNRAS.469..237P} (iron enrichment in the inner midplane, see also \citet[and reference therein]{2016MNRAS.457.1359P}), provided the dust enriched in Fe was rapidly accreted into planetesimals before further changes in the overall composition could take place. While this interpretation is speculative and would deserve more in-depth investigations, the new process of Fe enrichment stemming from density contrasts coupled with differences in fragmentation behavior suggests that it could have played an important role in explaining the abundance of iron in planetary materials.

In term of size-density sorting of chondritic components, our work on fragmentation adds several further constraints.  In order to allow  aerodynamic sorting during the fragmentation regime, our results suggest that  different chondritic components need to have the same fragmentation properties, but evidence suggests that this is not the case. To preserve the aerodynamical sorting (i) aggregation of chondritic material should have occurred when or where  fragmentation was not efficient ($v_{\rm rel}<v_{\rm frag}$), or  (ii) the single components in chondrites were highly resistant (very high $v_{\rm frag}$), or (iii) single components in chondrites already stopped growing and fragmenting at the time of accretion with their sorting dictated only by the gas drag.  As a consequence, aerodynamic sorting of metallic grains, sulfides and chondrules should have occurred following one of these three conditions. These conditions are compatible with the suggested location (around the snow line) and aggregation timescale (after $t\sim1.2$~Myr from CAI formation) proposed for the parent body formation of ordinary and carbonaceous chondrites  \citep{2015aste.book..635K,2014M&PS...49..772S}.

%If almost all the chondrites are characterized by a certain degree of aerodynamic sorting, the CB-CH-chondrites show instead interesting differences. These  chondrites contains large volume fraction  of fragments  ($\sim25\%$) and chondrules fragments ($\sim14\%$) \citep{1993GeCoA..57.2631B}. CH chondrites also contain large amount of metallic grains whose apparent diameter is larger than the chondrules \citep{1999JGR...10422053M}.  This is in contrast with  aerodynamical sorting where heavier grains should be smaller when compared with lighter grains. CH chondrites apparently accreted after a series of heating events and in turbulent environments \citep{2007M&PS...42..913G}. This suggests that accretion of these components may have occurred in a disc location where a strong fragmentation regime  was in place and where particles tends to be unsorted, as in planetesimals collisions \citep{2005Natur.436..989K}. Moreover, the fact that most of the  fragments found in CH chondrites belong to the silicate-rich chondrules and not to metal grains further confirm that different fragmentation properties may characterize metallic grains and silicates. 

\subsection{On the fragmentation thresholds and caveats}
\label{caveat}
In this work we used as $v_{\rm frag}$ of the considered species values taken from \citet{2014ApJ...783L..36Y} and values derived in Section~\ref{chemistry}. These values are based on theoretical and experimental evaluations. As already pointed out in the introduction, the fragmentation velocities for different species are very difficult to evaluate  as the resulting threshold is a function not only of the chemical composition of the dust but also a function of the grain physical properties such as their crystalline  or amorphous structure, size, porosity, shape \citep{2008ARA&A..46...21B,2009ApJ...702.1490W,2009MNRAS.393.1584T,2010A&A...513A..57Z,2013A&A...559A..62W,2013MNRAS.435.2371M,2014ApJ...783L..36Y}. 
However, the behaviour of fragmenting dust described in this work would still be valid, in the sense that if the fragmentation velocities and bulk densities of two different species (or aggregates) are known, their behaviour can be easily extrapolated from our results. Our finding can be extended to aggregates of mixed species as well. As such, in terms of aerodynamic sorting we find that if two aggregates have the same fragmentation properties (similar fragmentation threshold) they will aerodynamically sort. If they have different fragmentation properties  they will not sort aerodynamically until the disc environment (or aggregates) evolves very close to a pure-growth regime. 

In this work we considered, for simplicity, all the fragmentation velocities as constant, i.e. with no variation with time, grain size, or other parameters. If a variation of $v_{\rm frag}$ with the size of a given species is known, it could then be compared to the variation of $v_{\rm rel}$ with size (see equation~\ref{vrelvrel}), to infer at which stage any species would move into or away from a pure-growth regime according to its size and given disc conditions. Moreover, if species are ``resistant'' to fragmentation, they will vertically sort and radially drift  driven by their intrinsic density first, and their size later, as described in \citetalias{2017MNRAS.469..237P}. If species are ``sensitive'' to  fragmentation, it will be  $v_{\rm frag}$ that will dictate the dynamical behaviour, at early stages, when compared to other species. Furthermore, different $v_{\rm frag}$ and different disc models will change the timescales at which the transition between the fragmentation regime and the pure-growth regime could occur and where fragmentation would be more efficient. 

{We do not consider, for resolution limits and scope of the work, the erosion, disruption and bouncing that large grains/bodies (s$\sim10,100$ cm) can experience upon collision. According to the type of collision, the production of smaller grains with a large size distribution \citep{2010A&A...513A..56G,geretshauser2012simulation} can occur. As a consequence, a more complex situation can be produced where grains having different size but located in the same environment can experience new episodes of fragmentation and growth.

\section{Conclusions}
\label{conclusions}

In this work we studied the effects of growth and fragmentation in determining the behavior of a multi-phase dust. Fragmentation  changes  the chemical composition of the disc with values and trends that can actually diverge when compared to the case of pure-growth.  The chemical fractionation of dust via dynamical processes is sensitive to the $v_{\rm frag}$ of the considered species. As a consequence, fragmentation can affect the bulk composition of the planetesimals that may be accreted into planets as it could  change, for example, the rock/ice ratio. Our results suggest that the chemical fractionation observed in chondrite families and Mercury's enrichment in iron could be the result of  size-density sorting and different fragmentation properties of  dust grains.

Two species become aerodynamically (size-density) sorted in a fragmentation regime only if they have the same fragmentation properties. Eventually they will be size-density sorted at later stages when the pure-growth regime takes over. Our results suggest that  chondrite components that show a size-density sorting may have accreted into larger bodies in regions of the Solar Nebula, and/or  at a time where fragmentation was not efficient or not occurring at all, thus allowing the observed degree of sorting.  

Dust in the disc can evolve toward a pure-growth regime as $v_{\rm rel}$ is regulated by the Stokes number. Similarly to \citet[]{2017MNRAS.467.1984G,2017MNRAS.472.1162G} we  found that, when fragmentation is taken into account, drifting dust can pile up at  fragmentation fronts. Then, a runaway accretion of dust occurs as the accumulating grains transition to a pure-growth regime. Counter-intuitively, taking fragmentation into account can thus produce dust aggregates which are larger than those that result from a pure-growth simulation. As a consequence, we further confirm that fragmentation may be a key mechanism to overcome the radial-drift barrier in short timescales.

The fact that micron-size dust grains are seen in discs where planetesimals formation should be well underway suggests that discs surfaces and midplanes are in different fragmentation/growth regime and/or that dust and gas accretion occur in a differential way (active surface and dead midplane) as recently proposed.

We can speculate that if aggregates with different bulk composition or different sizes have different fragmentation properties, a formation of multiple fragmentation lines in discs where dust can pile up and efficiently form large aggregates can occur. Different fragmentation thresholds and different disc parameters can extend/reduce the effects of fragmentation. 

\section*{Acknowledgments}
\label{acknowledgments}

The authors are grateful to the LABEX Lyon Institute of Origins (ANR-10-LABX-0066) of the Universit\'e de Lyon for its financial support within the program "Investissements d'Avenir" (ANR-11-IDEX-0007) of the French government operated by the National Research Agency (ANR). FCP acknowledges the financial support of ANR-15-CE31-0004-1 (ANR CRADLE) and thanks Jonathan L\'eger for helping in managing the large amount of data which resulted from the simulations. JFG acknowledges funding from contract number ANR-16-CE31-0013 (Planet-Forming-Disks). All simulations were performed at the Common Computing Facility of LABEX LIO. The authors wish to thank the anonymous referee for their detailed comments and suggestions that  greatly improved the manuscript.

%%%%%%%%%%%%%%%%%%%%%%%%%%%%%%%%%%%%%%%%%%%%%%%%%%%%%%%%%%%%%%%%%%%%%%%%%%%%%%%%%%%%%%%%%%%%%%%%
\bibliographystyle{mnras}
\bibliography{biblio}

\appendix
\section{Testing resolutions}
\label{resolutiontest}
In this appendix we verify our resulting dust behaviour against a higher resolution (400,000 total particles). At the time of injection we have  39,877 silicate particles, 11,780 wustite particles, 12,146 sulfide particles and 136,197 ice particles for a total of 200,000 dust particles. The ratios are Fe/Si=0.60 and rock/ice=0.468.

In Figures~\ref{comparison01} and~\ref{comparison03} we compare the results for the G case presented in the main sections (250,000 particles) and the results retrieved from this new simulation of the G case with 400,000 particles. For completeness, we also add, for the G case, a shorter run with a lower resolution (150,000 particles). In this lower-res simulation we have  14,953 silicate-, 4,413 wustite-, 4,544 sulfide- and 51,090 ice-particle, returning a Fe/Si=0.60 and rock/ice=0.468.

In Figures~\ref{comparison04} and~\ref{comparison06} we compare the results for the GF case presented in the main sections (250,000 particles) and the results retrieved from this new simulation of the GF case with 400,000 particles. Given the longer computational time required these simulations run up to $t\sim2000$~yr.

Similarly to figures~\ref{fig9} and~\ref{fig7}, we report the time evolution of the total dust mass content normalised to the initial mass for the disc surface and midplane (Fig.~\ref{comparison01} and~\ref{comparison04}) and the time evolution  of the  size for single species in the disc surface and midplane (Fig.~\ref{comparison03} and~\ref{comparison06}). 

We find a very similar and consistent behaviour when comparing the low- and high-resolution of the two (G and GF) simulations. All the general trends and behaviour found in the main sections of this work  are preserved. This demonstrate that the resolution does not change the overall results and their interpretation.

Finally, we test our 250,000 particles GF simulation  against the resolution criterion ($h<c_{s}t_{s}$) proposed by \citet{2012MNRAS.420.2345L}. Results are reported in Fig.~\ref{comparison07}, showing that the criterion is satisfied.

\begin{figure*}
{\includegraphics[width=0.8\columnwidth]{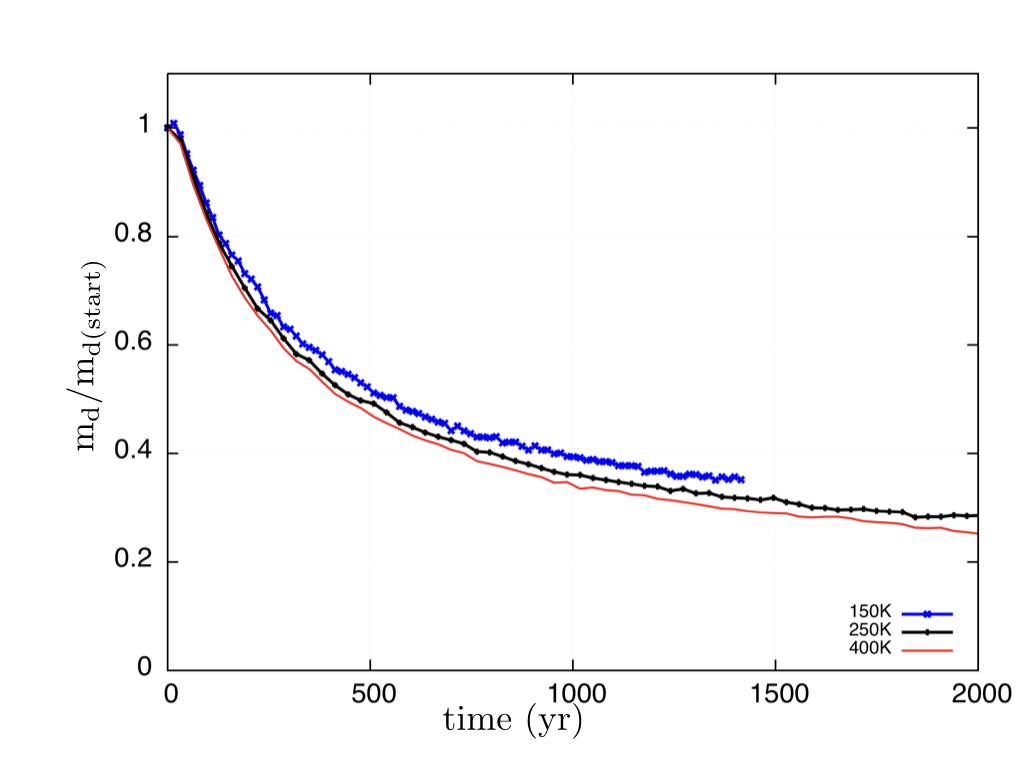}}
{\includegraphics[width=0.8\columnwidth]{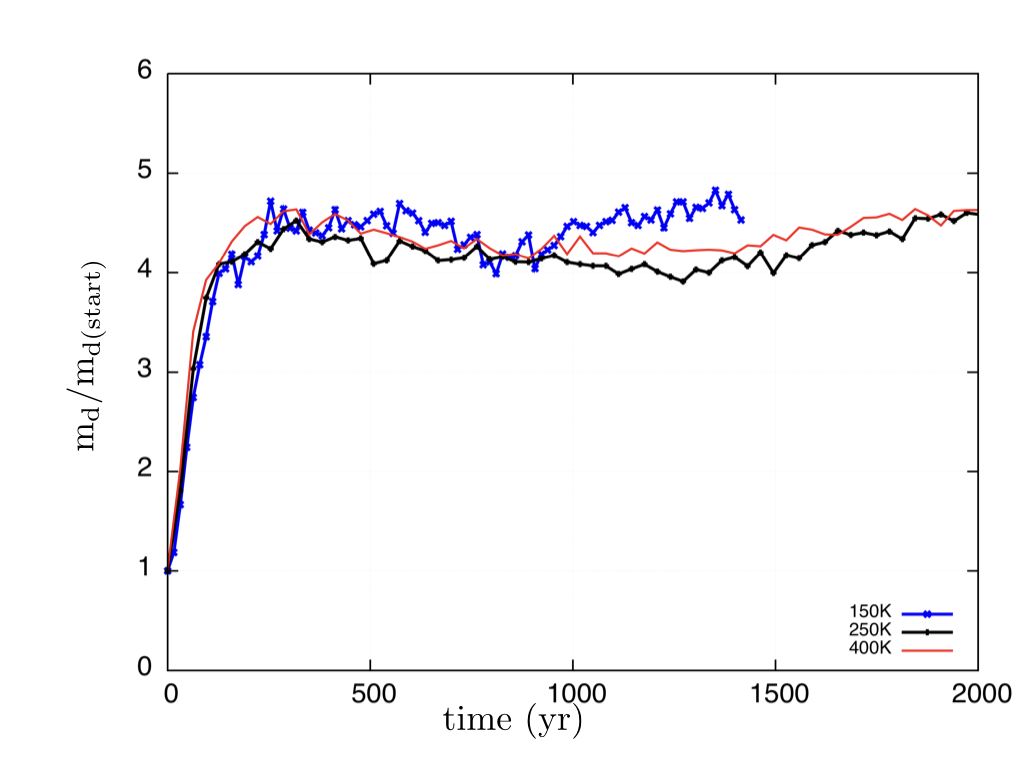}}\\
\caption{Time evolution of the total dust mass content compared to the initial mass in the disc surface (left) and midplane (right) for a G simulation with 400K particles (red line), 250K particles (black ticked line) and 150K particles (blue crossed-line). \label{comparison01}}
\end{figure*}

%\begin{figure*}
%{\includegraphics[width=0.95\columnwidth]{appendix003.png}}
%{\includegraphics[width=0.95\columnwidth]{appendix004.png}}\\
%\caption{Time evolution of the   mass content of single species compared to the initial mass in the disc surface (left-box) and midplane (right-box) for a G simulation with 400K particles (red line) and 250K particles (black ticked line). \label{comparison02}}
%\end{figure*}

\begin{figure*}
{\includegraphics[width=0.95\columnwidth]{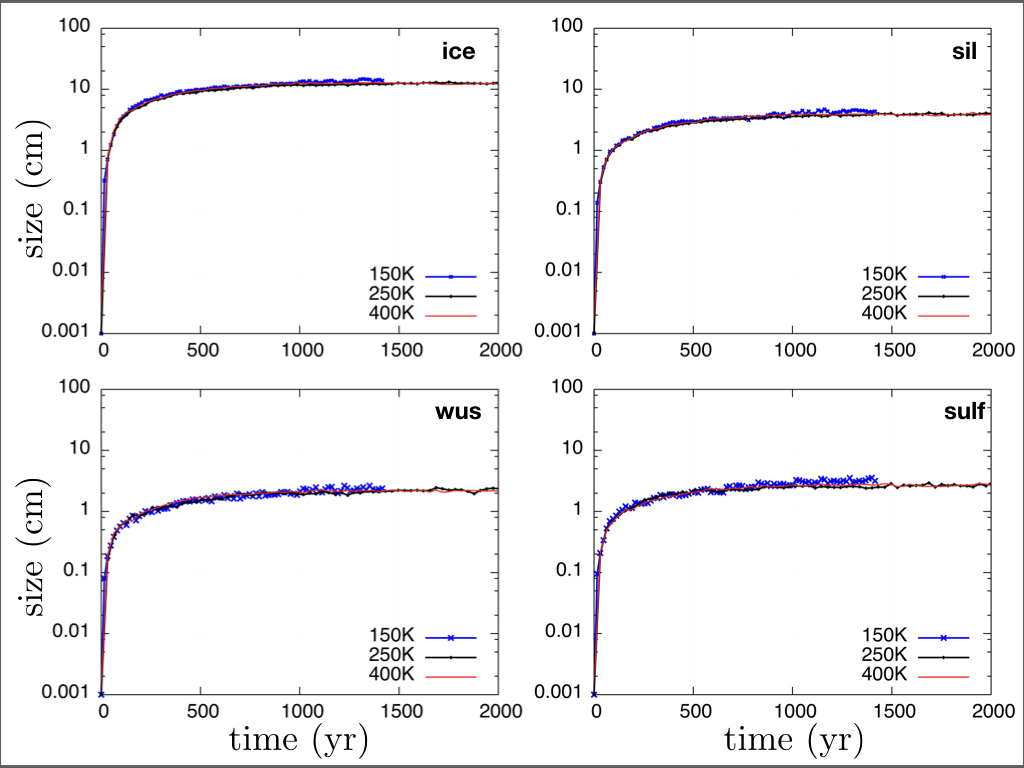}}
{\includegraphics[width=0.95\columnwidth]{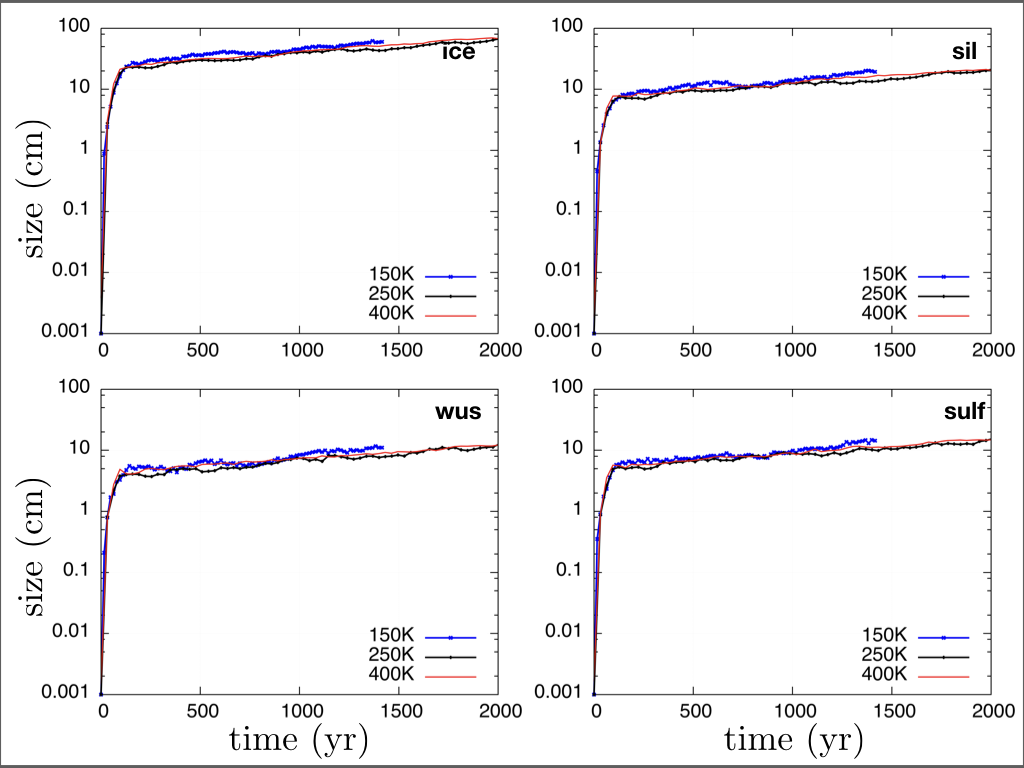}}\\
\caption{Time evolution of the  size for single species in the disc surface (left-box) and midplane (right-box) for a G simulation with 400K particles (red line),  250K particles (black ticked line) and 150K particles (blue crossed-line). \label{comparison03}}
\end{figure*}

\begin{figure*}
{\includegraphics[width=0.8\columnwidth]{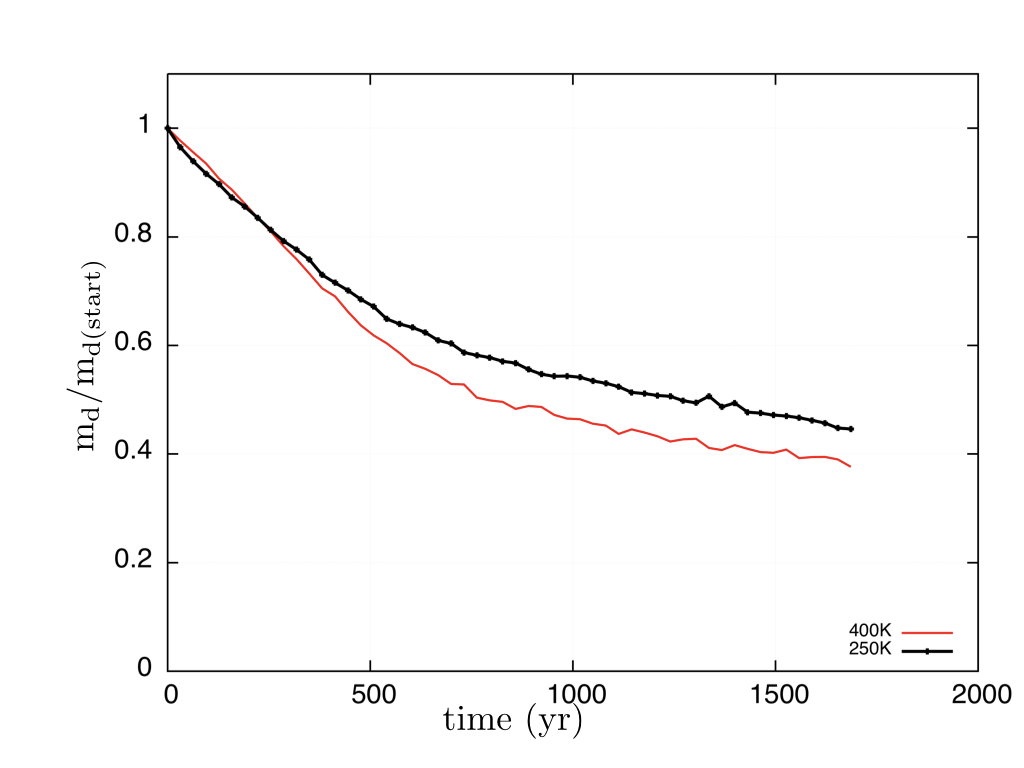}}
{\includegraphics[width=0.8\columnwidth]{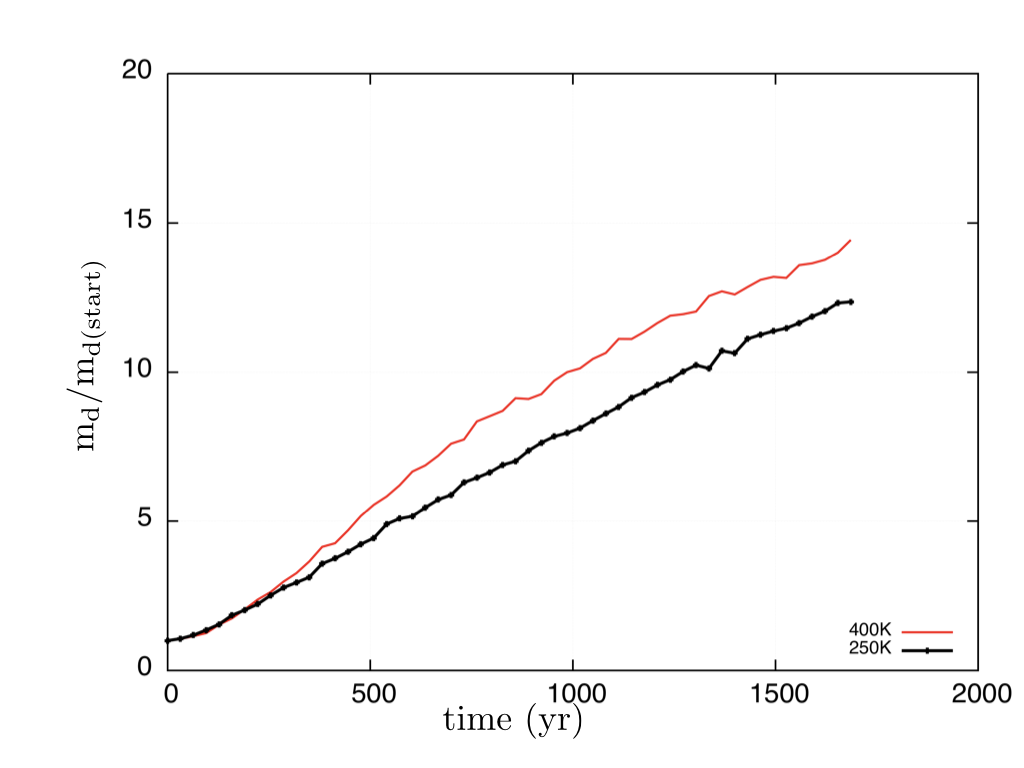}}\\
\caption{Time evolution of the total dust mass content compared to the initial mass in the disc surface (left) and midplane (right) for a GF simulation with 400K particles (red line) and 250K particles (black ticked line). \label{comparison04}}
\end{figure*}

%\begin{figure*}
%{\includegraphics[width=0.95\columnwidth]{appendix009.png}}
%{\includegraphics[width=0.95\columnwidth]{appendix010.png}}\\
%\caption{Time evolution of the   mass content of single species compared to the initial mass in the disc surface (left-box) and midplane (right-box) for a GF simulation with 400K particles (red line) and 250K particles (black ticked line). \label{comparison05}}
%\end{figure*}

\begin{figure*}
{\includegraphics[width=0.95\columnwidth]{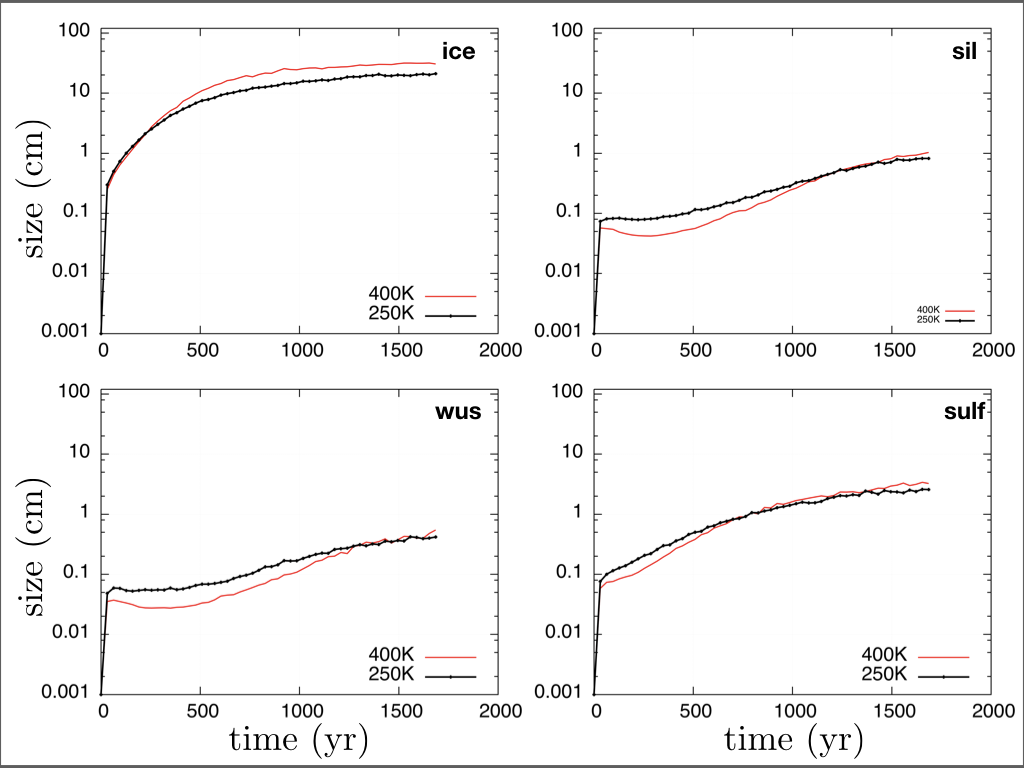}}
{\includegraphics[width=0.95\columnwidth]{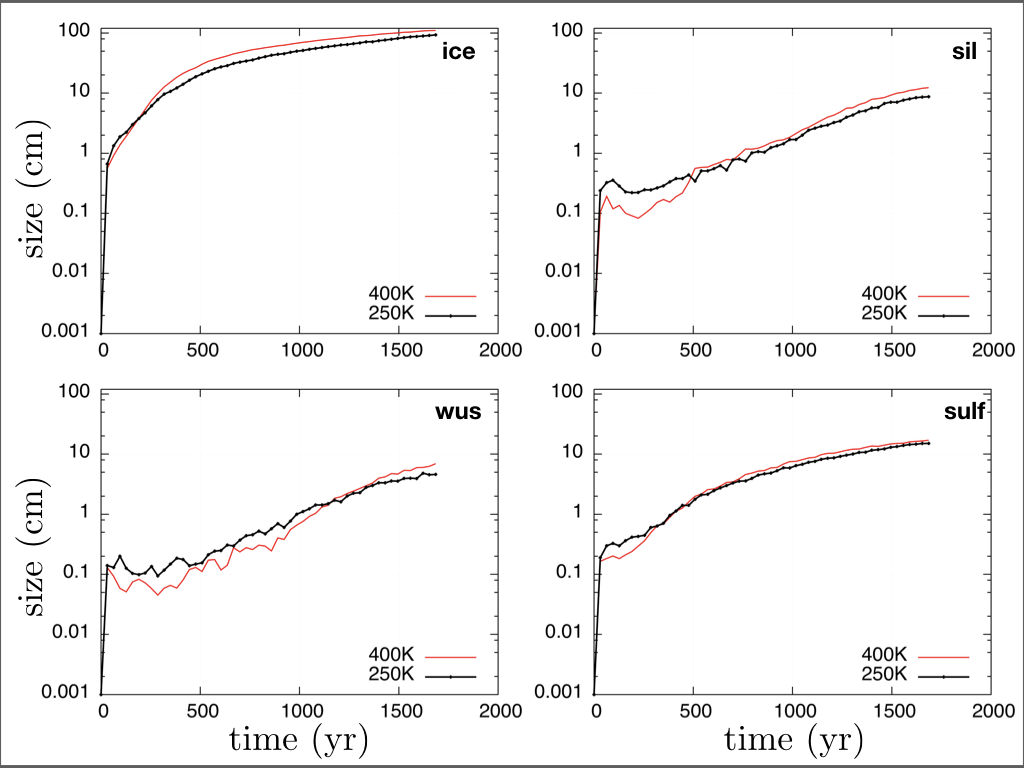}}\\
\caption{Time evolution of the  size for single species in the disc surface (left-box) and midplane (right-box) for a GF simulation with 400K particles (red line) and 250K particles (black ticked line). \label{comparison06}}
\end{figure*}

\begin{figure*}
{\includegraphics[width=0.8\columnwidth]{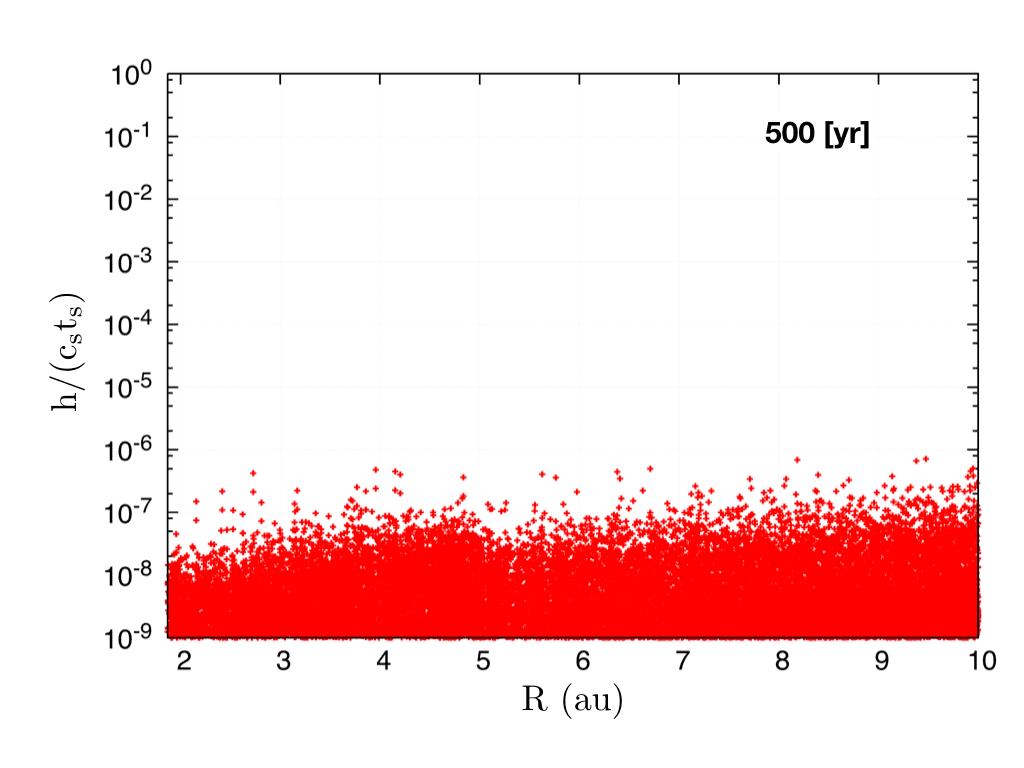}}
{\includegraphics[width=0.8\columnwidth]{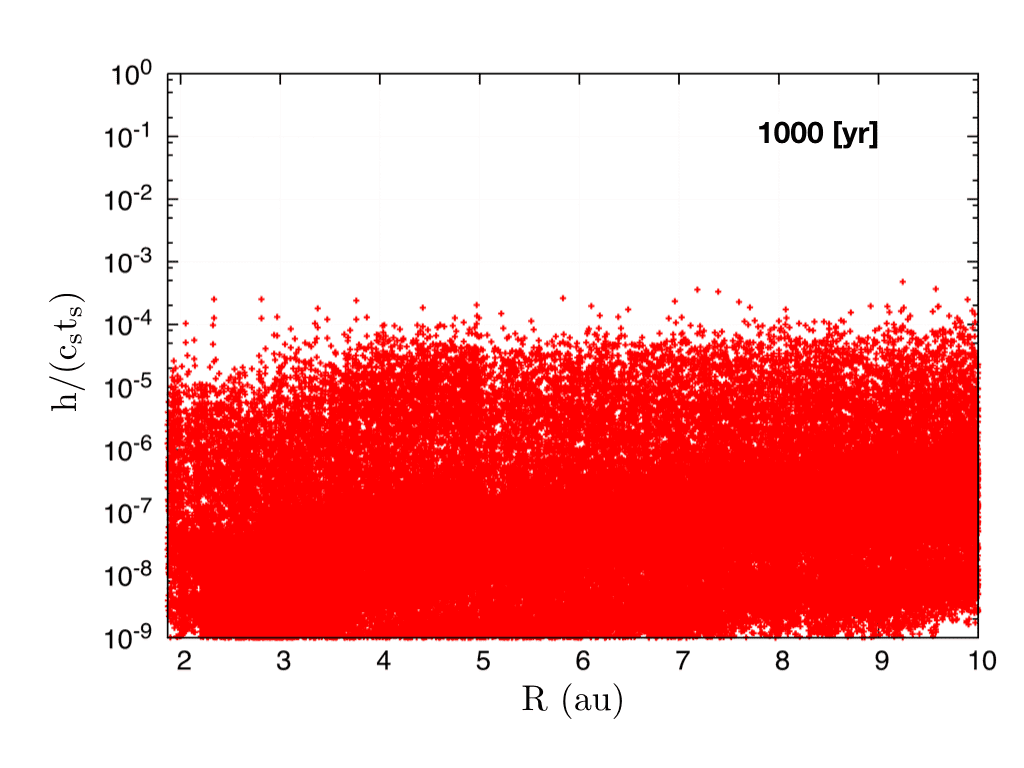}}\\
{\includegraphics[width=0.8\columnwidth]{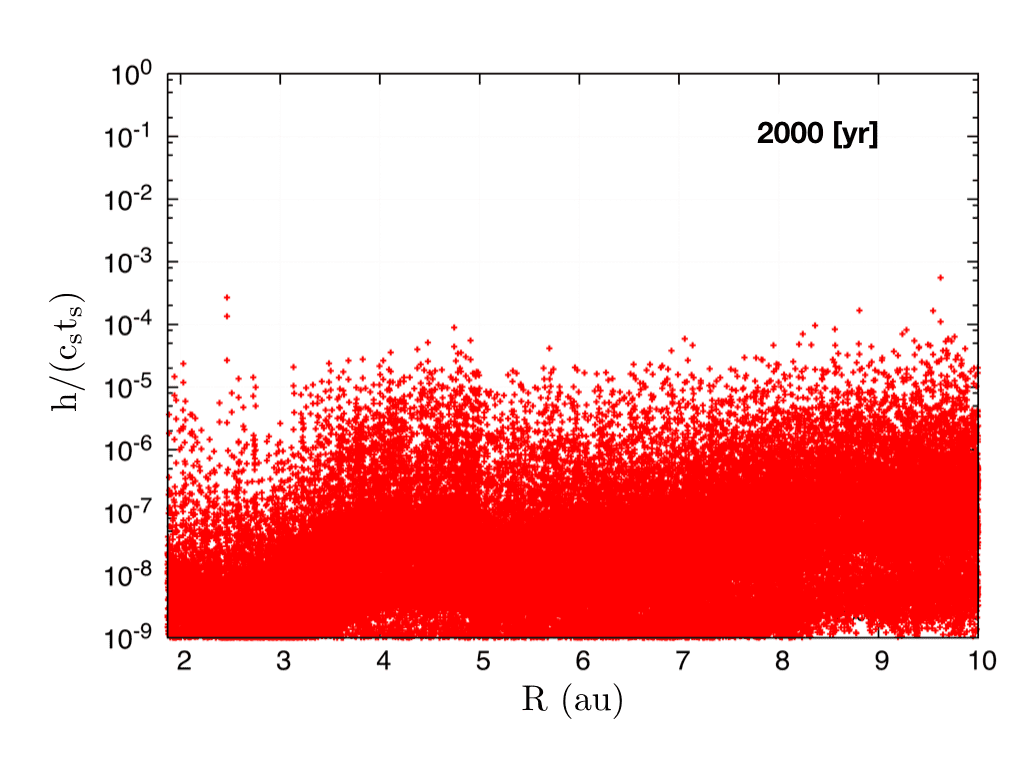}}
{\includegraphics[width=0.8\columnwidth]{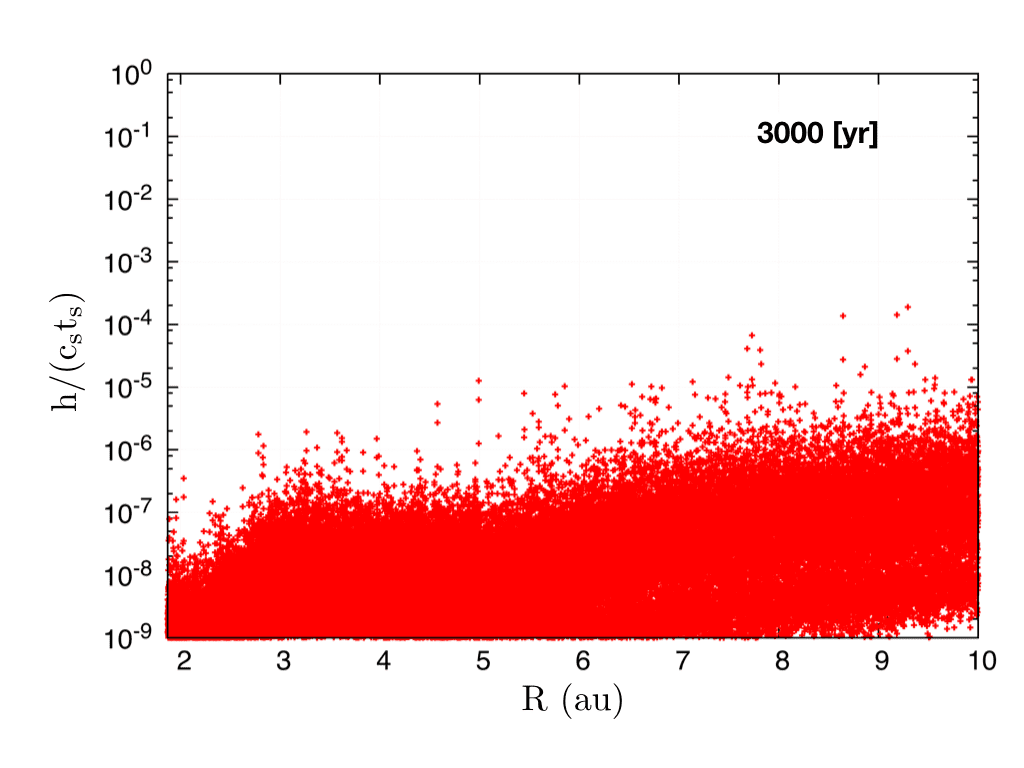}}\\
\caption{{\rm $h/c_{s}t_{s}$} as a function of radius, for GF at the four timesteps reported in the paper. The {\rm $h/c_{s}t_{s}$} for all partices is less then 1 and the resolution criterion by \citet{2012MNRAS.420.2345L} is satisfied.\label{comparison07}}
\end{figure*}

% Don't change these lines
\bsp	% typesetting comment
\label{lastpage}
\end{document}